%% file: manuscript.tex
\documentclass[a4paper,11pt]{article}
\pdfoutput=1 

\usepackage{jheppub} 
\usepackage[sort&compress]{natbib}
\usepackage[T1]{fontenc} 
\usepackage{subcaption}
\usepackage{amsfonts}
\usepackage{slashed}
\usepackage{bbold}
\usepackage{multirow}

\input{tikzdefs.tex}
\title{\boldmath A near-minimal leptoquark model for reconciling flavour anomalies and generating radiative neutrino masses }

\author{Innes Bigaran\footnote{Corresponding author.},}
\author{John Gargalionis,}
\author{Raymond R. Volkas}

\affiliation{ARC Centre of Excellence for Particle Physics at the
  Terascale,\\School of Physics, The University of Melbourne, Victoria 3010, Australia}

\emailAdd{innes.bigaran@unimelb.edu.au}
\emailAdd{garj@student.unimelb.edu.au}
\emailAdd{raymondv@unimelb.edu.au}

\abstract{We introduce two scalar leptoquarks, the SU$(2)_L$ isosinglet denoted $\phi\sim(\mathbf{3}, \mathbf{1}, -1/3)$ and the isotriplet $\varphi\sim(\mathbf{3}, \mathbf{3}, -1/3)$, to explain observed deviations from the standard model in semi-leptonic $B$-meson decays. We explore the regions of parameter space in which this model accommodates the persistent tensions in the decay observables $R_{D^{(*)}}$, $R_{K^{(*)}}$, and angular observables in $b\to s \mu\mu$ transitions. Additionally, we exploit the role of these exotics in existing models for one-loop neutrino mass generation derived from $\Delta L=2$ effective operators. Introducing the vector-like quark $\chi \sim (\mathbf{3}, \mathbf{2}, -5/6)$ necessary for lepton-number violation, we consider the contribution of both leptoquarks to the generation of radiative neutrino mass. We find that constraints permit simultaneously accommodating the flavour anomalies while also explaining the relative smallness of neutrino mass without the need for cancellation between leptoquark contributions. A characteristic prediction of our model is a rate of muon--electron conversion in nuclei fixed by the anomalies in $b \to s \mu \mu$ and neutrino mass; the COMETand Mu2e experiments will thus test and potentially falsify our scenario. The model also predicts signatures that will be tested at the LHC and Belle II.}

\begin{document} 
\maketitle
\flushbottom

\toccontinuoustrue


\section{Introduction}
\label{sec:intro}

The detection of neutrino oscillations establishes that neutrinos have small, but nonzero, masses and that the flavour and mass eigenstates do not coincide. However, the dynamical origin of the tiny scale of neutrino masses remains a mystery, as does any potential impact on the flavour structure of the Standard Model~(SM). A plausible explanation for the lightness of neutrinos is to \emph{not} explicitly introduce a tree-level mass term, but rather to engineer the generation of mass at loop-level: radiative models. We restrict our attention to models that induce a Majorana mass term and violate lepton-number by two units ($\Delta L = 2$). The magnitudes of these masses are naturally loop-suppressed\footnote{For a full review of radiative neutrino mass models see ref.~\cite{Cai:2017jrq}}. As such, this mechanism gives a neat explanation for the disparity between the sizes of neutrino masses relative to those of other SM fermions.
  
Extending the SM to permit neutrino-flavour violation invites a more thorough investigation of the flavour sector. The connection between neutrino physics and flavour physics is of interest given the capacity of precision measurements to explore parameter space for such  beyond-the-SM~(BSM) models. Indeed, a variety of measurements have hinted at violation of Lepton Flavour Universality (LFU) in precision observables: of particular interest are angular parameters~\cite{Aaij:2015oid, Sirunyan:2018jll} and branching ratios in the $b\to s$ transition, and in the ratios $R_{D^{(*)}}$~\cite{Aaij:2017uff,Aaij:2015yra,Lees:2012xj, Lees:2013uzd, Huschle:2015rga,Sato:2016svk, Hirose:2016wfn} and $R_{K^{(*)}}$~\cite{Aaij:2017vbb,Wehle:2016yoi,Aaij:2014ora}.  Such observables are often very sensitive to the virtual effects of exotics, such as those that are typically introduced in radiative neutrino mass models.

In this paper we address both of these problems by developing a radiative neutrino mass model that features BSM contributions to flavour observables. We  begin by outlining the anomalies we aim to address, before detailing a motivation for this model and a summary of related previous work.

\subsection{Neutrino masses and mixing}
Neutrino oscillations in vacuo depend on non-degenerate neutrino masses and lepton flavour mixing. The Pontecorvo-Maki-Nakagawa-Sakata (PMNS) leptonic mixing matrix $\mathbf{U}$ relates the neutrinos in their flavour and mass eigenstates. In the Majorana case, $\mathbf{U}$ can be written as the product of three rotations, the second of which depends on a phase, and a diagonal matrix of phases $\mathbf{P}$:
\begin{equation}
\mathbf{U} = 
\begin{pmatrix}
1 & 0 & 0 \\
0 & c_{23} & s_{23} \\
0 & -s_{23} & c_{23}
\end{pmatrix}
\begin{pmatrix}
c_{13} & 0 & s_{13}e^{-i\delta_{CP}} \\
0 & 1 & 0 \\
-s_{13}e^{i\delta_{CP}} & 0 & c_{13}
\end{pmatrix}
\begin{pmatrix}
c_{12} & s_{12} & 0 \\
-s_{12} & c_{12} & 0 \\
0 & 0 & 1
\end{pmatrix}
\mathbf{P},
\end{equation}
where $c_{ij} \equiv \cos \theta_{ij}$ and $s_{ij} \equiv \sin \theta_{ij}$. For the matrix $\mathbf{P}$ we adopt the convention
\begin{equation}
    \mathbf{P} = 
\begin{pmatrix}
e^{i\alpha_1} & 0 & 0 \\
0 & e^{i\alpha_2} & 0 \\
0 & 0 & 1
\end{pmatrix}.
\end{equation}
An overview of the most recent global fit to these oscillation parameters is given in table~\ref{table:2}, from the NuFIT collaboration~\cite{Esteban:2018azc}. These fits show a preference for the so-called \emph{normal} hierarchy of neutrino mass. This mass-ordering mimics the generational indices: that is to say, $m_1< m_2< m_3$, where $m_i$ represent mass-eigenvalues of the associated combinations of flavour states. An \emph{inverted} ordering is less preferred by fits to data, but still represents a viable alternative regime. We note that the Dirac phase $\delta_{CP}$ is poorly constrained by current data, and the Majorana phases $\alpha_{1,2}$ are entirely unconstrained. We exploit this freedom later in our analysis.

\begin{table}[t]
\begin{center}
\def\arraystretch{1.7} 
 \begin{tabular}{c| c }
 \hline
\textsc{Angular parameters} & \textsc{Squared-mass parameters} ($\text{eV}^2$) \\ [0.5ex] 
 \hline
 \hline
 $\sin^2(\theta_{12})=0.310^{+0.013}_{-0.012}$ & $\Delta m_{21}^2= 7.39^{+0.21}_{-0.20}\times 10^{-5} $\\ 
$\sin^2(\theta_{13})=0.02241^{+0.00065}_{-0.00017}$& $|\Delta m_{31}^2|= 2.525^{+0.033}_{-0.032}\times 10^{-3}  $ \\
$\sin^2(\theta_{23}) =0.580^{+0.017}_{-0.012} $ &  \\
$\delta_{CP} ={ 215^{+40}_{-29}}\; ^{\circ}$ & \\

\hline 
\end{tabular}
\end{center}
 \caption{ Summary of the average neutrino oscillation parameters from the NuFIT collaboration~\cite{Esteban:2018azc}, with the assumption of normal ordering. The angles, $\theta_{ij}$, can be taken  without loss of generality to be within the first quadrant. We implement these central values throughout this work. Majorana phases $\alpha_{1,2}$ are entirely unconstrained. }\label{table:2}
 \end{table}
 \normalsize

\subsection{Anomalies in flavour}

A number of flavour observables appear to indicate flavour-dependent coupling of BSM physics to SM particles. A strong hint of LFU violation exists in experimental measurements~\cite{Ricciardi:2016pmh, Ricciardi:2017lne} of semileptonic $B$-meson decays. A summary of the most recent experimental and theoretical averages for these observables may be found in table~\ref{table:3}.


\subsubsection{Neutral Current Processes}
The anomalies include those in the rare $b\to s$ flavour-changing neutral current (FCNC) transition. Recent measurements~\cite{Aaij:2017vbb,Wehle:2016yoi,Aaij:2014ora, Aaij:2019wad} have reinforced deviations from the SM in the ratios $ R_{K^{*}}$ and $R_{K}$:
\begin{equation}
 R_{K^{(*)}}= \frac{\Gamma( B\rightarrow K^{(*)} \mu^+ \mu^- )}{\Gamma( B\rightarrow K^{(*)} e^+ e^- )}.
\end{equation}
The errors on SM predictions for these ratios are at below the percent-level~\cite{Bordone:2016gaq}, strengthening the argument for serious consideration of the anomalies.
Other branching ratios of exclusive decays in $b \to s \mu \mu$ have also been measured to be in tension with the SM prediction~\cite{Aaij:2014pli, Aaij:2015esa}. Discrepancies are also present in
measurements of angular observables in $B\rightarrow K^{*} \mu^+ \mu^-$, and the most significant of these is seen in the quantity $P_5^\prime$~\cite{Khachatryan:2015isa, Aaij:2015oid, Aaboud:2018krd}. The CMS measurements of these angular observables are consistent with the SM values~\cite{Sirunyan:2017dhj}. Recent fits to the anomalous $b \to s$ data present large preferences for new physics  in operators contributing to $b \to s \mu \mu$~\cite{DAmico:2017mtc, Geng:2017svp, Capdevila:2017bsm, Altmannshofer:2017yso, Ciuchini:2017mik, Hiller:2017bzc, Aebischer:2019mlg, Ciuchini:2019usw, Kowalska:2019ley, Alok:2019ufo}.

\begin{table}[t!]
\begin{center}
\resizebox{\columnwidth}{!}{%
\def\arraystretch{1.9} 
 \begin{tabular}{c| c | c | c} 
 \hline
  & \textsc{SM prediction} & \textsc{Experimental value}& \textsc{Deviation}\\ [0.5ex] 
\hline \hline 
 $R_D$ & $0.299\pm 0.011$~\cite{Aoki:2019cca}  & $0.346 \pm 0.031$~\cite{Amhis:2016xyh} & \multirow{2}{*}{3.1$\sigma$~\cite{GiacomosTalk}}\\ 
$R_{D^*}$ & $0.252 \pm 0.003$~\cite{Aoki:2019cca,Fajfer:2012vx} & $0.301 \pm 0.013$~\cite{ Amhis:2016xyh} &  \\
\hline 
 $R_{K^*}$ (LHCb) &  \multirow{2}{*}{$1.00$~\cite{Aoki:2019cca}} &  $\begin{cases}
0.660^{+0.110}_{-0.070} \pm 0.024, & 0.045~{\text{GeV}^2 < q^2 <1.1~\text{GeV}^2}\\
0.685^{+0.113}_{-0.069} \pm 0.047, &  {1.1~\text{GeV}^2 < q^2 < 6~\text{GeV}^2 }
\end{cases}$~\cite{Aaij:2017vbb} &  $2.4\sigma$  \\
 $R_{K^*}$ (Belle) &  &  $\begin{cases}
0.90^{+0.27}_{-0.21} \pm 0.10, & 0.1~{\text{GeV}^2 < q^2 <8~\text{GeV}^2}\\
1.18^{+0.52}_{-0.32} \pm 0.10, &  {15~\text{GeV}^2 < q^2 < 19~\text{GeV}^2 }
\end{cases}$~\cite{Abdesselam:2019wac} &  $< 2\sigma$  \\
 \hline 
 $R_{K}$ & 1.00~\cite{Aoki:2019cca} & $0.846^{+0.060 \, +0.016}_{-0.054 \, -0.014},\hspace{0.3cm} 1.1~\text{GeV}^2< q^2<6.0~\text{GeV}^2$~\cite{Aaij:2019wad} &  $ 2.5\sigma$  \\ 
 \hline
\end{tabular}%
 }
\end{center}
 \caption{Summary of SM predictions and global averages with statistical and systematic uncertainties. Global averages for $R_D$,  $R_{D^*}$ and $R_K$ are sourced from the Heavy flavour Averaging Group (HFLAV)~\cite{Amhis:2016xyh}, and the LHCb measurement of $R_{K^*}$~\cite{Aaij:2017vbb}. Calculation of the averages and combined deviation from $R_D$ and $R_{D^*}$ will be discussed later in section~\ref{sec:bctaunu}.}
 \label{table:3}
 \end{table}
\normalsize

\subsubsection{Charged Current Processes}
\label{sec:chargedcurrentprocesses}
Persistent tension in the semi-leptonic transition $b \to c \tau \nu$ has been observed independently by many experimental collaborations: through the decay $B\rightarrow D \tau \nu$ by Belle and Babar~\cite{Lees:2012xj, Lees:2013uzd, Sato:2016svk, Hirose:2016wfn,Huschle:2015rga}; and $B\rightarrow D^{*} \tau \nu$ by Belle, Babar and LHCb~\cite{Aaij:2017uff,Aaij:2015yra,Lees:2012xj, Lees:2013uzd,Sato:2016svk, Hirose:2016wfn,Huschle:2015rga}. Each experiment has measured deviations from the SM in the following quantities: 
\begin{equation}
R_{D^{(*)}}= \frac{\Gamma( B\rightarrow D^{(*)} \tau \nu_{\tau} )}{\Gamma( B\rightarrow D^{(*)} \ell \nu_{\ell} )}; \hspace{2cm}\ell \in \{ e, \mu\}.
\end{equation}
Together these measurements amount to a deviation $\gtrsim 3 \sigma$~\cite{GiacomosTalk, Murgui:2019czp} from SM predictions, which can be significantly reduced in the presence of new physics~\cite{Fajfer:2012vx, Murgui:2019czp, Bardhan:2019ljo, Blanke:2018yud, Blanke:2019qrx}. 

Although the ratios $R_{D^{(*)}}$ are our primary concern in this work, we also introduce a number of other observables relevant to the charged current process that form the basis of predictions of our model. Specifically, in section~\ref{sec:results} we present the predicted values for the observables $R_{J/\psi}$, $f_L^{D^*}$, and various tau polarisation asymmetries. The first of these is the ratio of the tauonic mode to the muonic mode for $B \to J/\psi \ell \nu$,
\begin{equation}
    R_{J/\psi} \equiv \frac{\Gamma(B_c \to J/\psi \tau \nu)}{\Gamma(B_c \to J/\psi \mu \nu)},
\end{equation}
measured recently by LHCb to be $R_{J/\psi} = 0.71 \pm 0.17 \pm 0.18$~\cite{Aaij:2017tyk}. Although the ratio is also measured to be enhanced with respect to the SM prediction $R_{J/\psi}^{\text{SM}} \approx 0.25$--$0.29$~\cite{Anisimov:1998uk, Kiselev:2002vz, Ivanov:2006ni, Hernandez:2006gt, Huang:2007kb, Wang:2008xt, Issadykov:2018myx, Wen-Fei:2013uea, Alok:2017qsi, Azatov:2018knx, Hu:2019qcn, Leljak:2019eyw, Azizi:2019aaf}, the central value of the measurement shows a very large effect that cannot be well-accommodated with BSM contributions~\cite{Murgui:2019czp}, although the error bars are very large. The observable $f_L^{D^*}$, the longitudinal polarisation of the $D^*$ in $B \to D^* \tau \nu$, also differs from the SM expectation by $\sim 1.6 \sigma$:
\begin{equation}
    f_L^{D^*} = 0.60 \pm 0.08 \pm 0.04,
\end{equation}
as measured by the Belle collaboration~\cite{Abdesselam:2019wbt}, and has been shown to have good discriminating power for BSM explanations of $R_{D^{(*)}}$. The third class of observables we consider are tau polarisation asymmetries (see ref.~\cite{Asadi:2018sym} for a detailed discussion in the context of explaining $R_{D^{(*)}}$). The polarisation asymmetry in the longitudinal direction of the $\tau$ in the $D^*$ mode has also recently been measured by Belle~\cite{Hirose:2016wfn}:
\begin{equation}
\mathcal{P}_{\tau}^{*} = -0.38 \pm 0.51^{+0.21}_{-0.16} .
\end{equation}
Although the errors are large, the projected Belle II sensitivity at $50 \text{ ab}^{-1}$ for the same observable in the $D$ mode is estimated at about $3\%$~\cite{Alonso:2017ktd}, and we expect the $\mathcal{P}_\tau^{*}$ to be measured even more precisely at Belle II.

\subsection{ Anomalous leptonic magnetic moments}

Precise measurements of the deviation in the semi-classical value of the muon gyromagnetic ratio, $g_\mu=2$, have demonstrated an inconsistency. This is parameterised by the quantity
\begin{align} a_{\mu} \equiv \frac{g_\mu-2}{2}.\end{align} There is a persistent deviation between the SM prediction and the experimentally measured value~\cite{Chapelain:2017syu, Blum:2013xva},
\begin{equation}
\Delta a_\mu= a_\mu^{\text{exp}} - a_\mu^{\text{SM}}= (286\pm63\pm 43) \times 10^{-11},
\end{equation}
corresponding to a  $3.6\sigma$ anomaly. The error values refer to the experimental and theoretical prediction errors, respectively. Similarly, recent experimental results have indicated a deviation from the SM for the electron anomalous magnetic moment, of $2.5\sigma$ significance~\cite{articleParker}.  The leading candidates to explain these anomalies involve flavour-dependent, loop-level, BSM effects~\cite{Blum:2013xva}.

\subsection{Neutrino mass and the one-leptoquark solution}

Effective $\Delta L=2$ interactions involving SM fields were systematically studied by Babu and Leung~\cite{Babu:2001ex} up to mass-dimension~(D) eleven. By opening-up such operators at tree-level, and looping-off external fields, neutrino mass is generated at loop-level: radiative neutrino mass generation. Ref.~\cite{Cai:2014kra} investigated the $D=7$ operators in further detail, assessing the viability of minimal UV-completions for yielding neutrino masses consistent with the observed values. They identified the particle content of such completions, and explored the explicit phenomenology of one particular model: a completion of $\mathcal{O}_{3b}= (L^i Q^j)(L^k d^c)H^l \epsilon_{ij} \epsilon_{kl}$, involving the introduction of a scalar leptoquark~(LQ) field, $\phi$, and an exotic vector-like quark, $\chi$ (`\textsc{Model 2}' in table~\ref{table:1}).

It is important to note that generating neutrino mass in `\textsc{Model 2}' relies explicitly on mixing of the vector-like exotic with the SM $b$-quark. A direct consequence of this is that the mixing parameters are heavily constrained by measurements of $b$ couplings and associated observables~\cite{Aguilar-Saavedra:2013qpa, AguilarSaavedra:2009es}. 
A similar radiative neutrino mass model (`\textsc{Model 1}' in table~\ref{table:1}) containing the isotriplet, instead of the isosinglet, was also identified by ref.~\cite{Cai:2014kra}, although the implications of this model were not thoroughly explored. 

\small
\begin{table}[t!]
\begin{center}
\small
\def\arraystretch{1.5}
\begin{tabular}{ c | c}
\hline
\textsc{Model 1} & \textsc{Model 2} \\
\hline
\hline 
$\varphi\sim (\mathbf{3}, \mathbf{3}, -1/3)$ & $\phi\sim (\mathbf{3}, \mathbf{1}, -1/3)$\\
$\chi \sim (\mathbf{3}, \mathbf{2}, -5/6)$ & $\chi \sim (\mathbf{3}, \mathbf{2}, -5/6)$\\
\hline
\end{tabular}
\caption{ Particle content of two key radiative models identified by ref.~\cite{Cai:2014kra}. The tuple entries refer to transformation properties under $SU(3)_C \otimes SU(2)_{L} \otimes U(1)_Y$, and here we adopt the hypercharge convention $Q=I_3+Y$. } 
\label{table:1}
\end{center}
\end{table}
\normalsize

The isosinglet leptoquark, $\phi$, also features in a study by Bauer and Neubert~\cite{Bauer:2015knc} as a simple explanation for the flavour anomalies $R_{D}$, $R_{D^{*}}$, and the anomalous $b\to s$ data. This model was further studied in refs.~\cite{Becirevic:2016oho, Cai:2017wry, Buttazzo:2017ixm, Angelescu:2018tyl}, where its viability as a combined explanation for the flavour anomalies was evaluated in more detail. An idiosyncrasy of the model is that the charged current $R_{D^{(*)}}$ anomalies are mediated at tree-level, while the $b \to s$ anomalies are explained by box diagrams with the leptoquark in the loop. We have noted the role of this LQ in models of radiative neutrino mass (see, e.g. \cite{Babu:2010vp, Angel:2013hla, Cai:2014kra, Popov:2016fzr}), and the connection between radiative neutrino mass and the flavour anomalies has also been explored more broadly in the literature~\cite{Cai:2017wry, Hati:2018fzc, Singirala:2018mio, Cheung:2017efc, Pas:2015hca,
Dorsner:2017ufx,
Deppisch:2016qqd, Datta:2019tuj, Guo:2017gxp, Popov:2019tyc}. The authors of ref.~\cite{Cai:2017wry} studied the overlap between the flavour anomalies and a two-loop radiative neutrino mass model containing the leptoquark $\phi$. In that model mild tensions exist in explaining both $R_{D^{(*)}}$ and $R_{K^{(*)}}$; at best, it was found that this model could reconcile these anomalies together to within a $2\sigma$ region, if the model was restricted to the minimal particle content -- however the central values cannot both be met. If, however, the isosinglet leptoquark did not contribute to the $b \to s$ anomalies, it was found that a combined explanation of neutrino mass, $(g-2)_\mu$ and $R_{D^{(*)}}$ was possible in the minimal model. 

\subsection{A motivation for near-minimality}

Leptoquarks as BSM candidates have experienced a resurgence of interest in recent years. While use of scalar and vector LQs in constructing models of LFU-violation has a long history, the new measurements, particularly of anomalies, have provided additional motivation for these exotics. Additionally, they can play a pivotal role in models for explaining other SM problems, as detailed in ref.~\cite{Dorsner:2016wpm}. It is also interesting that many LQ models are motivated by unification, as they offer a direct portal between the quark and lepton sectors, and this is one way of motivating non-minimal phenomenological models of the anomalies (see, e.g. Ref.~\cite{Becirevic:2018afm}).

The restricted success of the one-leptoquark solution of Bauer and Neubert motivates us to explore next-to-minimal models to explain these anomalies together. Noting the relevance for flavour observables of introducing $b$-quark mixing to the models presented in table~\ref{table:1}, vector-like fermion extensions to the SM are particularly intriguing. Further, it is known that the interactions of the isotriplet\footnote{The triplet $\varphi$ and singlet $\phi$ are also referred to in the literature as $S_3$ and $S_1$, respectively -- particularly in the review in ref.~\cite{Dorsner:2016wpm}.}  $\varphi$~\cite{Hiller:2014yaa,Kumar:2018era} contribute to $b \to s \mu \mu $ transitions at tree-level, whereas the isosinglet scalar leptoquark only generates loop level contributions, making it much easier to generate a significant BSM correction. We therefore propose a merger of `\textsc{Model 1}'  and `\textsc{Model 2}', to capture the beneficial features of both.

Neutrino oscillations imply a violation of family lepton number, whereas the flavour measurements imply a violation of LFU. This work will aim to explore the connection between these two phenomena, and extend upon earlier work in the field to construct a non-minimal model with a broader scope for explaining deviations from the SM of experimental observations.  

The remainder of this work will be structured as follows. Section~\ref{sec:model} will outline the mathematical structure of the model as a completion of a dimension-7 effective operator for radiative neutrino mass generation. Section~\ref{sec:anomalies} will develop the calculation and framework of contributions to the aforementioned flavour anomalies. Section~\ref{sec:constraints} will proceed to investigate additional relevant flavour constraints and observables, establishing the phenomenology of this model. Section~\ref{sec:results} contains the results from an investigation of parameter space, implementing aforementioned constraints. Section~\ref{sec:con} contains a discussion of implications and prospects of this work.

\section{The Model}
\label{sec:model}
 
Combining `\textsc{Model 1}'  and `\textsc{Model 2}' we arrive at the BSM field content:
\begin{align}
\chi_L \sim(\textbf{3}, \textbf{2}, -5/6)\sim (\chi_1,\chi_2)^T_{L},\label{chi} \;\;\text{and} \;\; \chi_R\sim(\textbf{3}, \textbf{2}, -5/6)\sim (\chi_1,\chi_2)^T_{R},\\
 \phi \sim ({\textbf{3}}, \textbf{1}, -1/3),\hspace{1cm}\text{and} \hspace{0.5cm} \varphi \sim  (\textbf{3}, \textbf{3}, -1/3)\sim (\varphi_1, \varphi_2, \varphi_3)^{T}.
\end{align}
 In the SM gauge basis, the complete set of Yukawa interactions between SM and BSM fields are described by the following Lagrangian:
\begin{align} \mathcal{L}_{BSM} \equiv \mathcal{L}_{\text{int}} - \mathcal{V},
\end{align}
 with the BSM portion of the scalar potential, $\mathcal{V}$, given by:
 \begin{equation}
\begin{aligned}
 \mathcal{V} = \;\; & m_\phi^2|\phi|^2+m_\varphi^2 |\varphi|^2+\lambda_{H\phi} |H|^2 |\phi|^2 +\frac{1}{2}\lambda_\phi |\phi|^4  +\lambda_{\varphi_1} |\varphi|^4  \\ 
&+ \lambda_{\varphi \phi} |\varphi|^2 |\phi|^2 + \lambda_{H\varphi_1} |\varphi|^2 |H|^2+\lambda_{H\varphi_2} [\varphi H]_{\mathbf{2}}[\varphi H]_{\mathbf{2}}+ (\lambda_m H \phi H^\dagger  \varphi^\dagger +\text{h.c}),\label{pot2}
 \end{aligned}
 \end{equation}
 where the notation `$[\;\;]_{\mathbf{i}}$' means that bracketed fields are combined through a tensor product to produce the $\mathbf{i}$-dimensional representation of SU(2)$_L$. Additionally, the set of SM-BSM Yukawa interactions is given by the interaction lagrangian, $\mathcal{L}_{\text{int}}$:
 \begin{equation}
 \begin{aligned}
 \mathcal{L}_{\text{int}} =  m_Q \overline{\chi}\chi+ Y_d  \overline{d_R} H \chi_L  +  &\left(\lambda_{\phi L}\phi^\dagger -\lambda_{\varphi L}\varphi^\dagger\right) \overline{L_L^c} Q_L\\
& + \lambda_R \overline{e_R^c} u_R \phi^\dagger + \left(\lambda_{\chi\phi} \phi-\lambda_{\chi\varphi}\varphi \right)\overline{\chi_R} L_L +\text{h.c} \label{lagrange}
 \end{aligned}
 \end{equation}
We may introduce indices $\{i,j\}$ on Yukawa couplings, to reference the relevant fermion flavour\footnote{Note that these generational indices will occasionally be replaced with their associated particle symbols (e.g. $y_{23} \mapsto y_{\mu b}$). }, for example: 
\begin{equation}
 \begin{aligned}\lambda_{\varphi L}\varphi^\dagger\overline{L_L^c} Q_L \mapsto \lambda_{\varphi L}^{ij}\varphi^\dagger\overline{{L^i_L}^c} Q^j_L. \end{aligned}  \end{equation}
 For clarity we often omit flavour indices, however when present we adopt the convention that for terms with two flavour indices, the first\footnote{Although this is not often the convention in the literature, we choose this so that \emph{Lepto-Quark} can act as a mnemonic, maintaining convention with ref.~\cite{Cai:2014kra}.} refers to the lepton and the second to the quark.

Note that we have imposed global $U(1)_B$ baryon-number conservation by switching off possible di-quark couplings for both leptoquarks. This circumvents bounds from proton stability~\cite{Dorsner:2016wpm,Kovalenko:2002eh}.  

\subsection{Vector-like bottom partner and associated mixing}

An appealing feature of this model, particularly for resolving anomalies in $B$-physics, is mixing of the vector-like fermion $\chi$ with the SM down-type quark. To avoid obvious constraints that would arise from mixing with the lighter flavours, we restrict it to the third generation $b$-quark. This mixing is a result of the term:
\begin{equation}
\mathcal{L}_{int} \supset Y_{d, i}\overline{\hat{d}^i_{R}} \left(H^+\hat{\chi}_2 - H^0 \hat{\chi}_1\right),
\end{equation}
where we have introduced the notation ` $\hat{\;}$ ' to represent the gauge-basis fermion eigenstates. For the remainder of this work we will refer to the only non-zero element of $Y_d$ in this model as $Y_b \equiv Y_{d,3}$.

A feature of the one-loop neutrino mass models derived from dimension-7 operators is that the neutrino mass matrix is proportional to the mass matrix of the SM fermion in the loop~\cite{Cai:2014kra}. In this case, the coupling of the vector-like quark to the $b$ quark dominates, and for this reason we restrict the consideration of the $\hat{\chi}_1$ mixing with the down-type quarks to the $b$ quark. This mixing is chiral and in general quantified by two mixing angles, $\theta_L$ and $\theta_R$, for the left- and right-chiral sectors, respectively:
\begin{align}
\begin{pmatrix}
b_{L/R}\\
{\chi_1}_{L/R}
\end{pmatrix}=
V_{L/R}^\dagger
\begin{pmatrix}
\hat{b}_{L/R}\\
\hat{\chi_1}_{L/R}
\end{pmatrix}, \;\; \text{where} \;\;
V_{L/R} \equiv \begin{pmatrix}
\cos\theta_{L/R} & \sin \theta_{L/R}\\
-\sin \theta_{L/R} & \cos\theta_{L/R}\\
\end{pmatrix}.
\end{align}

The associated mass matrix is defined through
\begin{equation}
\mathcal{L} \;\;\supset \;\;
\begin{pmatrix}
\overline{\hat{b}_L} & \overline{\hat{\chi_1}}_L
\end{pmatrix}
\mathbf{M}
\begin{pmatrix}
\hat{b}_R\\
\hat{\chi_1}_R
\end{pmatrix}
\;\; \text{where}\;\; \mathbf{M}\equiv
\begin{pmatrix}
m_{\hat{b}} & 0\\
m_{b\chi} & m_\chi\\
\end{pmatrix},
\end{equation}
\begin{equation}
\text{with}\hspace{0.5cm} \langle H_0 \rangle \equiv \frac{v}{\sqrt{2}}= 174\text{ GeV},\hspace{0.5cm}  m_{b\chi}\equiv -\langle H_0 \rangle Y_{b},\hspace{0.5cm} m_{\hat{b}}= \langle H_0 \rangle y_b,
\end{equation}
where $y_b$ is the b-quark Higgs Yukawa coupling.
Through singular-value decomposition, the unitary rotation matrices, $V_L$ and $V_R$, rotate the fields into the mass basis:
\begin{equation}
\mathcal{L} \;\;\supset \;\;
\begin{pmatrix}
\overline{\hat{b}_L} & \overline{\hat{\chi_1}}_L
\end{pmatrix}
V_L V_L^\dagger
\mathbf{M}
V_R V_R^\dagger
\begin{pmatrix}
\hat{b}_R\\
\hat{\chi_1}_R
\end{pmatrix}=
\begin{pmatrix}
\overline{b}_L & \overline{\chi_1}_L
\end{pmatrix}
 V_L^\dagger
\mathbf{M}
V_R
\begin{pmatrix}
b_R\\
{\chi_1}_R
\end{pmatrix}.
\end{equation}
We bring the matrices to diagonal form by alternately left- and right-multiplying $V_L^\dagger \mathbf{M} V_R$ with its hermitian conjugate: 
\begin{align}
(V_L^\dagger \mathbf{M} V_R) (V_L^\dagger \mathbf{M} V_R)^\dagger = V_L^\dagger \mathbf{M}\mathbf{M}^\dagger V_L =\text{diag}(m_b^2, m_{\chi_1}^2), \\
(V_L^\dagger \mathbf{M} V_R)^\dagger (V_L^\dagger \mathbf{M} V_R) = V_R^\dagger \mathbf{M}^\dagger\mathbf{M} V_R=\text{diag}(m_b^2, m_{\chi_1}^2).
\end{align}
The diagonalisation produces the following expressions for the mass and mixing parameters, in the limit that $m_\chi \gg m_{bB}$:
\begin{align}
m_b^2= m_{\hat{b}}^2-\frac{m_{\hat{b}}^2 m_{b\chi}^2}{m_\chi^2- m_{\hat{b}}^2},\hspace{2cm}
m_{\chi_1}^2= m_{\chi}^2+ \frac{m_{\chi}^2 m_{b\chi}^2}{m_\chi^2- m_{\hat{b}}^2},
\end{align}

\begin{equation}
\label{eq:bmixing}
\sin\theta_L= \frac{m_{b\chi}m_{\hat{b}}}{m_{\chi}^2-m_{\hat{b}}^2},\hspace{2cm}
\sin\theta_R= \frac{m_{b\chi} m_\chi}{m_{\chi}^2-m_{\hat{b}}^2}.
\end{equation}
\normalsize

\subsection{Structure of eigenbasis mapping}

To calculate flavour observables, we first transform into the charged-fermion mass eigenbasis. Beginning with the SM fields, where $\{i,j,l\}$ represent flavour indices, and defining $\mathfrak{L}_a$, $\mathfrak{R}_b$ as unitary rotations\footnote{The SM quark-sector Cabibbo-Kobayashi-Maskawa (CKM) matrix is defined in terms of these rotations as $\mathbf{V} \equiv \mathfrak{L}_{u}^\dagger\mathfrak{L}_{d}$ under this convention.} between the gauge and mass bases, we have that
\begin{equation}
\begin{aligned}
&\hat{d}_{R,i}\mapsto [\mathfrak{R}_d]_{il} {\hat{d}}_{R, l},\hspace{2cm} u_{R,i}\mapsto[\mathfrak{R}_u]_{il}{u}_{R,l},\hspace{2cm} e_{R,i}\mapsto[\mathfrak{R}_e]_{il} {e}_{R,l},\\
&\hat{d}_{L,i}\mapsto[\mathfrak{L}_d]_{il} {\hat{d}}_{L,l},\hspace{2.15cm}e_{L,i}\mapsto[\mathfrak{L}_e]_{il} {e}_{L,l},\hspace{2.2cm} \nu_{L,i}\mapsto [\mathfrak{L}_e]_{il}\breve{\nu}_{L,l},\\
&u_{L,i}\mapsto[\mathfrak{L}_u]_{il}{u}_{L,l}.
\end{aligned}
\end{equation}
\normalsize
Above, the ` $\hat{\;}$ ' on the down-type quarks and exotics indicate that they are yet to be fully rotated into the mass basis -- i.e. the mixing is yet to be incorporated. Here $\breve{\nu}$ represents the neutrino weak-eigenstate, which is related to the neutrino mass-eigenstate basis via the usual PMNS matrix. We parameterise the PMNS matrix by the central values quoted in the NuFit collaboration 2018 global fit (table~\ref{table:2})~\cite{Esteban:2018azc}, and in our numerical analysis we scan over values for the Majorana phases.

Redefining the coupling constants to absorb this transformation, the interaction Lagrangian becomes: 
  \begin{equation}
 \begin{split}
 \mathcal{L}_{int} = \;\;&
\left(y^{\chi \varphi}\overline{\chi_{2,R}}\breve{\nu}_{ L}+x^{L\varphi}\overline{e^{C}_L} \hat{d}_{L} \right){\varphi}_3^{\dagger}+\left(y^{\chi \varphi}\overline{\hat{\chi}_{1,R}}e_L- y^{L\varphi}\overline{\breve{\nu}^{C}_L} u_{L}\right)\varphi_1^{ \dagger}\\
&+ \frac{1}{\sqrt{2}}y^{\chi \varphi}\left(\overline{\hat{\chi}_{1,R}}\breve{\nu}_{ L}-\overline{\chi_{2,R}}e_{L}\right)\varphi_2+\frac{1}{\sqrt{2}}\left(y^{L\varphi}\overline{e^{C}_L} u_{L}+x^{L\varphi}\overline{\breve{\nu}^{C}_L} \hat{d}_{L}\right)\varphi_2^{  \dagger}\\
  & +\left(x^{L\phi} \overline{\breve{\nu}^C_L} \hat{d}_L - y^{L\phi} \overline{e^C_L} u_L+y^{R\phi}\overline{e^C_{R}} u_{R}\right) \phi^\dagger+y^{\chi \phi}\left(\overline{\hat{\chi}_{1,R}}\breve{\nu}_L+\overline{\chi_{2,R}} e_L\right) \phi + \text{h.c.} \label{lagrangian-1}
 \end{split}
 \end{equation}
The correspondences between the couplings are given by the following redefinitions: 
 \begin{equation}
\begin{aligned}
y^{\chi \phi} \equiv \lambda_{\chi \phi} \mathfrak{L}_e,\\
y^{\chi \varphi} \equiv \lambda_{\chi \varphi} \mathfrak{L}_e,\\
y^{L\varphi} \equiv \lambda_{\varphi L} \mathfrak{L}_e^T \mathfrak{L}_u,\\
y^{L\phi} \equiv \lambda_{\phi L} \mathfrak{L}_e^T \mathfrak{L}_u,\\
y^{R \phi} \equiv \lambda_R \mathfrak{R}_e^T \mathfrak{R}_u.
\end{aligned}
\end{equation}
in terms of unphysical mixing matrices. The matrices $y^{L \varphi}$ and $y^{L \phi}$ are related by the physical CKM matrix to $x^{L \varphi}$ and $x^{L \phi}$, respectively, so they are not independent:
\begin{align}
\label{eq:ckmmixing}
x^{L \varphi} \equiv y^{L \varphi} \mathbf{V}, \;\; \text{and}\;\; x^{L \phi} \equiv y^{L \phi} \mathbf{V}.
\end{align}
The following holds for the mass-mixing of SM with BSM fields:
\begin{equation}
\begin{aligned}
 \hat{d}_{i, R/L} &= d_{i, R/L}; \;\; i=1,2,\hspace{2cm}
\hat{d}_{3, R/L}= c_{\theta_{R/L}} b_{R/L} +s_{\theta_{R/L}} {\chi_1}_{R/L},\\
\hat{\chi_2}_{R/L}&={\chi_2}_{R/L},\hspace{3.6cm} \hat{\chi_1}_{R/L}=-s_{\theta_{R/L}} b_{R/L} +c_{\theta_{R/L}} {\chi_1}_{R/L}.\end{aligned}\end{equation}
Note that we have denoted the mass-eigenstate by $\chi_1$ rather than the oft-used `$B$' to avoid later confusion when discussing $B$-meson decays.

At this point, it is important to explicitly note that some quartic field couplings in the scalar potential, $\mathcal{V}$ (eq.~\eqref{pot2}), will  be set to zero, for simplicity, in subsequent calculations -- in particular, those which generate $\varphi$-$\phi$ mixing and mass-splitting between triplet components\footnote{A discussion of the structure of this scalar mixing may be found in appendix \ref{appendix1}.}, $\lambda_m$ and $\lambda_{H\varphi _2}$ in ~\ref{pot2}. For the remainder of this work we will take that each of the isotriplet components is degenerate in mass, $m_\varphi$.

\begin{figure}[t]
\begin{centering}
   \begin{minipage}{.4\textwidth}
        \centering
    \begin{tikzpicture}
      \scriptsize
      \node[vtx, label=-120:$\nu_i$] (i1) at (0,0) {};
      \node[vtx] (v1) at (1.7,0) {};
      \node[vtx] (v2) at (3.7,0) {};
       \node[vtx, label=90:$\varphi^\dagger/\phi^\dagger$] (i3) at (2.7,1) {};
      \node[vtx, label=-60:$\nu_j^C$] (o1) at (5.4,0) {};
      \node[vtx, label=-90:$b/{\chi_1}$] (i4) at (2.7,0) {};
      \graph[use existing nodes]{
      i1 --[fermion] v1;
      v1--[fermion]v2;
      o1 --[fermion] v2;
      };  
       \draw[thick, dashed, postaction={decorate, decoration={markings, mark=at position .5 with {\arrow[>=latex]{<}}}}] (1.7,0) arc (180:0:1);
    \end{tikzpicture}
    \end{minipage}
      \hspace{1cm}
    \begin{minipage}{0.4\textwidth}
        \centering
    \begin{tikzpicture}[scale=0.9]
      \scriptsize
      \node[vtx, label=-120:$\nu_i$] (i1) at (0,0) {};
      \node[vtx] (v1) at (1.7,0) {};
      \node[vtx] (v2) at (3.7,0) {};
      \node[vtx, label=-60:$\nu_j^C$] (o1) at (5.4,0) {};
       \node[vtx, label=90:$\varphi/\phi$] (i3) at (2.7,1) {};
        \node[vtx, label=-90:$\overline{b}/\overline{\chi_1}$] (i4) at (2.7,0) {};
      \graph[use existing nodes]{
      i1 --[fermion] v1;
      v2--[fermion]v1;
      o1 --[fermion] v2;
      };  
       \draw[thick, dashed,postaction={decorate, decoration={markings, mark=at position .55 with {\arrow[]{latex}}}}]  (1.7,0) arc (180:0:1);
    \end{tikzpicture}   \end{minipage}
    \caption{Dominant one-loop contributions to radiative neutrino mass generation, completions of operator $\mathcal{O}_3$. Arrow direction indicates the flow of fermion-number.}\label{fig:1}
    \end{centering}
\end{figure}
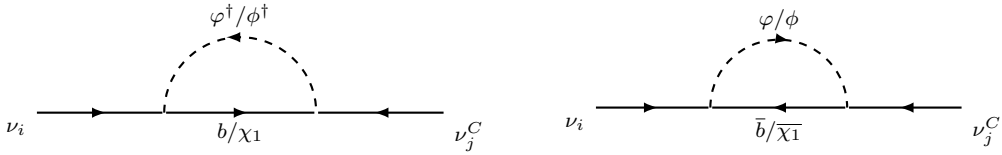
 \subsection{Massive neutrinos: a tale of two regimes}
 \label{sec:neutreg}
The leading-order contributions to neutrino masses are from the one-loop diagrams in figure~\ref{fig:1}. After EWSB, we calculate the radiatively generated neutrino mass matrix, in the limit that $m_b \ll m_{\chi}$, to be: 
\begin{equation}
\begin{aligned}
(m_\nu)_{ij}&= \frac{3m_{\chi_1} m_b }{16\pi^2}m_{b\chi}\times \\
&\quad \left[ 2(x_{i3}^{L\phi}y_j^{\chi\phi}+y_i^{\chi\phi*}x_{j3}^{L\phi*})\frac{ \ln(\frac{m_{\chi_1}}{m_\phi})}{m_{\chi_1}^2-m_\phi^2}+(x_{i3}^{L\varphi}y^{\chi\varphi}_j+y_i^{\chi\varphi*}x_{j3}^{L\varphi*})\frac{\ln(\frac{m_{\chi_1}}{m_\varphi})}{m_{\chi_1}^2-m_\varphi^2}\right], \label{neutrino_mass}
\end{aligned}
\end{equation}
where the relative factor of two arises from the Clebsch-Gordan coefficients in eq.~\ref{lagrangian-1}. 

 Upon inspection of eq.~\ref{neutrino_mass}, we note that there are two copies of similarly structured contributions: one from $\varphi$, and one from $\phi$. In our analysis we consider, for simplicity, two distinct phenomenological regimes:
\begin{description}
\item[Regime 1:] The contribution to the neutrino masses comes solely from the isotriplet leptoquark $\varphi$ (corresponding to \textsc{Model 1}).
\item[Regime 2:] The contribution to the neutrino masses comes solely from the isosinglet LQ $\phi$ (corresponding to \textsc{Model 2}).
\end{description}
In each regime, the alternate LQ is still important for the flavour anomalies, but contributions to neutrino mass generation are subdominant. We leave a full exploration of the overlap of these phenomenological regimes, and the implications of LQ mixing, to future work.

\subsubsection{Implementing a Casas-Ibarra-like parameterisation}

Deriving the physical neutrino masses $\{ m_1, m_2, m_3\}$ requires that we diagonalise eq.\eqref{neutrino_mass}, via the standard PMNS matrix $\mathbf{U}=(\mathbf{u_1}, \mathbf{u_2}, \mathbf{u_3})$ :
\begin{align}
\text{diag}(m_1, m_2, m_3)=U^{T}_{il} (m_\nu)_{lk}U_{kj},\label{dia}
\end{align}
Expanding this relationship, we can express the neutrino mass matrix as a \emph{rank-2} matrix in terms of the low-energy parameters $U_{ij}$ and $m_i$:
\begin{align}
\mathbf{m_\nu}= \mathbf{U}^{\dagger}\text{diag}(m_1, m_2, m_3)\mathbf{U}^*= m_2\mathbf{u_2^\dagger u_2^*}+ m_k\mathbf{u_k^\dagger u_k^*}.\label{umass}
\end{align}

The value of $k$ depends on the hierarchy assumption for SM neutrinos. In each case, the lightest neutrino is taken to be massless in order to obtain the rank-two flavour structure of each term in eq.~\eqref{neutrino_mass}. For definiteness, we adopt normal ordering and set $k=3$ in what follows. We use the notation $\eta \in \{ \phi, \varphi\}$ to denote the LQ that contributes to neutrino mass generation:
\begin{align}
\mathbf{m_\nu}= m_0 \left(\mathbf{x_{L\eta}}\mathbf{y^\dagger_{\chi \eta}} + \mathbf{y_{\chi \eta}}\mathbf{x^\dagger_{L\eta}}\right), \hspace{1cm}
m_0= (1+\delta)\frac{3m_{b\chi}m_b m_{\chi_1}}{16\pi^2({m_{\chi_1}^2-m_{\eta}^2})} \ln\left(\frac{m_{\chi_1}}{m_{\eta}}\right),\label{CI}
\end{align}
where $\delta=0$ for $\eta \equiv \varphi$, and $\delta=1$ for $\eta \equiv \phi$. The matrices $\mathbf{x_{L\eta}} $ and $ \mathbf{y_{\chi \eta}}$ are column matrices of Yukawa couplings. Under the specified assumptions, couplings of the form
\begin{equation}
     \mathbf{x_{L\eta}} \sim x^{L \eta}_{j3}\;\; \text{and}\;\; \mathbf{y_{\chi\eta}} \sim  y^{\chi \eta}_i,\hspace{1cm}  i,j\in \{e, \mu, \tau\},
\end{equation}
are mutually constrained by neutrino oscillation measurements. 

We adopt a Casas--Ibarra-like procedure~\cite{Casas:2001sr} and parameterise our ignorance of the coupling constants by introducing a parameter $\zeta \in \mathbb{C}$ through re-expressing eq.~\eqref{CI} as
\begin{align}
\mathbf{m_\nu}= \frac{m_0}{2}\left[ \left(\frac{\mathbf{x_{L\eta}}}{\zeta}+\zeta\mathbf{ y^\dagger_{\chi\eta}} \right)\left(\frac{\mathbf{x_{L\eta}}}{\zeta}+\zeta\mathbf{ y^\dagger_{\chi\eta}} \right)^T -\left( \frac{\mathbf{x_{L\eta}}}{\zeta}- \zeta\mathbf{ y^\dagger_{\chi\eta}}\right)\left(\frac{\mathbf{x_{L\eta}}}{\zeta}- \zeta\mathbf{ y^\dagger_{\chi\eta}}  \right)^T\right],\label{zeta1}
\end{align}
where, upon expansion, $\zeta$ cancels out. This means that the neutrino oscillation parameters can be fitted for any value of $\zeta$. Matching eq.~\eqref{zeta1} to eq.~\eqref{umass}, we obtain expressions for the coupling matrices in terms of $\zeta$, $m_i$ and $\mathbf{u}_i$:
\begin{subequations}
\label{eq:yuknu}
\begin{align}
\label{eq:yuknux}
\mathbf{x_{L\eta}} &= \frac{\zeta}{\sqrt{2m_0}}(\sqrt{m_2}\mathbf{u_2}^*+i\sqrt{m_3}\mathbf{u_3}^*) , \\
\mathbf{y_{\chi\eta}^\dagger} &= \frac{1}{\zeta\sqrt{2m_0}}(\sqrt{m_2}\mathbf{u_2}^*-i\sqrt{m_3}\mathbf{u_3}^*). \label{eq:yuknuy}
\end{align}
\end{subequations}
Parameterising the Yukawa couplings in terms of $\zeta$ enables us to efficiently scan over the parameter space that agrees with the oscillation measurements. The coupling values are inputted at the high-energy scale -- consistent with the energy-scale of the LQ mediator. 

Upon assuming normal ordering as detailed above, the measurements of physical mass-squared values can be translated into measurements of $m_i^2$ (table \ref{table:2}):
\begin{align}
|\Delta m_{21}^2|\approx m_2^2= 7.39^{+0.21}_{-0.20}\times 10^{-5}\;\text{eV}^2, \hspace{0.5cm} |\Delta m_{31}^2|\approx m_3^2 =2.525^{+0.033}_{-0.032}\times 10^{-3}\;\text{eV}^2 .
\end{align}
To demonstrate that our model fits neutrino mass, we simply need to show that we can assign reasonable values of $\zeta$ and $m_0$, within appropriate limits defined by flavour-violating processes and perturbativity of the generated couplings. Imposing a perturbativity bound $p$ on the couplings constrains $\zeta$ and $m_0$ such that $\forall j$:
\begin{equation}
\begin{aligned}
\label{eq:zetaperta}
\left|x_{j3}^{L \eta}\right| \leq p  &\implies
\left|{\zeta}\right|\left|(\sqrt{m_2}U_{j2}^*+i\sqrt{m_3}U_{j3}^*)\right|\leq \sqrt{2m_0}\;p , \\
|y_{j3}^{\chi \eta}| \leq p  &\implies
 \frac{1}{|\zeta|}\left|(\sqrt{m_2}U_{j2}^*-i\sqrt{m_3}U_{j3}^*)\right|\leq \sqrt{2m_0}\;p. 
\end{aligned}
\end{equation}
 We impose these constraints with $p = \sqrt{4 \pi}$ in our analysis. The remaining couplings in this model remain free parameters to be assigned values in accordance with constraints in subsequent chapters.

\section{Ameliorating Anomalies}
\label{sec:anomalies}

In section \ref{sec:neutreg} we identified the couplings in each mass regime which are fixed by the Casas-Ibarra parameterisation. To address the remaining goals of this model, it remains to calculate the corrections to the anomalous processes outlined in section \ref{sec:intro}: $ R_{K^{(*)}},R_{D^{(*)}}$ and $(g-2)_\mu$. To parameterise the BSM contributions to these processes we frame our constraints in terms of effective operators, $\mathcal{O}_{i}$, weighted by the Wilson coefficients $C_i$, such that the effective lagrangian at a particular energy scale is given by: 
\begin{align}
\mathcal{L}_{\text{effective}}= \sum_{i}(C^{\text{BSM}}_i+C^{\text{SM}}_i)  \mathcal{O}_i\label{reference_eff}
\end{align}
The set $\{ \mathcal{O}_i\}$ represents an operator basis that encompasses the interactions of this model at low energy, usually corresponding to below the mass scale of the BSM mediator. There are a number of commonly used EFT bases, so we will be careful to specify the basis of interest and coefficient normalisation as we proceed with our discussion of constraints. Where these are relevant for application in computational procedures, they will be referenced in accordance with the Wilson Coefficient exchange format (WCxf)~\cite{Aebischer:2017ugx}.

\subsection{\texorpdfstring{$b \to s \mu \mu$: $R_{K}$ and $R_{K^{*}}$}{bsmumu: Rk and RKstar}}

\begin{figure}[t!]
\centering
 \begin{tikzpicture}
  \node[vtx, label=180:$b$] (i1) at (0,1) {};
  \node[vtx, label=0:$\mu^+$] (i2) at (3,2) {};
  \coordinate[] (v1) at (1.5,1) {};
  \coordinate[] (v2) at (3.0,0) {};
  \node[vtx, label=0:$\mu^-$] (o1) at (4.5,1) {};
  \node[vtx, label=0:$s$] (o2) at (4.5,-1) {};
  \graph[use existing nodes]{
     i1 --[fermion] v1;
     i2 --[fermion] v1;
     v2 --[fermion] o2 ;
     v2--[fermion] o1;
     v1 --[cscalar, edge label'=$\varphi_3$] v2;
  };
  \end{tikzpicture}
 \caption{Dominant model contribution to the neutral current $b\to s\mu \mu$ process}\label{bsmumu}
\end{figure}
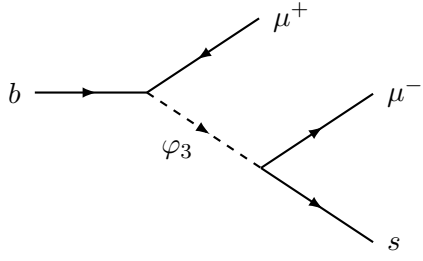

The leading-order contribution from our model to the $b\to s\mu\mu$ transition is given by the isotriplet, $\varphi$, via the tree-level diagram shown in figure~\ref{bsmumu}. The isosinglet $\phi$ also contributes a one-loop box contribution to this process, as detailed in refs.~\cite{Bauer:2015knc,Cai:2017wry}.

Using identities summarised in appendix \ref{appendix2}, the diagram in figure~\ref{bsmumu} corresponds to a BSM contribution to the effective operator $\mathcal{O}^{\mu\mu}_{LL}$: 
\begin{equation}
    \mathcal{O}^{\mu\mu}_{LL}=(\overline{s}\gamma^\mu  P_L b)(\overline{\mu}\gamma_\mu P_L \mu ).
\end{equation}
 Typically, fits to the available experimental data on the $b\to s$ decays~\cite{Altmannshofer:2017fio} involve the chiral-basis containing $\mathcal{O}^{\mu\mu}_{LL}$, and the related operator $\mathcal{O}^{\mu\mu}_{LR}$: 
$$\mathcal{O}^{\mu\mu}_{LR}=(\overline{s}\gamma^\mu P_L b)(\overline{\mu}\gamma_\mu P_R \mu ).$$
These find a good fit to the data so long as the following expressions are satisfied~\cite{Aebischer:2019mlg}:
\begin{equation}
{C}^{\mu\mu}_{LL}\approx  -1.06 \times \left( \frac{4 G_F}{\sqrt{2}} V_{tb} V_{ts}^* \frac{\alpha}{4\pi}\right), \hspace{1cm} {C}^{\mu\mu}_{LR}\approx  0,
\end{equation}
where the multiplier corresponds to normalisation for this particular EFT, quoted in this way to ensure consistency with eq.~\eqref{reference_eff}. This choice of coefficients eliminates the tensions in $R_{K^{(*)}}$ and significantly improves the SM fit to all of the $b \to s$ data, with a total pull from the SM of $6.5 \sigma$~\cite{Aebischer:2019mlg}.
 
As proposed in ref.~\cite{Bauer:2015knc}, the isosinglet-only model provides a one-loop contribution to these operators, and in general generates non-zero values for both ${C}^{\mu\mu}_{LL}$ and ${C}^{\mu\mu}_{LR}$. In this case, fitting ${C}^{\mu\mu}_{LR} \approx 0$ requires positing a suppression of the appropriate right-handed Yukawas $\mathbf{y}^{R\phi}$. In contrast, the leading-order contribution generated by the $SU(2)$-triplet $\varphi$ is at tree-level, and has  ${C}^{\mu\mu}_{LR} \approx 0$ at the high scale. 

For consistency with the literature, we translate the quoted operators to the so-called `\texttt{flavio}' basis for the Weak Effective Theory (WET), as outlined in ref.~\cite{Aebischer:2017ugx}. Implementing the Fierz transforms as quoted in Appendix B, the operator mapping for $\mathcal{O}^{\mu\mu}_{LL}$ is:
\begin{equation}
\mathcal{O}^{\mu\mu}_{LL} = \frac{1}{2} \left(\mathcal{O}^{\mu\mu}_{9}-\mathcal{O}^{\mu\mu}_{10} \right),
\end{equation}
with the operators $\mathcal{O}^{\mu\mu}_{9}$ and $\mathcal{O}^{\mu\mu}_{10}$ defined as: 
\begin{align}
\mathcal{O}_9^{\mu\mu} &\equiv \frac{4 G_F}{\sqrt{2}} V_{tb} V_{ts}^* \frac{\alpha}{4\pi} \left(\overline{s} \gamma^\mu P_L b \right)\left( \overline{\mu} \gamma_\mu \mu \right),\\
\mathcal{O}_{10}^{\mu\mu} &\equiv \frac{4 G_F}{\sqrt{2}} V_{tb} V_{ts}^* \frac{\alpha}{4\pi} \left(\overline{s} \gamma^\mu P_L b \right)\left( \overline{\mu} \gamma_\mu \gamma_5 \mu \right).
\end{align}
We label the corresponding BSM contributions\footnote{ From here onwards, reference to the Wilson coefficients will be in the context of BSM contributions. To avoid overcrowding of sub- and super-scripts, they will not explicitly include a superscript `BSM'.} to these operators ${C}^{\mu\mu}_9$ and ${C}^{\mu\mu}_{10}$. They are generated at tree-level by the $\varphi$ LQ with coefficients
\begin{align}
\label{eq:c9c10}
{C}^{\mu\mu}_9= -{C}^{\mu\mu}_{10} = -\frac{c_{\theta_L} \pi}{2\sqrt{2}G_F\alpha}\frac{1}{V_{tb}V_{ts}^*}\frac{x_{23}^{L\varphi}x_{22}^{L\varphi*}}{m_\varphi^2}.
\end{align}
Analogous to the fits quoted above, we take the central value for ${C}^{\mu\mu}_9 =-{C}^{\mu\mu}_{10}$ as 
\begin{equation}
\label{c9c10eq}
{C}^{\mu\mu}_9 =-{C}^{\mu\mu}_{10} \approx -0.53,
\end{equation}
with a one-sigma region of  $[-0.62, -0.45]$~\cite{Aebischer:2019mlg}. In this model, this corresponds to a central value of:
\begin{equation}
c_{\theta_L} |x_{23}^{L\varphi} x_{22}^{L\varphi*}|\approx  \left(\frac{m_\varphi}{24 \text{ TeV}}\right)^2.
\end{equation}
Note that the couplings above are derived from purely real values for the Wilson coefficients, and most fits assume real values for the Wilson coefficients. To ensure a ${C}^{\mu\mu}_9=-{C}^{\mu\mu}_{10}$ consistent with these, we fix the value of $C^{\mu\mu}_9$ such that $\text{Im}~C^{\mu\mu}_9 \approx 0$. In Regime 2, this corresponds to scanning over real-only values for the two free-parameters $x_{22}^{L\varphi*}$ and $x_{23}^{L\varphi}$, whereas in Regime 1 this corresponds to constraints on the Casas-Ibarra parameterisation. These will be discussed in section~\ref{sec:leptonic}.

\subsection{\texorpdfstring{$b \to c \tau \nu$: $R_{D}$ and $R_{D^{*}}$}{bctaunu: RD and RDstar}} \label{sec:bctaunu}

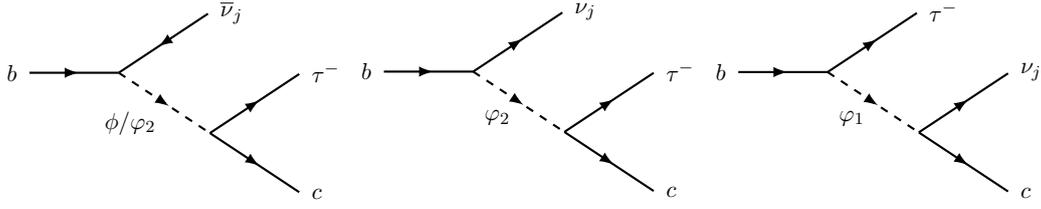
\begin{figure}[t!]
\begin{centering}
   \begin{minipage}{.3\textwidth}
\begin{tikzpicture}[thick,scale=0.8, every node/.style={transform shape}]
  \node[vtx, label=180:$b$] (i1) at (0,1) {};
  \node[vtx, label=0:$\overline{\nu}_j$] (i2) at (3,2) {};
  \coordinate[] (v1) at (1.5,1) {};
  \coordinate[] (v2) at (3.0,0) {};
  \node[vtx, label=0:$\tau^-$] (o1) at (4.5,1) {};
  \node[vtx, label=0:$c$] (o2) at (4.5,-1) {};
  \graph[use existing nodes]{
     i1 --[fermion] v1;
     i2 --[fermion] v1;
     v2 --[fermion] o2;
     v2 --[fermion] o1;
     v1 --[cscalar, edge label'=$\phi/\varphi_2$] v2;
  };
  \end{tikzpicture}
    \end{minipage}
       \begin{minipage}{.3\textwidth}
\begin{tikzpicture}[thick,scale=0.8, every node/.style={transform shape}]
  \node[vtx, label=180:$b$] (i1) at (0,1) {};
  \node[vtx, label=0:${\nu}_j$] (i2) at (3,2) {};
  \coordinate[] (v1) at (1.5,1) {};
  \coordinate[] (v2) at (3.0,0) {};
  \node[vtx, label=0:$\tau^-$] (o1) at (4.5,1) {};
  \node[vtx, label=0:$c$] (o2) at (4.5,-1) {};
  \graph[use existing nodes]{
     i1 --[fermion] v1;
     v1 --[fermion] i2;
     v2 --[fermion] o2;
     v2 --[fermion] o1;
     v1 --[cscalar, edge label'=$\varphi_2$] v2;
  };
  \end{tikzpicture}
    \end{minipage}
           \begin{minipage}{0.3\textwidth}
\begin{tikzpicture}[thick,scale=0.8, every node/.style={transform shape}]
  \node[vtx, label=180:$b$] (i1) at (0,1) {};
  \node[vtx, label=0:$\tau^-$] (i2) at (3,2) {};
  \coordinate[] (v1) at (1.5,1) {};
  \coordinate[] (v2) at (3.0,0) {};
  \node[vtx, label=0:${\nu}_j$] (o1) at (4.5,1) {};
  \node[vtx, label=0:$c$] (o2) at (4.5,-1) {};
  \graph[use existing nodes]{
     i1 --[fermion] v1;
     v1 --[fermion] i2;
     v2 --[fermion] o2;
     v2 --[fermion] o1;
     v1 --[cscalar, edge label'=$\varphi_1$] v2;
  };
  \end{tikzpicture}
    \end{minipage}
        \caption{Dominant model contributions to the charged-current $b\to c\tau \nu$ process, \emph{not} assuming a global conservation of lepton flavour, where $j \in \{e, \mu, \tau \}$.}\label{bctaunu}
\end{centering}
\end{figure}

The contribution from this model to the $b\to c \tau \nu$ transition is described by three diagrams, as illustrated in figure~\ref{bctaunu}. The dominant BSM effects manifest in contributions to the following\footnote{The Fierz transformations used to derive these operators from the diagram in figure~\ref{bctaunu} are given in appendix \ref{appendix2}. } effective operators, expressed in the \texttt{flavio} basis for the Weak Effective Theory (WET):
\begin{align}
\mathcal{O}_{S_L,j}& \equiv -\frac{4G_F}{\sqrt{2}}V_{cb}(\overline{c} P_L b)(\overline{\tau}P_L \nu^j),\\
\mathcal{O}_{V_L,j}& \equiv -\frac{4G_F}{\sqrt{2}}V_{cb}(\overline{c}\gamma^\mu P_L b)(\overline{\tau}\gamma_\mu P_L\nu^j),\\
\mathcal{O}_{T,j}& \equiv -\frac{4G_F}{\sqrt{2}}V_{cb}(\overline{c} \sigma^{\mu\nu} P_L b)(\overline{\tau}\sigma_{\mu\nu}P_L\nu^j)
\end{align}
The neutrino index, here denoted $j$, can run over all three flavours -- the leptoquark interactions need not conserve lepton flavour, and final-state neutrino flavour is rarely a direct observable, particularly in collider studies. Consequentially, the BSM contributions to the operator coefficients, calculated at the LQ mass scale, are as follows:\begin{align}
{C}_{S_L,j}=& \frac{\sqrt{2}c_{\theta_L}}{4G_FV_{cb}}\frac{1}{2}\left( \frac{x_{j3}^{L\phi}y_{32}^{R\phi*}}{m_\phi^2} \right), \label{eq:csl}\\
{C}_{V_L,j}=& \frac{\sqrt{2}c_{\theta_L}}{4G_FV_{cb}}\frac{1}{2}\left(\frac{x_{j3}^{L\phi}y_{32}^{L\phi*}}{m_\phi^2}- \frac{x_{j3}^{L\varphi}y_{32}^{L\varphi*}}{2m_\varphi^2} \right),\label{eq:cvl}\\
{C}_{T,j}=&-\frac{1}{4}{C}_{S_L,j}. \label{eq:ct}
\end{align}
We will often drop the neutrino-flavour index $j$ to avoid unnecessary clutter when $j=3$, the case that leads to constructive interference with the SM contribution.

In the limit of small mixing between $\chi$ and the SM $b$-quark, these Wilson coefficients are generated solely from the leftmost diagram in figure~\ref{bctaunu}. The other diagrams are subject to suppression\footnote{In both cases the topmost vertex originates from a nonzero coupling $y^{\chi \varphi}_{i}$, therefore the effective coupling to the $b$-quark after SM-BSM mixing is proportional to $\sin \theta_R$.  }  by $\sin \theta_R \approx 0$. (See section~\ref{sec:bBmix} for a discussion of the constraints on mixing parameters).

Whilst the QCD Ward identity implies that the vector coupling ${C}_{V_L,j}$ does not run with energy scale, the relationship (multiplier) between the tensor and scalar couplings will change dramatically with energy-scale running. This will be accounted for in subsequent calculations.

The values required for these Wilson coefficients to give a good fit to $R_D$ and $R_{D^*}$ have been studied in the literature, typically under the assumption of lepton-flavour conservation. Existing fits~\cite{Cai:2014kra} suggest a good match to data can be attained with contributions to the vector operator $C_{V_L}$. Contributions in the direction $C_{S_L} = -4 C_{T}$~\cite{Cai:2017wry,Angelescu:2018tyl,Tanaka:2012nw} can also provide a good fit to data, and this approach is subject to fewer constraints (See section~\ref{sec:constraints}).

\begin{figure}
    \centering
    \includegraphics[width=0.49\linewidth]{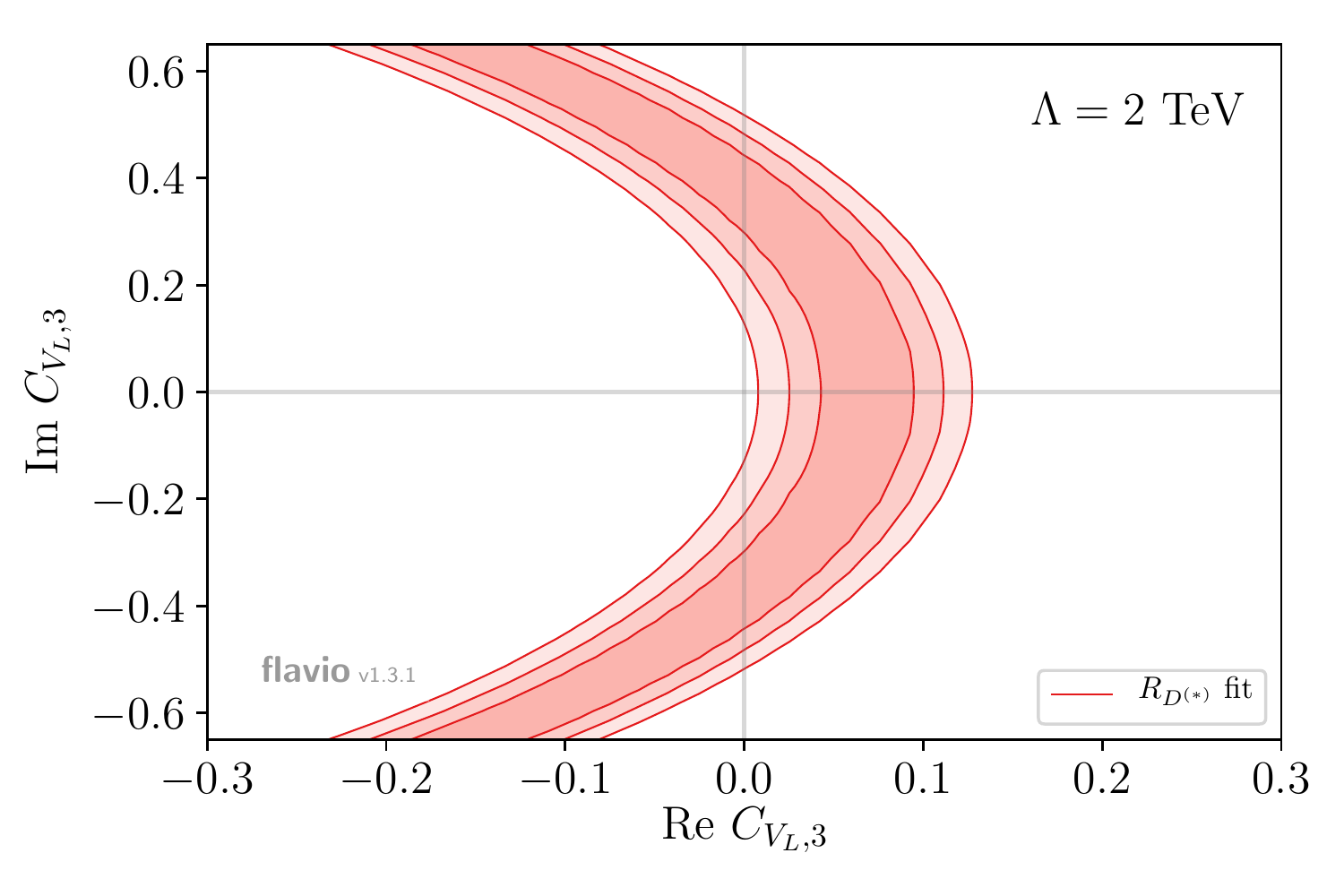}
    \includegraphics[width=0.49\linewidth]{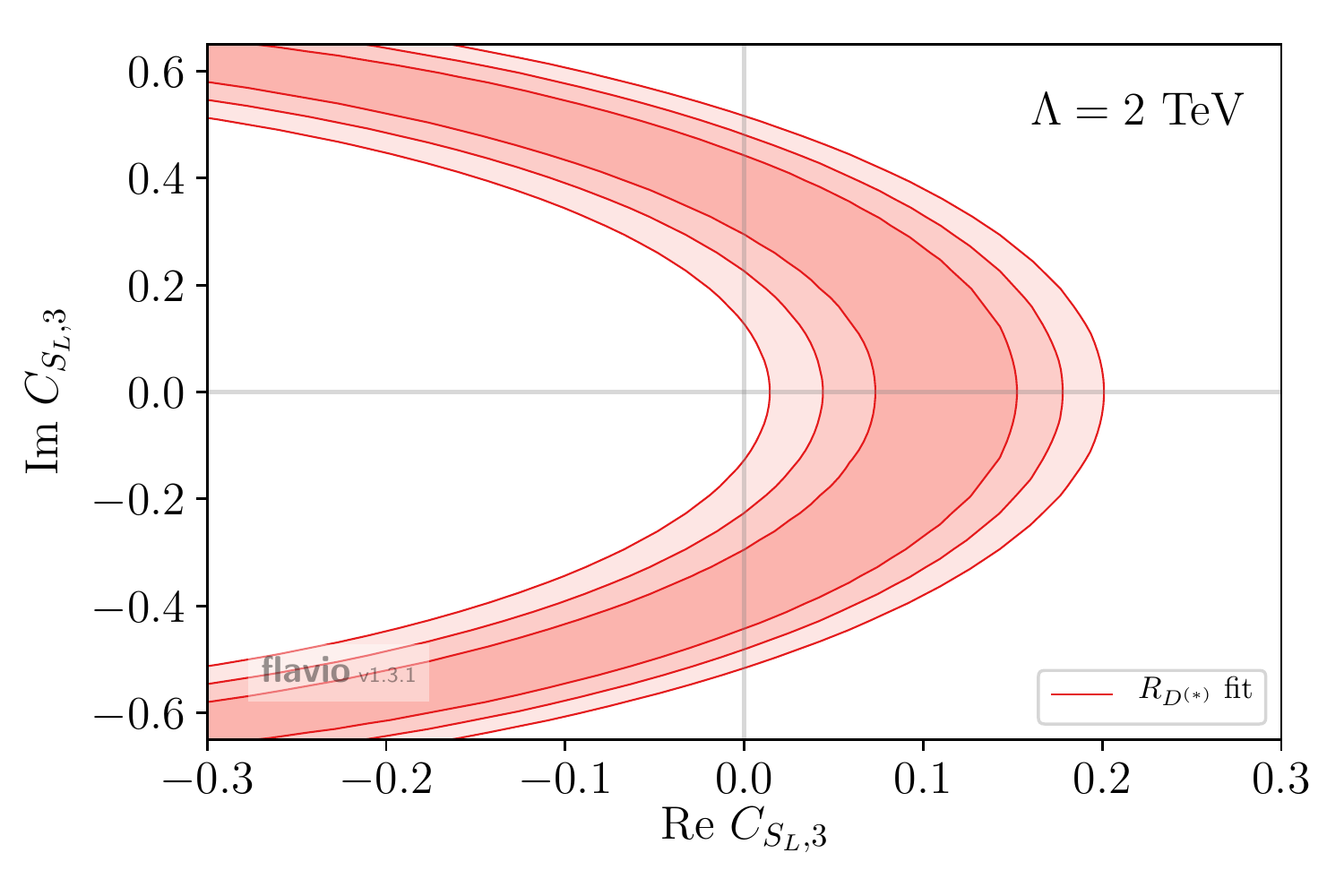}
    \includegraphics[width=0.49\linewidth]{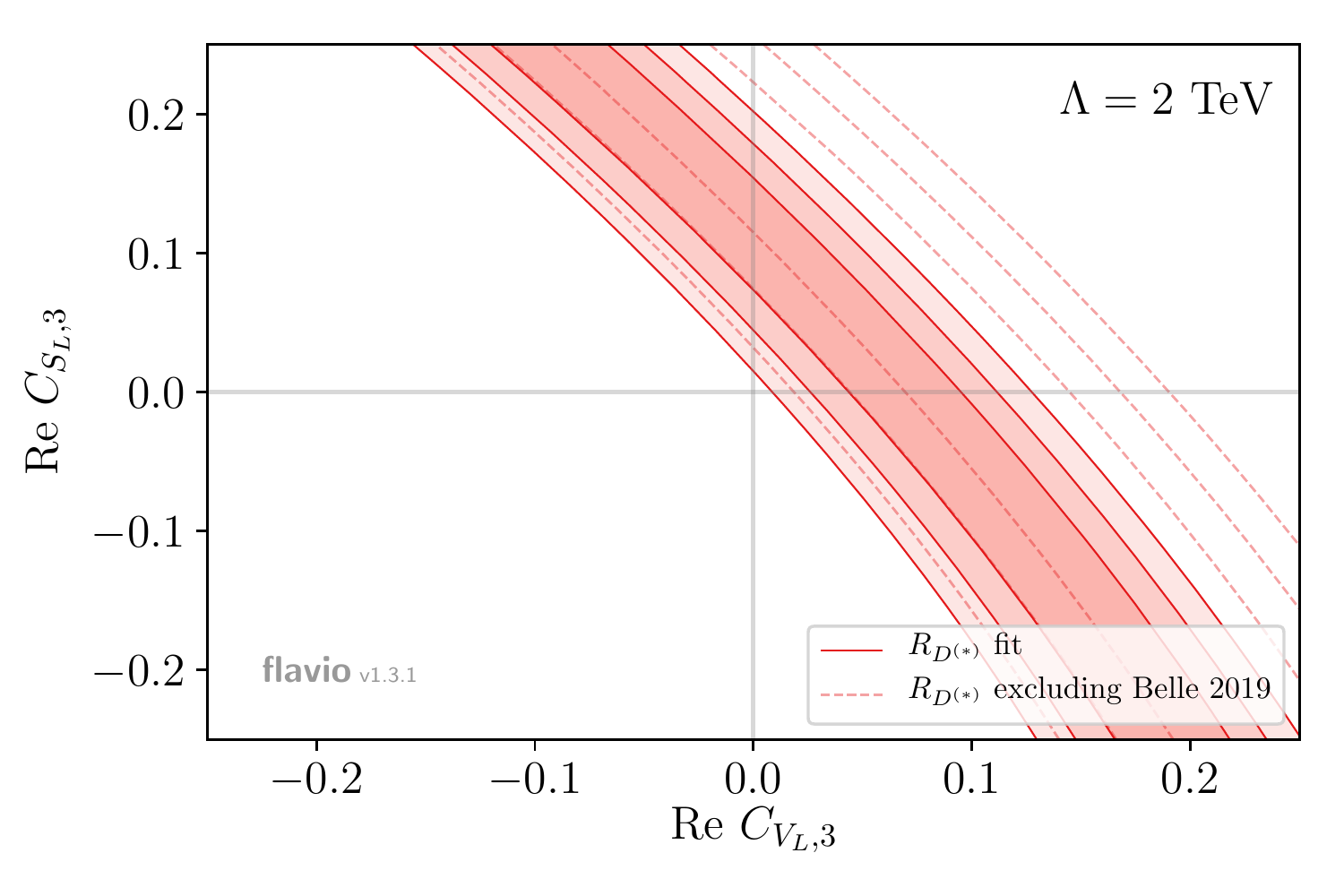}
    \caption{The results of our fit to $R_D$ and $R_{D^*}$ including the new Belle measurement~\cite{GiacomosTalk}. Contours show the $1$, $2$ and $3\sigma$ regions of the fit, dashed lines show the fit results without the recent Belle measurement. The scalar and tensor coefficients are run to the $b$-quark mass scale from $2 \text{ TeV}$. See Table~\ref{tab:fitresults} for central values and the text for more details.}
    \label{fig:fit}
\end{figure}

We incorporate the new Belle combined measurement~\cite{GiacomosTalk} into a fit of all measurements of $R_{D}$ and $R_{D^*}$ using the fitting software \texttt{flavio}~\cite{Straub:2018kue}\footnote{We note that our fit does not include the measurements of $f_L^{D^*}$ and $R_{J/\psi}$, since errors here are still large. Instead, we take the central values from our fits and discuss predictions for these observables in section~\ref{sec:results}.}. The fit contours are shown in figure~\ref{fig:fit}, with the fit excluding the new Belle measurement shown with dashed contours to indicate its effect. We find the best-fit point 
\begin{equation}
({C}_{V_L}, {C}_{S_L}) \approx (-0.18, 0.36),
\end{equation}
for the 2D fit to $\text{Re} C_{V_L}$ and $\text{Re} C_{S_L} (\Lambda) = -4 C_T(\Lambda)$ at $\Lambda = 2 \text{ TeV}$. We also fit to $C_{V_L}$ with $C_{S_L}(\Lambda)=0$ and vice versa. These results are summarised in table~\ref{tab:fitresults}. We comment here that in this model, for the isosinglet LQ $\phi$ contributing to the direction $C_{S_L}(\Lambda) = - 4 C_T(\Lambda)$, the vector operator will also always be non-zero. This follows from the relation in eq.~\eqref{eq:ckmmixing}. The leading contribution where only $x_{33}^{L\phi}$ is non-zero is suppressed by $|V_{ts}| \approx 0.04$, but this contribution can still be sizeable if $x_{33}^{L\phi}$ is chosen to be large. A short discussion of this scenario is included in our phenomenological analysis (see section~\ref{sec:vectorOp}).
\begin{table}[t]
    \centering
    \begin{tabular}{c||c|c|c}
       Fit & Best fit & $1\sigma$ region & $2\sigma$ region \\
       \hline
       $C_{V_L}$  & $0.069$ & $[0.044, 0.094]$ & $[0.026, 0.11]$ \\
       $C_{S_L}$ & $0.14$ & $[0.077, 0.15]$ & $[0.047, 0.18]$ \\
       $(C_{V_L}, C_{S_L})$ & $(-0.18, 0.36)$ & --- & --- \\
    \end{tabular}
    \caption{The results of our fit to $R_D$ and $R_{D^*}$ including the new Belle combined measurement~\cite{GiacomosTalk}. The first row shows the best fit point and $\sigma$-regions fitting to $C_{V_L}$ with all other operator coefficients vanishing. The second row shows the same for $C_{S_L}(\Lambda) = -4 C_T(\Lambda)$ for $\Lambda = 2 \text{ TeV}$ and all other coefficients set to zero. The third row shows the best fit point for a 2D fit to $\text{Re}\, C_{V_L}$ and $\text{Re}\, C_{S_L}(\Lambda)= -4 \text{Re}\, C_T(\Lambda)$, again for $\Lambda = 2 \text{ TeV}$.}
    \label{tab:fitresults}
\end{table}

\newpage
\subsection{Leptonic magnetic moments: \texorpdfstring{$(g-2)_\ell$}{gminus2mu}}
 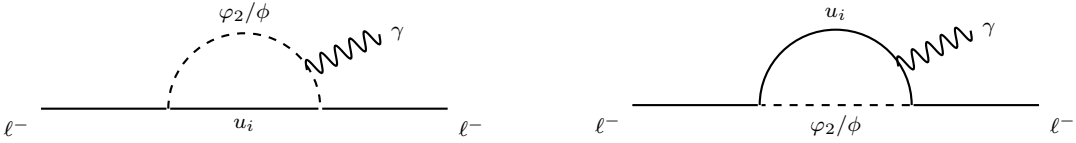
\begin{figure}[t]
\centering
    \begin{tikzpicture}
      \scriptsize
      \node[vtx, label=-120:$\ell^-$] (i1) at (0,0) {};
      \node[vtx] (v1) at (1.7,0) {};
      \node[vtx] (v2) at (3.7,0) {};
       \node[vtx] (v3) at (3.5,0.5) {};
        \node[vtx,label=0:$\gamma$] (v4) at (4.5,1) {};
      \node[vtx, label=-60:$\ell^-$] (o1) at (5.4,0) {};
       \node[vtx, label=90:$\varphi_2/\phi$] (i3) at (2.7,1) {};
        \node[vtx, label=-90:$u_i$] (i4) at (2.7,0) {};
      \graph[use existing nodes]{
      i1 --[thick] v1;
      v1--[thick]v2;
      o1 --[thick] v2;
      v3 --[photon]v4;
      };  
       \draw[thick, dashed]  (1.7,0) arc (180:0:1);
    \end{tikzpicture}  
    \hspace{1cm}
    \centering
    \begin{tikzpicture}
      \scriptsize
      \node[vtx, label=-120:$\ell^-$] (i1) at (0,0) {};
      \node[vtx] (v1) at (1.7,0) {};
      \node[vtx] (v2) at (3.7,0) {};
       \node[vtx] (v3) at (3.5,0.5) {};
        \node[vtx,label=0:$\gamma$] (v4) at (4.5,1) {};
      \node[vtx, label=-60:$\ell^-$] (o1) at (5.4,0) {};
       \node[vtx, label=90:$u_i$] (i3) at (2.7,1) {};
        \node[vtx, label=-90:$\varphi_2/\phi$] (i4) at (2.7,0) {};
      \graph[use existing nodes]{
      i1 -- [thick]v1;
      v2--[scalar]v1;
      o1 --[thick] v2;
      v3 --[photon]v4;
      };  
       \draw[thick]  (1.7,0) arc (180:0:1);
    \end{tikzpicture}
    \caption{ Dominant contributions to the lepton magnetic moment from BSM content in this model. Fermion lines are omitted as there are multiple valid assignments possible for these topologies, each of which will be considered in calculations.}  \label{fig:6}
    \end{figure}
There are two viable leading-order diagrams for this correction (figure~\ref{fig:6}). Since contributions from these diagrams are well-established in the literature~\cite{Bauer:2015knc, Dorsner:2016wpm} we simply quote and interpret the results. For the muon magnetic moment, in the limit that $m_\varphi^2, m_\phi^2 \gg m_t^2$, the contributions are given by:
\begin{equation}
a_\mu^\phi= \sum_i \frac{m_\mu m_{u_i}}{4\pi^2 m_\phi^2}\left(\frac{7}{4}- \ln\frac{m_\phi^2}{m_{u_i}^2}\right)\text{Re}(y^{R\phi}_{2i} y^{L\phi}_{2i}) - \frac{m_\mu^2}{32\pi^2 m_\phi^2} \left[\sum_i |y^{L\phi}_{2i}|^2+ \sum_i |y^{R\phi}_{2i}|^2 \right] \label{momentgeneral}
\end{equation}

\begin{equation}
\label{eq:amu}
a_\mu^\varphi=  - \frac{m_\mu^2}{32\pi^2 m_\varphi^2}\sum_i |y^{L\varphi}_{2i}|^2
\end{equation}
As $m_\mu \ll m_{t}$, the like-chirality, terms are small --- leading to the requirement of non-vanishing right-chiral couplings to obtain an adequate fit to the anomaly. We are left with the contribution generated by $\phi$ in this limit:

\begin{equation}
a_\mu^\phi\sim \frac{m_\mu m_{t}}{4\pi^2 m_\phi^2}\left(\frac{7}{4}- \text{ln}\frac{m_\phi^2}{m_{t}^2}\right)\text{Re}(y^{R\phi}_{23} y^{L\phi}_{23}).
\end{equation}

\noindent By a similar argument, we arrive at the contribution to the electron anomalous magnetic moment: 
\begin{equation}
a_e^\phi\sim \frac{m_e m_{t}}{4\pi^2 m_\phi^2}\left(\frac{7}{4}- \text{ln}\frac{m_\phi^2}{m_{t}^2}\right)\text{Re}(y^{R\phi}_{13} y^{L\phi}_{13}).
\end{equation}

Through the nature of generating these contributions from both right- and left-handed couplings, it is a key characteristic of this model that opposite sign contributions can be generated for the electron and muon\footnote{As can be seen in equation (\ref{momentgeneral}), the like-chiral corrections allow no freedom with assigning a direction to the corrections, as they are proportional to the modulus-squared of a Yukawa coupling. }. In both neutrino mass regimes \emph{at least} one of the two necessary Yukawa couplings is a free parameter unconstrained by contributions to the other anomalies outlined in this section. For the remainder of this work we will focus on the parameter space required to correct the muon anomalous magnetic moment, as it is the more persistent and significant discrepancy. We leave further discussion of corrections to the electron magnetic moment in this model to future work.

\section{Constraints}
\label{sec:constraints}

Below we discuss the constraints relevant to our model and the limits we require in our subsequent analysis. We restrict our main discussion to what we consider to be the minimal scenario to explain the $B$ anomalies and $(g-2)_\mu$. Here, the isotriplet LQ $\varphi$ explains the neutral current anomalies, while the $\text{SU}(2)$ singlet $\phi$ explains the charged current anomalies with contributions to the scalar, tensor and vector operators. Minimally, this implies non-zero values for $x^{L\phi}_{33}$ and $y^{R\phi}_{32}$. The top-mass enhancement evident in eq.~\eqref{eq:amu} means that only small values for the product of $y_{23}^{R\phi}$ and $y_{23}^{L\phi} = x_{23}^{L\phi}$ are required to explain the anomalous magnetic moment of the muon.

The leptoquark that participates in the neutrino mass generation must have a non-zero Yukawa coupling to the electron, and this is the most import phenomenological consequence for the constraints we consider. This, together with the relation in eq.~\eqref{eq:ckmmixing}, means that constraints from processes involving the first generation of SM fermions cannot be avoided completely. In fact, the hierarchy present in the leptoquark couplings to charged leptons is fixed by measured PMNS matrix elements, while the couplings to light quarks are suppressed by CKM matrix elements. Explicitly
\begin{align}
y^{L\eta}_{ij} &= x^{L\eta}_{i3} V_{j3}^* \nonumber\\
               &= V_{j3}^* \frac{\zeta}{\sqrt{2m_0}} \left( \sqrt{m_2} u_{i2}^* + i \sqrt{m_3} u_{j3}^* \right). \label{eq:yxckm}
\end{align}
Of course, the Lagrangian in eq.~\eqref{lagrangian-1} contains many more parameters than these. For simplicity, we turn off any couplings not immediately related to the anomalies or neutrino mass. In reality these need only be small enough to respect any limits placed on them by experiment\footnote{ Note that constraints from neutrinoless double beta decay are not explicitly considered in this analysis. The contributions are CKM suppressed and the couplings involved are exactly those involved in neutrino mass generation. As such, BSM contributions from this model to this process are negligible.  }.

The choice of the minimal set of Yukawa couplings depends on the choice of neutrino-mass regime. In this section, expressions are given in generality assuming $x_{13}^{L\phi} \neq 0$ and $x_{13}^{L\varphi} \neq 0$. If one chooses a particular regime, then the Yukawa coupling to the electron of the leptoquark that does not participate in the neutrino mass can be switched off, and we make this choice according to the principle of minimality discussed above. Thus only one leptoquark will have a Yukawa coupling to the electron at a time.  Of course an additional phenomenological consequence of choosing a neutrino-mass regime is the absence of $\Delta L = 2$ interactions for one of the leptoquarks, $\eta$. This can be achieved by turning off the associated couplings $y_{i}^{\chi \eta}$. In the absence of LQ mixing, the term is not generated at any order since the interactions of $\eta$ now conserve lepton number.

Below we summarise these comments with concrete Yukawa-coupling textures. The constraints presented in this section assume the following set of non-zero Yukawa couplings:
\begin{equation}
\mathbf{x}^{L\phi} = \begin{pmatrix} 0 & 0 & x_{13}^{L\phi} \\ 
                                     0 & 0& x_{23}^{L\phi} \\
                                     0 & 0 & x_{33}^{L\phi}
                                     \end{pmatrix}, \quad
\mathbf{y}^{R\phi} = \begin{pmatrix} 0 & 0 & 0 \\ 
                                     0 & 0& y_{23}^{R\phi} \\
                                     0 & y_{32}^{R\phi} & 0 
                                     \end{pmatrix}
                                     \quad \text{ and } \quad
\mathbf{x}^{L\varphi} = \begin{pmatrix} 0 & 0 & x_{13}^{L\varphi} \\ 
                                     0 & x_{22}^{L\varphi} & x_{23}^{L\varphi} \\
                                     0 & 0 & x_{33}^{L\varphi}
                                     \end{pmatrix}.
\end{equation}
However, if the restriction is made to regime 1, it is understood that $x_{13}^{L\phi}, x_{23}^{L\phi} = 0$ and $y_i^{\chi \phi} = 0$, while regime 2 implies $x_{13}^{L\varphi}, x_{33}^{L\varphi} = 0$ and $y_i^{\chi \varphi} = 0$.  We do discuss other Yukawa-coupling textures throughout this section where appropriate. Notably, we comment briefly on explaining $R_{D^{(*)}}$ with contributions only to the vector operator $C_{V_L}$, and the constraints associated with this scenario are presented in this section as well.

This parameter space is explored in the context of the constraints implied by fits to the flavour anomalies and neutrino mass. We use a suite of computational machinery for most of the calculations, and this setup is discussed in section~\ref{sec:computer}.  Where appropriate we explicitly write out the dominant contributions to observables where we consider that this provides useful insight. Some observables are also calculated separate to these methods, and these are also discussed in detail below.

\subsection{Calculation pipeline}
\label{sec:computer}

Using \texttt{SARAH}~\cite{Porod:2014xia,Porod:2011nf} we construct the model from the Lagrangian upward, using inbuilt machinery to encode the algebraic structure of the fields, associated global symmetries and mixing. \texttt{SARAH} generates an output module for use with \texttt{SPheno}~\cite{Porod:2011nf}, which can calculate the Wilson coefficients, decay rates and a subset of flavour observables, defined by \texttt{FlavorKit}~\cite{Porod:2014xia}, for a particular assignment of model parameters. A full discussion of the underlying machinery and symbioses of these programs can be found in ref.~\cite{Vicente:2015zba}.

In addition to the above, \texttt{Flavio}~\cite{Straub:2018kue} was utilised to process manually calculated Wilson coefficient dictionaries where appropriate, or to take as input the Wilson coefficient \texttt{.json} files outputted by \texttt{SPheno}. This enabled us access to a broader class of flavour observable calculations than would have been otherwise possible. The running of Wilson coefficients in \texttt{Flavio} is implemented using the \texttt{Wilson} package~\cite{Aebischer:2018bkb}.

By implementing this combination of computational machinery, it was possible to construct an efficient parameter scan over a large number of dimensions. In doing so, it was possible to determine the regions of parameter space that establish this model as viable for its desired purpose: reconciling the anomalies, generating radiative neutrino mass, and satisfying the most compelling experimental constraints, as we discuss in this section.
\begin{figure*}[b]
    \centering
    \begin{subfigure}[t]{0.5\textwidth}
        \centering
        \includegraphics[scale=0.54]{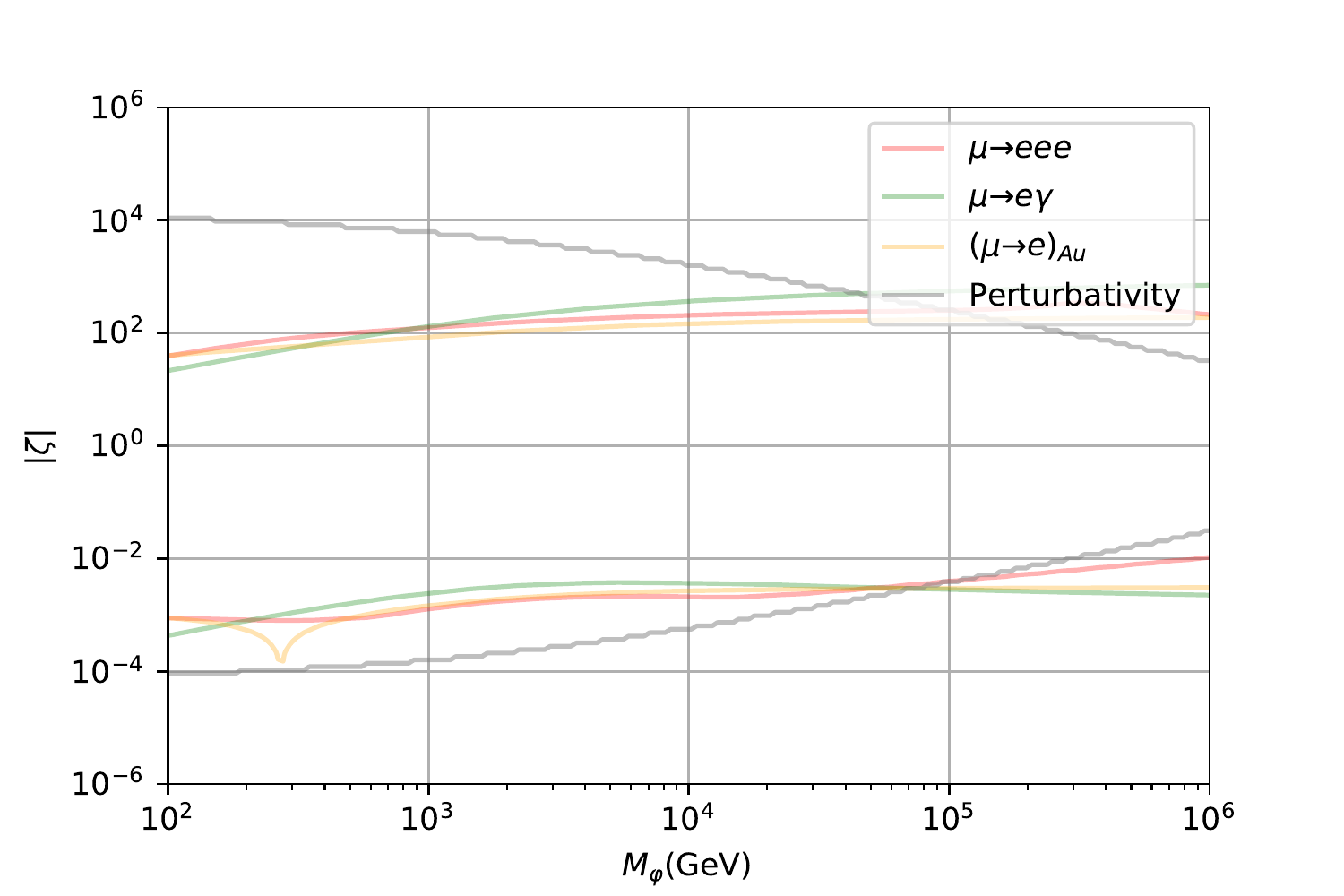}
        \caption{}
    \end{subfigure}~\begin{subfigure}[t]{0.5\textwidth}  
    \centering
          \includegraphics[scale=0.54]{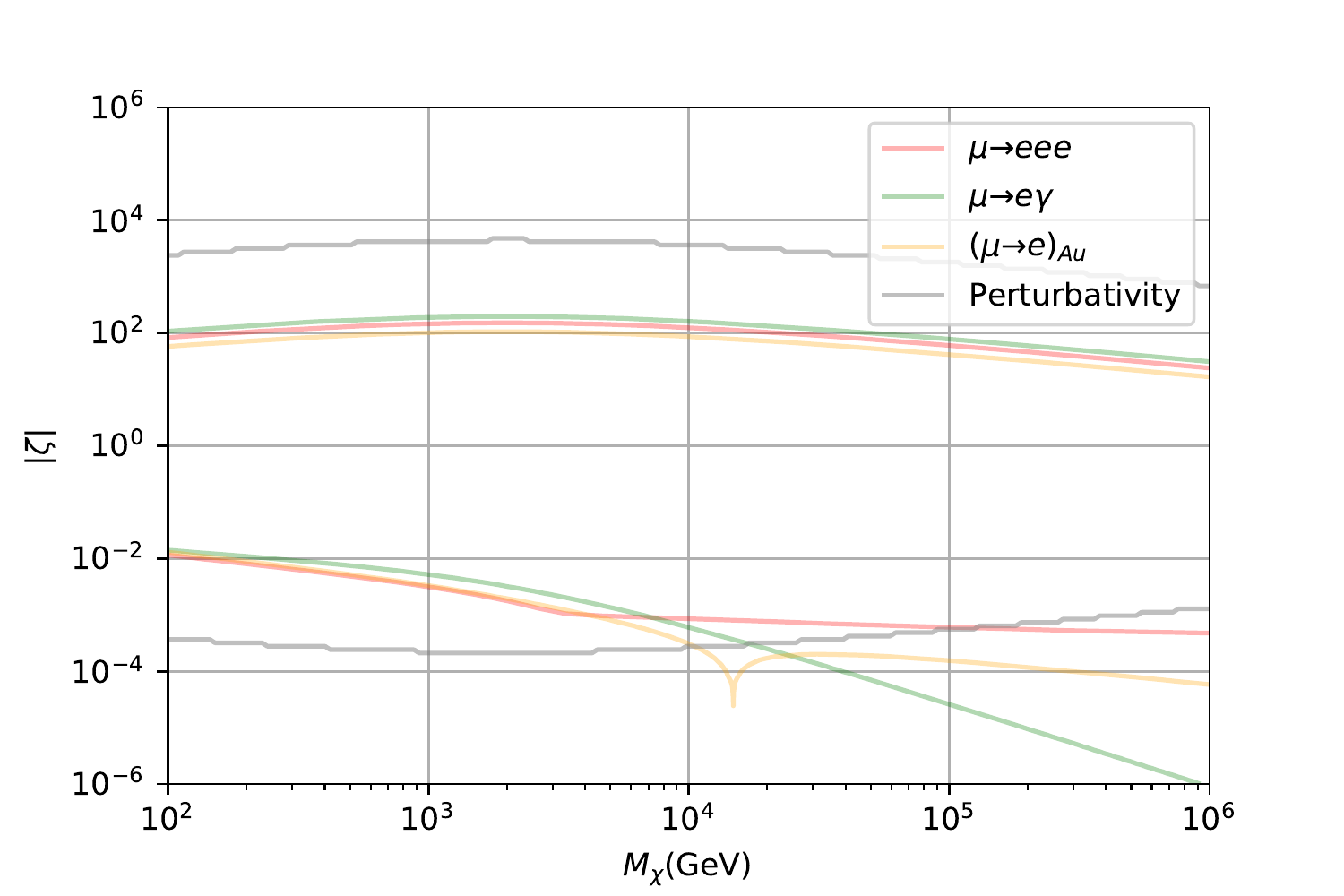}
        \caption{}
    \end{subfigure}
    \caption{\textbf{Regime 1} Constraints on $\zeta$ for varied isotriplet $\varphi$ LQ mass (a), and vector-like quark mass (b); in each case the alternate mass is fixed
    at 2~TeV and the isosinglet LQ couplings are switched off.   Allowed points for each constraint lie between the two same-coloured lines. The `dip' in both graphs is due to an accidental cancellation by virtue of parameter choice.}\label{region2}
 \end{figure*}
\begin{figure*}[t]
    \centering
    \begin{subfigure}[t]{0.5\textwidth}
        \centering
        \includegraphics[scale=0.54]{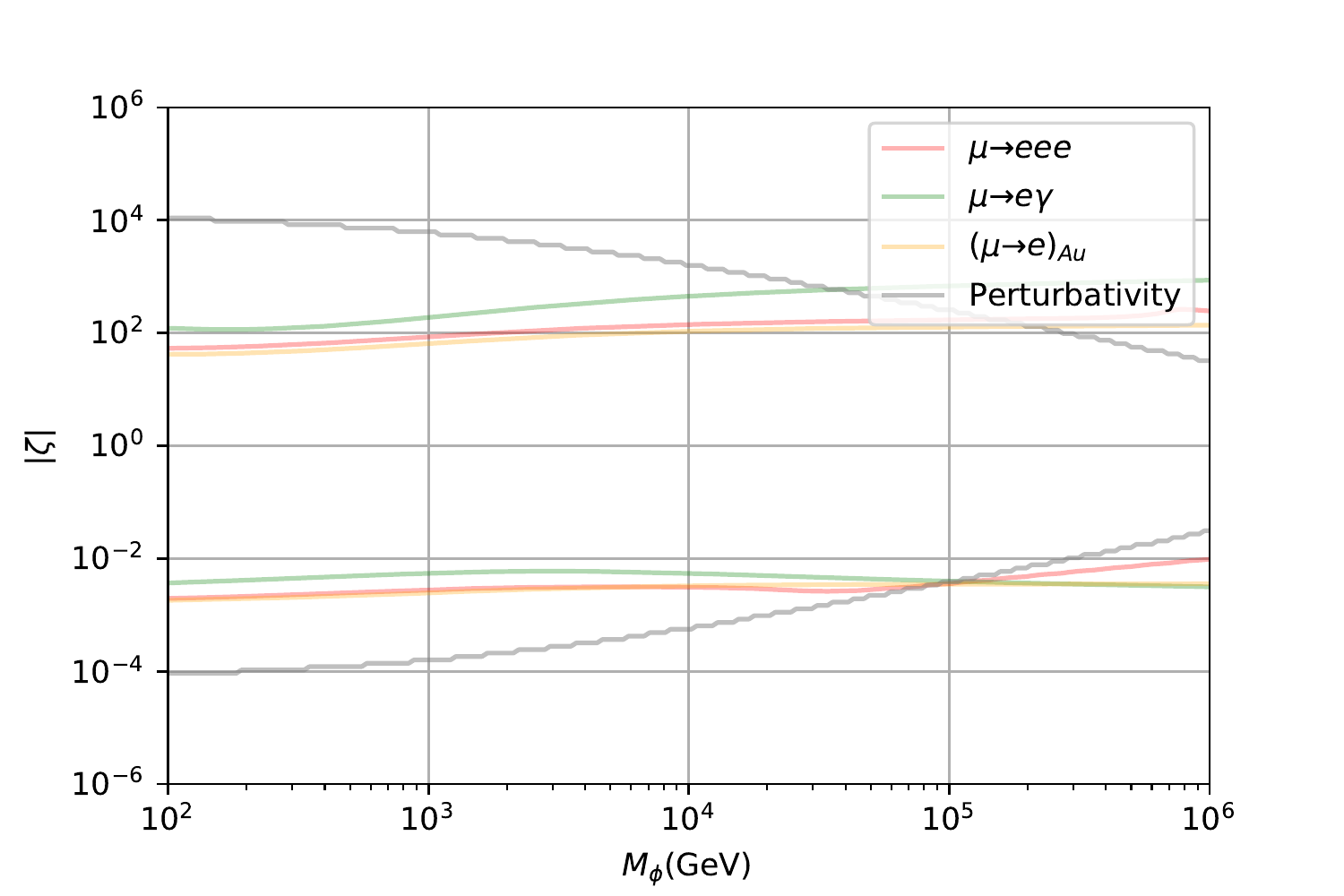}
        \caption{}
    \end{subfigure}~\begin{subfigure}[t]{0.5\textwidth}  
    \centering
          \includegraphics[scale=0.54]{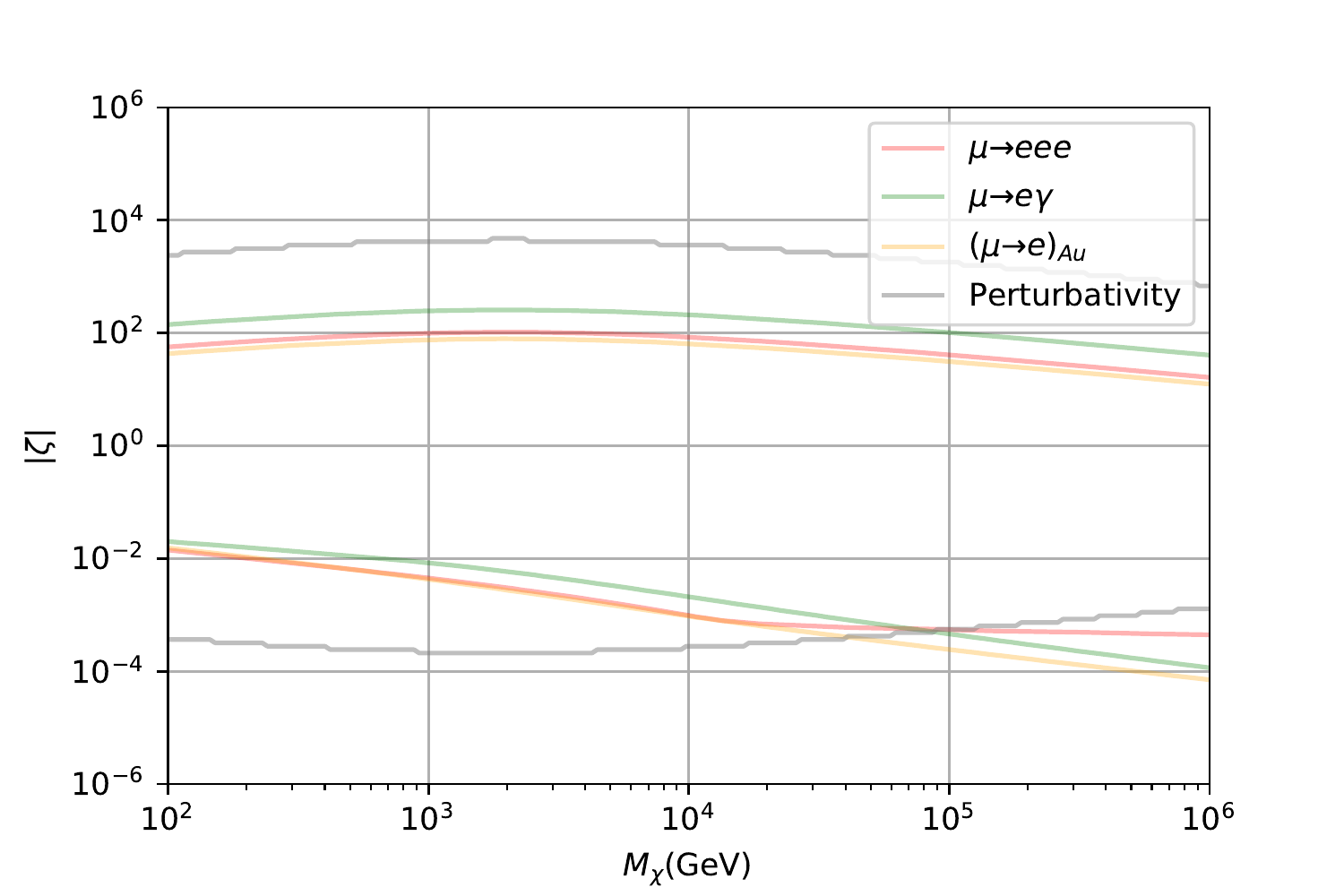}
        \caption{}
    \end{subfigure}
    \caption{\textbf{Regime 2} Constraints on $\zeta$ for varied singlet $\phi$ LQ mass (a), and vector-like quark mass (b); in each case the alternate mass is fixed at 2~TeV and the isotriplet LQ couplings are switched off. Allowed points for each constraint lie between the two same-coloured lines.}
    \label{region1}
\end{figure*}

\subsubsection{Validation and benchmark regions}
To begin, we consider this model solely as a radiative neutrino mass model, \emph{only} switching on the couplings generated by the Casas-Ibarra parameter. Truncating parameter space so as to only include those important for radiative neutrino mass generation, we may validate our calculations against the scans performed in ref.~\cite{Cai:2014kra} for the $\phi$-$\chi$ model (\textsc{Model 2} in table~\ref{table:1}). In figures \ref{region2} and \ref{region1}, we depict the two parameter regimes fixed by the Casas-Ibarra procedure: neutrino mass generated by $\varphi$ (Regime 1), or by $\phi$ (Regime 2).  This is proof-of-principle to validate this computational setup, and more generally to determine benchmark regions of parameter space for subsequent scans. Throughout this section  we have implemented the perturbativity bound that the magnitude of each \emph{physical} Yukawa coupling is less than $\sqrt{4\pi}$.

Ref.~\cite{Angel:2013hla} identify the most important flavour-changing processes as the LFV decays: $\mu \to e \gamma$, $\mu \to e e e$, as well as $\mu-e$ conversion in nuclei, all of which depend on the LQ couplings fixed by $\zeta$. We have scanned over the magnitude of the Casas-Ibarra parameter, $|\zeta|$. The observables in question are proportional to the product of two LQ couplings generated by the Casas-Ibarra parameterisation, and therefore, $\propto |\zeta|^2$. Any phase information is irrelevant for these calculations. 

For Regime 2, where $\phi$ generates neutrino mass, these plots represent replication of those in ref.~\cite{Cai:2014kra}, with updated measurements and using the framework outlined above. The minimal contrast between these two results acts as initial validation of the calculations in \texttt{SPheno/SARAH} for these LFV processes. Although we have not imposed any further constraints in these scans, nevertheless from figure~\ref{region2} and figure~\ref{region1} we can begin refining parameter space by noting that $|\zeta|\in (10^{-2}, 10^{2})$ roughly defines the region which is capable of accommodating these LFV constraints with perturbative couplings. This is consistent with the bounds prescribed earlier in eq.~\eqref{eq:zetaperta}.

\subsection{Collider bounds}

\begin{figure}[t!]
\centering
\includegraphics[width=11cm]{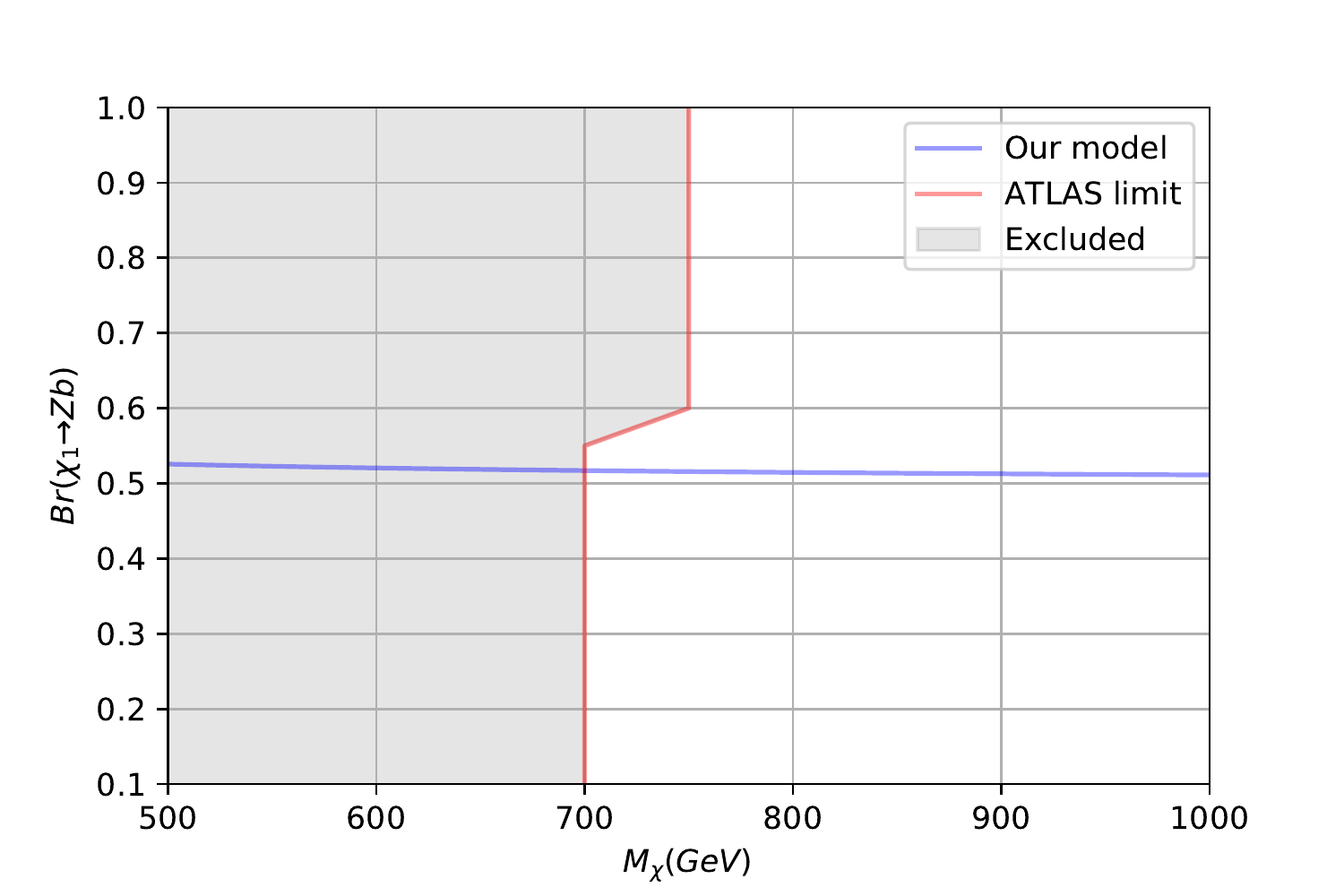}
\caption{Projected ATLAS limit \cite{Aaboud:2018pii} on vector-like quark mass from $\chi_1 \to Z b$ demonstrates a lower limit on $m_\chi$ , with fixed mixing Yukawa $Y_{b}=1.$} \label{branch}
\end{figure}

Many collider searches already exist for third-generation quark `partners', as they are referred to in many SUSY models. Here, we focus on those presented in the context of each vector-like quark component as a part of the $\chi$ isodoublet. Contributions to the decay branching fraction limit were found to be sensitive only to the BSM-SM quark mixing and the mass of the exotic, and the following analysis considers a mixing Yukawa $Y_{b}=1.$

The branching fraction of the exotic bottom-partner $\chi_1$ can be expanded as per the following;
\begin{equation}
\text{Br}(\chi_1 \to Z b)+ \text{Br}(\chi_1 \to H b) \sim 1.
\end{equation}
The third possible decay channel, $\chi_1\to W t$, is highly suppressed by the small mixing between the exotic and SM quark sectors assumed for this channel. Taking the constraints from the ATLAS collaboration, and calculating the the corresponding branching fraction using \texttt{SPheno}, the resultant limit on the vector-like quark mass can be read off the graph in figure~\ref{branch}: $$m_{\chi_1} \gtrsim 700~\text{GeV}. $$This represents a very low bound from decays of $\chi_1$, and decays of $\chi_2$ impose a slightly stronger constraint. The ATLAS collaboration imply a lower bound on the mass of $\chi_2$ (or `$Y$' as it is referred to) of:
\begin{equation}
m_{\chi_2} \gtrsim 1350 \text{ GeV}.
\end{equation}
The masses of each component of the isodoublet $\chi$ are of comparable orders of magnitude, as we will see in section~\ref{sec:bBmix}. Therefore, we adopt the more conservative bound for $m_{\chi}=m_{\chi_2}\sim m_{\chi_1}$ from $\chi_2$ decay in subsequent discussion.

 Similarly to the vector-like quarks, approximate bounds on the leptoquark masses can be inferred from collider searches. The most recent analyses from ATLAS~\cite{Aaboud:2019bye} and CMS~\cite{Sirunyan:2018nkj, Sirunyan:2018kzh} place model-independent lower bounds on third-generation leptoquarks at
 \begin{equation}
  \label{masseta}
    m_\eta \gtrsim 800 \text{ GeV}.
 \end{equation}
 In the limiting cases where the branching ratio of the LQ is mostly to charged leptons or neutrinos, the bounds are generally $\sim 1 \text{ TeV}$, with the most stringent bound $m_\eta \gtrsim 1.2 \text{ TeV}$ coming from $bb + \text{MET}$ searches. 

In our model couplings of third-generation quarks to the muon are unavoidable. Here, 
 limits from $tt\mu\mu$ searches can exclude leptoquark masses below $1.3 \text{ TeV}$, assuming $\text{Br}(\eta \to t \mu) \approx 100\%$~\cite{ Sirunyan:2018ruf}. Additionally, since $\varphi$ must couple the strange quark to the muon, dimuon--dijet searches are also potentially relevant. In this case, the limits can be as large as $m_\eta \gtrsim 1.5 \text{ TeV}$~\cite{Aaboud:2019jcc, Sirunyan:2018ryt}.

\begin{figure}[t] 
  \centering
  \includegraphics[width=0.7\textwidth]{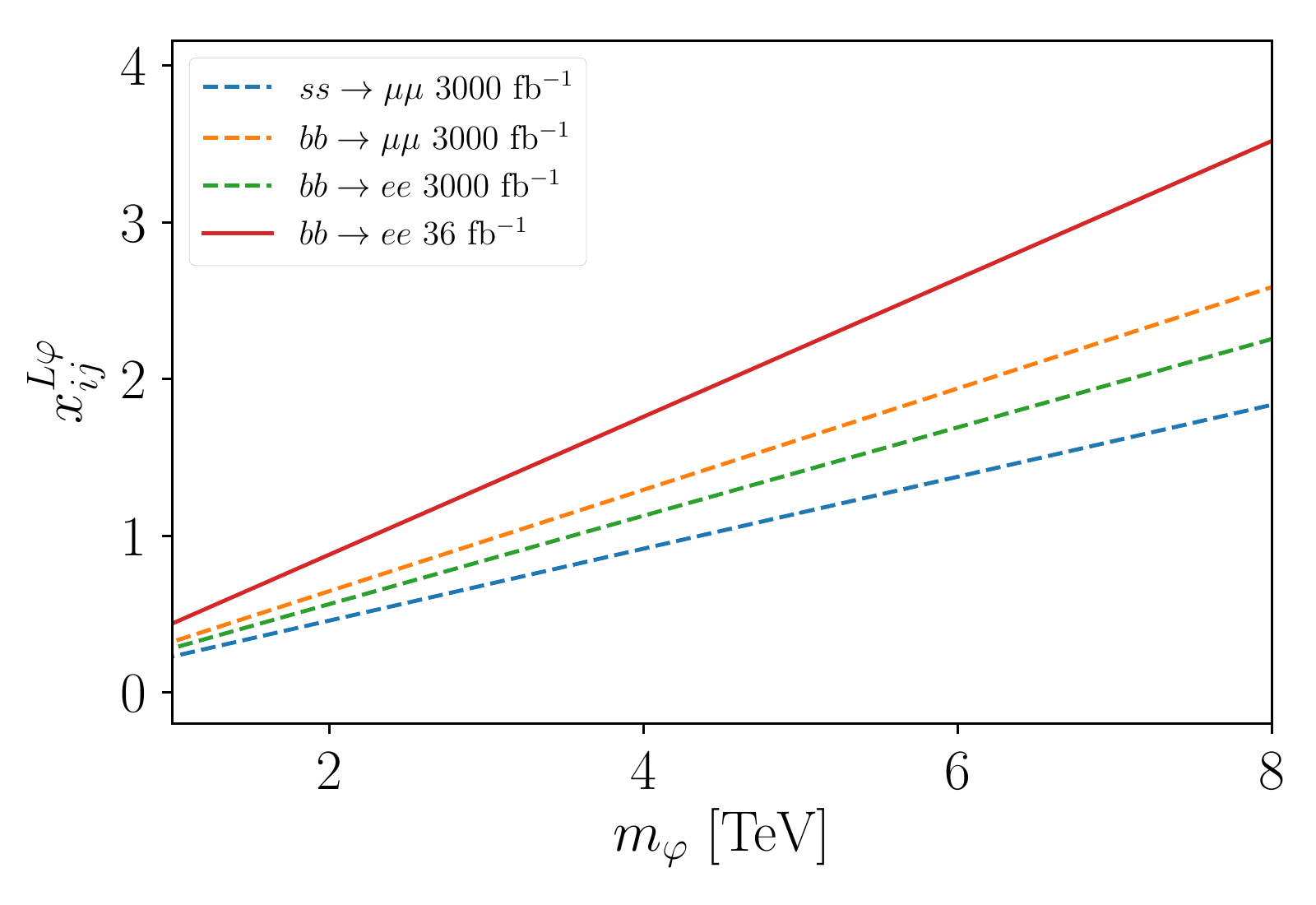}
  \caption{The figure shows the current (solid) and projected (dashed) upper limits on the couplings of the $\varphi$ LQ to down-type quarks and charged leptons $x^{L\varphi}_{ij}$. The limits are from LHC searches in $pp \to \ell \ell$ high-$p_T$ tails at $13 \text{ TeV}$ from ATLAS~\cite{ATLAS-CONF-2017-027}, derived from ref.~\cite{Greljo:2017vvb}. The Yukawa coupling being constrained depends on the process. For example, $ss \to \mu \mu$ will constrain $x_{22}^{L\varphi}$.}
  \label{fig:lhcbounds}
\end{figure}

High-$p_T$ dilepton production through the $\varphi$ leptoquark has also been shown to provide interesting constraints and signatures for the leptoquarks in our model~\cite{Angelescu:2018tyl}. The leptoquark contributes to the processes $pp \to \ell \ell$ through tree-level $t$-channel graphs whose effects can alter the tail of the differential cross-sections for $pp \to \ell \ell$. We take the limits from ref.~\cite{Angelescu:2018tyl} for the muon and tau modes derived from $36 \text{ fb}^{-1}$ of ATLAS data at $13 \text{ TeV}$~\cite{Aaboud:2017sjh, Aaboud:2017buh}. We derive bounds on $bb \to ee$ and extract the $3000 \text{ fb}^{-1}$ ATLAS sensitivity for the electron and muon modes from ref.~\cite{Greljo:2017vvb}. These bounds are shown in figure~\ref{fig:lhcbounds}. We find that the limits on $cc \to ee$ and $uu \to ee$ give less stringent bounds on $|\zeta|$, and thus we do not include them in our numerical scans.

\subsection{Limits on the \texorpdfstring{$b$}{b}-quark/vector-like quark mixing}
\label{sec:bBmix}
Of central importance in this model is the mixing generated by the terms $Y_{b} \bar{b}_R H \chi_L + \text{h.c.}$ between the $b$ quark and the vector-like quark $\chi$. This mixing is a necessary ingredient for the violation of lepton number by $\varphi$, and plays a governing role in the overall scale of the neutrino mass $m_0$ according to eq.~\eqref{CI}. Its size also dictates the extent to which $\Delta L=2$ neutrino final states are important to consider, for example in $B \to K^{(*)} \nu \nu$.

The mixing of the $b$ with $\chi$ leads to new contributions to the oblique electroweak parameters $S$ and $T$. These have been measured to high precision by LEP~\cite{ALEPH:2005ab}. The mixing also leads to an alteration of the $Zbb$ coupling at tree-level, for which global electroweak fits have suggested a small deviation from the SM value, e.g.~\cite{Ciuchini:2013pca}. The dominant contributions to these effects are encapsulated in the effective dimension-6 Lagrangian generated by the heavy $\chi$ at the scales probed by experiment:
\begin{equation}
\mathcal{L}^{(6)}_{\chi} \supset \frac{C^{33}_{Hd}}{m_\chi^2} (H^\dagger i \overset{\leftrightarrow}{D}_\mu H) (\bar{b}_R \gamma^\mu b_R) + \frac{C^{33}_{dH}}{m_\chi^2} y_b (H^\dagger H) (\bar{Q}_3 b_R H).
\end{equation}
The first operator modifies the electroweak precision observables discussed above and the second affects Higgs measurements and is currently poorly constrained. We take the $2\sigma$ bounds on the operator coefficient $C_{Hd}^{33}$ from the electroweak fit performed in ref.~\cite{Ciuchini:2013pca} $C_{Hd}^{33} \in [-0.38, 0.03]$ to derive
\begin{equation}
    \label{eq:bmixinglimit}
    |Y_{b}| \in [0.25, 0.87] \left( \frac{m_\chi}{\text{TeV}} \right).
\end{equation}
This implies the bounds $\theta_R \in [0.06, 0.21]$ at $95\%$ confidence, with central value $\theta_R \approx 0.16$. This agrees with ref.~\cite{Aguilar-Saavedra:2013qpa} which studied the effects of the doublet $\chi$ and other vector-like quarks. The relation $\theta_L \approx \frac{m_b}{m_\chi} \theta_R$ from eq.~\eqref{eq:bmixing} implies the $\cos\theta_L$ factors appearing in eq.~\eqref{eq:c9c10} and eq.~\eqref{eq:csl}--\eqref{eq:ct} do not suppress the contributions to the anomalous observables\footnote{ This result is a stronger, and more general, constraint than that quoted from direct searches in ref.~\cite{Aaboud:2018ifs}, which suggests a $95\%$ confidence interval of $\sin \theta_R \in [0.17,0.55]$ for $m_\chi \sim 800$ GeV.}. Restricting this mixing to be small consequentially reduces the mass-splitting between the components of the exotic doublet $\chi$, such that $m_{\chi} \sim m_{\chi_1} \sim m_{\chi_2} $ remains a valid approximation.

\subsection{Leptonic decays}
\label{sec:leptonic}
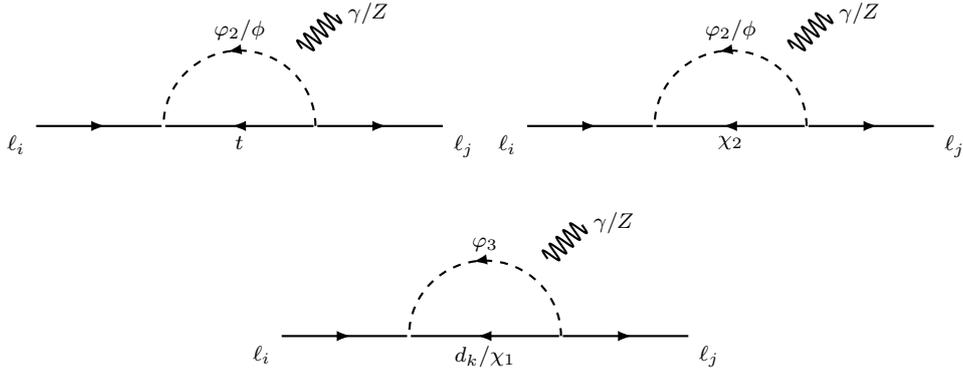
\begin{figure}[t!]
\centering
    \begin{tikzpicture}
      \scriptsize
      \node[vtx, label=-120:$\ell_i$] (i1) at (0,0) {};
      \node[vtx] (v1) at (1.7,0) {};
      \node[vtx] (v2) at (3.7,0) {};
       \node[vtx] (v3) at (3.5,1) {};
        \node[vtx,label=0:$\gamma/Z$] (v4) at (4,1.5) {};
      \node[vtx, label=-60:$\ell_j$] (o1) at (5.4,0) {};
       \node[vtx, label=90:$\varphi_2 /\phi$] (i3) at (2.7,1) {};
        \node[vtx, label=-90:$t$] (i4) at (2.7,0) {};
      \graph[use existing nodes]{
       i1 --[fermion] v1;
      v2--[fermion]v1;
      v2 --[fermion] o1;
      v3 --[photon]v4;
      };  
         \draw[thick, dashed,postaction={decorate, decoration={markings, mark=at position .55 with {\arrow[]{latex}}}}]  (3.7,0) arc (0:180:1);
    \end{tikzpicture}  
\vspace{0.5cm}
    \centering
    \begin{tikzpicture}
      \scriptsize
      \node[vtx, label=-120:$\ell_i$] (i1) at (0,0) {};
      \node[vtx] (v1) at (1.7,0) {};
      \node[vtx] (v2) at (3.7,0) {};
     \node[vtx] (v3) at (3.5,1) {};
        \node[vtx,label=0:$\gamma/Z$] (v4) at (4,1.5) {};
      \node[vtx, label=-60:$\ell_j$] (o1) at (5.4,0) {};
       \node[vtx, label=90:$\varphi_2/\phi$] (i3) at (2.7,1) {};
        \node[vtx, label=-90:$\chi_2$] (i4) at (2.7,0) {};
      \graph[use existing nodes]{
     i1 --[fermion] v1;
      v2--[fermion]v1;
      v2 --[fermion] o1;
      v3 --[photon]v4;
      };  
         \draw[thick, dashed,postaction={decorate, decoration={markings, mark=at position .55 with {\arrow[]{latex}}}}]  (3.7,0) arc (0:180:1);
    \end{tikzpicture}  

  \begin{tikzpicture}
      \scriptsize
      \node[vtx, label=-120:$\ell_i$] (i1) at (0,0) {};
      \node[vtx] (v1) at (1.7,0) {};
      \node[vtx] (v2) at (3.7,0) {};
       \node[vtx] (v3) at (3.5,1) {};
        \node[vtx,label=0:$\gamma/Z$] (v4) at (4,1.5) {};
      \node[vtx, label=-60:$\ell_j$] (o1) at (5.4,0) {};
       \node[vtx, label=90:$\varphi_3$] (i3) at (2.7,1) {};
        \node[vtx, label=-90:$d_k/\chi_1$] (i4) at (2.7,0) {};
      \graph[use existing nodes]{
      i1 --[fermion] v1;
      v2--[fermion]v1;
      v2 --[fermion] o1;
      v3 --[photon]v4;
      };  
       \draw[thick, dashed,postaction={decorate, decoration={markings, mark=at position .55 with {\arrow[]{latex}}}}]  (3.7,0) arc (0:180:1);
    \end{tikzpicture} 
    \centering
    \caption{Contributions to $\ell_i \to \ell_j \gamma$ processes,  where additional splitting of the boson line gives contributions to $\ell_i \to \ell_j\ell_k \overline{\ell_k}$. The photon/$Z$ can be attached to any of the four lines.   }  \label{fig:mueg}
    \end{figure}

\begin{table}[t]
  \begin{center}
  \def\arraystretch{1.3} 
    \begin{tabular}{|c|c|} 
    \hline
      \textsc{Process}  & \textsc{Limits} \\
      \hline \hline
       $\text{Br}(\mu \to e \gamma)$ & $< 4.2 \times 10^{-13}$   \\
       $\text{Br}(\mu \to 3 e$) &  $< 1.0 \times 10^{-12}$  \\
       $\frac{\sigma(\mu \text{Au}\to e\text{Au})}{\sigma(\mu \text{Au}\to \text{capture})}$ &  $< 7.0 \times 10^{-13}$  \\
       $\text{Br}(\tau \to e \gamma)$ & $< 3.3 \times 10^{-8}$   \\
       $\text{Br}(\tau \to \mu \gamma)$ &$< 4.4 \times 10^{-8}$    \\
       $\text{Br}(\tau \to 3\mu)$ &  $< 2.1 \times 10^{-8}$  \\
       $\text{Br}(\tau \to 3 e)$ &  $< 2.7 \times 10^{-8}$  \\
      \hline
    \end{tabular}
  \end{center}
    \caption{Values given without citation are taken from ref.~\cite{PhysRevD.98.030001}.} \label{table:muonelectron}
    \end{table}
The leptoquarks contribute to the LFV decays  $\ell_i \to \ell_j \gamma$ and $\ell_i \to \ell_j\ell_k \overline{\ell}_k$ at loop level through diagrams like those depicted in figure~\ref{fig:mueg}, as well as $H$ penguins and box diagrams for the latter. The strongest of these constraints on the couplings are from first- and second-generation LFV processes. The experimental bounds on these processes can be found in table~\ref{table:muonelectron}.

\begin{figure}[t!]
\begin{center}
 \begin{tikzpicture}
  \node[vtx, label=180:$\bar{f}$] (i1) at (0,2.5) {};
  \node[vtx, label=0:$\ell_i^-$] (i2) at (3.6,2.5) {};
  \coordinate[] (v1) at (1.8,2) {};
  \coordinate[] (v2) at (1.8,0) {};
  \coordinate[label=0:$\varphi/\phi$] (v6) at (1.8,1) {};
  \node[vtx, label=180:$\ell_j^-$] (o1) at (0,-1) {};
  \node[vtx, label=0:$\bar{f}$] (o2) at (3.6,-1) {};
  \graph[use existing nodes]{
     i1 --[antifermion] v1;
     v1 --[fermion] i2;
     v2 --[antifermion] o2 ;
     o1--[fermion] v2;
     v1 --[cscalar] v2;
  }; 
  \end{tikzpicture}
    \caption{Demonstrative topology of the leading-order LQ contribution to muon-electron conversion in nuclei, where $f \in \{u,d\}$.}\label{fig:mueAu}
      \end{center}
\end{figure}
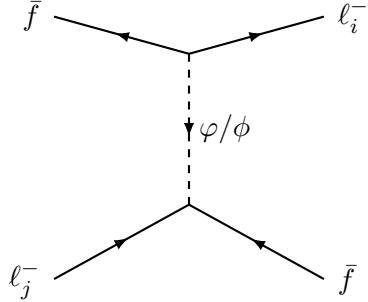

The related process of muon--electron conversion in nuclei is mediated at tree-level by the leptoquark participating in the neutrino mass generation (figure~\ref{fig:mueAu}), although there are also loop-level topologies similar to those for $\ell_i \to \ell_j \ell_k \overline{\ell}_k$ with quarks in place of the same-flavour lepton-antilepton pair. We presume that a coherent conversion process dominates, \emph{i.e} the final state nucleon is the same as the initial state, therefore the coupling to the sea-quarks\footnote{This includes possible contribution from the vector-like quark $\chi$, which will be ignored for the remainder of this discussion.} is negligible~\cite{Kitano:2002mt}. The tree-level contributions to muon-electron conversion depicted by figure~\ref{fig:mueAu} will lead to vector, scalar and tensor effective operator contributions, via Fiertz transformation (appendix \ref{appendix2}). These will be discussed for the two neutrino mass regimes below.

\subsubsection{Ruling-out Regime 2}
\label{sec:reg2}
For neutrino mass as generated in Regime 2 ($\phi$ couplings fixed by Casas-Ibarra), contributions to muon-electron conversion are dominated by tree-level processes via the LQ mediator $\phi$. Assuming the right-handed couplings to first- and second- generation quarks and leptons are switched-off, the dominant contribution is from the effective interaction:
\begin{align}
    \mathcal{L}_{\text{effective}}^{\mu-e, \phi}&\sim  \frac{1}{2 m_\phi^2} y_{11}^{L\phi*}y^{L\phi}_{21}\left(\overline{e} \gamma_{\mu} P_L \mu\right)\left(\overline{u} \gamma^{\mu} P_L u \right).
\end{align}
LQ couplings between charged leptons and up-type quarks are unavoidably generated by CKM mixing: \begin{equation}
    y^{L\phi}_{ij} = x^{L\phi}_{i3}V_{j3}^{*}+ x^{L\phi}_{i2}V_{j2}^{*}+x^{L\phi}_{i1}V_{j1}^{*}.
\end{equation}

Ref.~\cite{Kosmas:2001mv} contains a study of the effective Lagrangian approach to constraining these contributions, and provides model-independent limits which we will interpret here. The strongest constraints from muon-electron conversion in nuclei presently come from measurements involving gold~(Au) nuclei. In the absence of accidental cancellation, the experimental constraint in table~\ref{table:muonelectron}, results in the following bound on the dominant contribution to this measurement:
\begin{align}
 \label{eq:zetamuebound}
 \frac{ |x^{L\phi*}_{13}x^{L\phi}_{23}|}{m_\phi^2} \lesssim 2.76 \times 10^{-8}\; \text{GeV}^{-2}.
\end{align}
Since both Yukawa couplings involved are fixed by the Casas-Ibarra procedure, the bound can be interpreted in terms of $|\zeta|$. We find\footnote{Throughout this discussion we have assumed $m_0$ is positive-definite.}
\begin{equation}
\label{eq:zetabound}
|\zeta|^2 \lesssim 1.59\left( \frac{m_\phi}{\text{TeV}}\right)^2 \left( \frac{m_0}{0.05 \text{ eV}}\right).
\end{equation}

To contrast with other constraints, we begin by parameterising the Casas-Ibarra parameter via $\zeta =(1+i
\kappa)\text{Re}\zeta$, giving $|\zeta|= |\text{Re}\zeta|\sqrt{1+\kappa^2}$, where $\kappa \in \mathbb{R}$. The only viable scenario for this leptoquark to explain the charged current anomalies alone is for it to contribute to the scalar and tensor operators~\cite{Cai:2017wry, Angelescu:2018tyl, Crivellin:2019qnh}. As discussed in section~\ref{sec:bctaunu}, there are always unavoidable contributions to the vector operator $C_{V_L}$ present as well once $x_{33}$ is non-zero. Our fit results suggest
\begin{equation}
    \text{Re}\, C_{S_L} \approx 0.14,\;\;\;
    \text{Im}\, C_{S_L} \approx 0,
\end{equation}
which implies, assuming $c_{\theta_L} \sim 1$ and $y_{32}^{R\phi} \in \mathbb{R}$, and inputting numerical values for $U_{ij}$:
\begin{align}
\label{eq:Cs}
\frac{\sqrt{2}}{4G_FV_{cb}}\frac{1}{2}\left(\frac{y_{32}^{R\phi*}}{m_\phi^2} \right) \left(\frac{0.05 \;\text{eV}}{m_0}\right)^{1/2}(0.22 - 0.45 \kappa)\text{Re}\zeta \sim 0.14.
\end{align}

\begin{figure}
\centering
\includegraphics[scale=0.8]{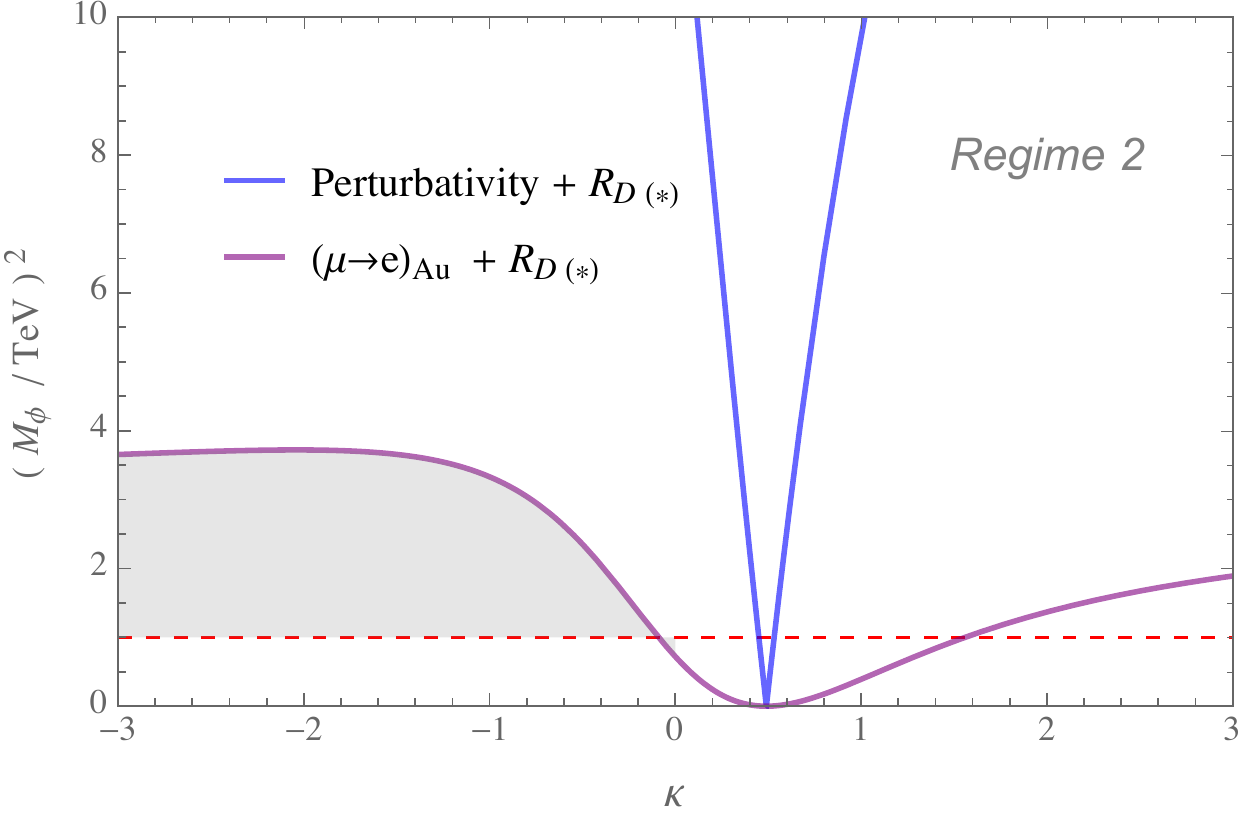}
\caption{Constraints on neutrino mass Regime 2 ($\phi$ generating neutrino mass). Coloured lines represent upper bounds on the $LQ$ mass-squared, and the  shaded region represents the allowed points when also considering suppression of imaginary component of respective Wilson coefficient $C_{S_L}$. The red-dashed line in each indicates the approximate lower mass-bound from collider constraints.  }\label{regime2:plot}
\end{figure}

Combining this with the above bound on Re$(\zeta)^2$ (eq.\eqref{eq:zetabound}), and saturating the perturbativity bound with $|y_{32}^{R\phi*}|=\sqrt{4\pi}$, we obtain
\begin{align}
\label{proof1}
\left(\frac{m_\phi}{\text{TeV}}\right)^2 \lesssim 3 \frac{(0.49-\kappa)^2}{(1+\kappa^2)}.
\end{align}
Re-expressing the upper perturbativity constraint given in eq.\eqref{eq:zetaperta} in terms of numerical inputs gives
\begin{align}
\label{pertalpha}
\text{Re}\,\zeta^2 \lesssim \frac{44}{(1+\kappa^2)}\left( \frac{m_0}{0.05\text{ eV}}\right),
\end{align}
which, when combined with eq.\eqref{eq:Cs}, reduces to:
\begin{align}
\label{proof2}
\left(\frac{m_\phi}{\text{TeV}}\right)^2 \lesssim 27\frac{ |0.49-\kappa|}{\sqrt{1+\kappa^2}}. 
\end{align}
These two upper bounds are shown in figure~\ref{regime2:plot} as a function of $\kappa$. In order to ensure that $\text{Im\,} C_{S_L} \approx 0,$ we can rephrase this requirement as needing a suppression of $\text{Im}\, C_{S_L}$ relative to $\text{Re}\, C_{S_L}$, that is to say a \emph{minimal} requirement of:
\begin{align}
 \left|\frac{\text{Re}\, C_{S_L}}{\text{Im}\, C_{S_L}}\right| \sim \left|\frac{1-2\kappa}{1+2\kappa}
\right|\gtrsim 1.
\end{align}
 To satisfy all three of these constraints already reduces the available parameter space significantly (the grey-shaded region is allowed). These are necessary, but not sufficient, conditions for these parameters to satisfy in order to ensure that these constraints are met. We may note from figure~\ref{regime2:plot} that these constraints alone indicate a very small allowed mass region of $1~\text{TeV}\lesssim m_\varphi\lesssim 2~\text{TeV}$.

\subsubsection{Constraints on Regime 1}

Following the calculation procedure above, we find for Regime 2 that the effective interaction contributing to muon-electron conversion is given by:
\begin{align}
    \mathcal{L}_{\text{effective}}^{\mu-e, \varphi}&\sim \frac{ c_{\theta_L}}{4m_\varphi^2} y_{11}^{L\varphi*}y^{L\varphi}_{21}\left(\overline{e} \gamma_\mu P_L \mu \right)\left(\overline{u} \gamma^\mu P_R u \right),
\end{align}
Implementing the bound determined in ref.~\cite{Kosmas:2001mv}, as above, constraints from muon-electron conversion in Gold nuclei give the following:
\begin{align}
 \frac{|x_{13}^{L\varphi*}x^{L\varphi}_{23}|}{m_\varphi^2}  \lesssim 5.51 \times 10^{-8}\; \text{GeV}^{-2},
\end{align}
which implies, via the Casas-Ibarra parameterisation, and using the notation consistent with the previous section:
\begin{align}
\label{muevar}
\text{Re}\zeta^2 \lesssim \frac{0.35}{(1+\kappa^2)} \left(\frac{m_{\varphi}}{\text{TeV}}\right)^2\left(\frac{|m_0|}{0.05 \;\text{eV}}\right).  
\end{align}

Requiring agreement with earlier discussed value for $\text{Re}C_9=-\text{Re}C_{10}$ fit eq.~\eqref{c9c10eq} leads to the following central value: 
\begin{align}
\label{c9fitHere}
\text{Re}\zeta^2 \sim \frac{( 1.7\times 10^{-5})}{(0.46+\kappa)^2(x_{22}^{L\varphi})^2} \left( \frac{m_\varphi}{\text{TeV}}\right)^4 \left( \frac{|m_0|}{0.05 \text{ eV}}\right)
\end{align}
Combining eq.~\eqref{c9fitHere} and eq.~\eqref{muevar}, and assuming $c_{\theta_L}~1$ whilst saturating the perturbativity bound for $x_{22}^{L\varphi}$, gives the constraint:
\begin{align}
\label{proof3}
\left( \frac{m_\varphi}{\text{TeV}}\right)^2\lesssim 2.6\times 10^{5}\;\frac{(0.46+\kappa)^2}{(1+\kappa^2)}
\end{align}
Also, re-expressing the perturbativity constraint, combining eq.~\eqref{pertalpha} and eq.~\eqref{c9fitHere}, gives: 
\begin{align}
\label{proof4}
\left( \frac{m_\varphi}{\text{TeV}}\right)^2\lesssim 5.7\times 10^3\;\frac{|0.46+\kappa|}{\sqrt{1+\kappa^2}}.
\end{align}

These provide a necessary, but not sufficient, bound on these parameters to satisfy the specified constraints. Simply from contrasting the size of the pre-factors between equations \ref{proof3} and \ref{proof4},\ref{proof1} and \ref{proof2}, the unconstrained parameter space for mass-squared in Regime 1 is significantly greater than for Regime 2. This is justification for concentrating solely on Regime 1 for the remainder of this work.

\subsection{\texorpdfstring{$Z$}{Z} decays}
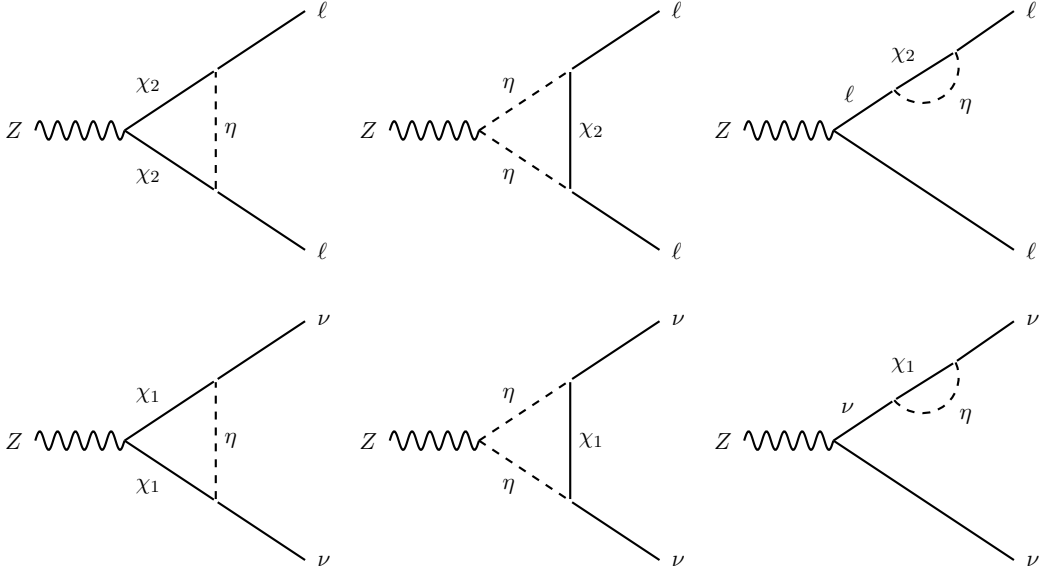
\begin{figure}[t]
\begin{minipage}{0.3\textwidth}
\begin{tikzpicture}[thick,scale=0.8, every node/.style={transform shape}]
  \node[vtx, label=180:$Z$] (i1) at (0,1) {};
  \node[vtx] (i2) at (3,2) {};
  \coordinate[] (v1) at (1.5,1) {};
  \node[vtx] (v2) at (3.0,0) {};
  \node[vtx,label=0:$\ell$] (o1) at (4.5,3) {};
  \node[vtx, label=0:$\ell$] (o2) at (4.5,-1) {};
  \graph[use existing nodes]{
     i1 --[photon] v1;
     v1 --[majorana,edge label=$\chi_2$] i2;
     v2 --[majorana] o2;
     i2 --[majorana] o1;
     v2 --[majorana, edge label=$\chi_2$] v1;
     v2--[scalar,edge label'=$\eta$]i2;
  };
  \end{tikzpicture}
\end{minipage}
\begin{minipage}{0.3\textwidth}
\begin{tikzpicture}[thick,scale=0.8, every node/.style={transform shape}]
  \node[vtx, label=180:$Z$] (i1) at (0,1) {};
  \node[vtx] (i2) at (3,2) {};
  \coordinate[] (v1) at (1.5,1) {};
  \node[vtx] (v2) at (3.0,0) {};
  \node[vtx,label=0:$\ell$] (o1) at (4.5,3) {};
  \node[vtx, label=0:$\ell$] (o2) at (4.5,-1) {};
  \graph[use existing nodes]{
     i1 --[photon] v1;
     v1 --[scalar,edge label=$\eta$] i2;
     v2 --[majorana] o2;
     i2 --[majorana] o1;
     v1 --[scalar, edge label'=$\eta$] v2;
     v2--[majorana,edge label'=$\chi_2$]i2;
  };
  \end{tikzpicture}
\end{minipage}
\begin{minipage}{0.3\textwidth}
\begin{tikzpicture}[thick,scale=0.8, every node/.style={transform shape}]
  \node[vtx, label=180:$Z$] (i1) at (0,1) {};
  \coordinate[] (v1) at (1.5,1) {};
  \node[vtx] (v2) at (3,0) {};
  \node[vtx] (a) at (2.5,1.67) {};
  \node[vtx, label=0:$\eta$] (k) at (3.4, 1.4) {};
    \node[vtx] (b) at (3.5,2.3) {};
  \node[vtx, label=0:$\ell$] (o2) at (4.5,-1) {};
  \node[vtx, label=0:$\ell$] (o1) at (4.5,3) {};
  \graph[use existing nodes]{
     i1 --[photon] v1;
     b--[majorana, edge label'= $\chi_2$]a;
     v1--[majorana, edge label= $\ell$]a;
     o2--[majorana]v1;
     b--[majorana]o1;
  };
    \draw[thick, dashed]  (2.5,1.67) arc (-140:30:0.6);
  \end{tikzpicture}
\end{minipage}
\vspace{0.5cm}

\begin{minipage}{0.3\textwidth}
\begin{tikzpicture}[thick,scale=0.8, every node/.style={transform shape}]
  \node[vtx, label=180:$Z$] (i1) at (0,1) {};
  \node[vtx] (i2) at (3,2) {};
  \coordinate[] (v1) at (1.5,1) {};
  \node[vtx] (v2) at (3.0,0) {};
  \node[vtx,label=0:$\nu$] (o1) at (4.5,3) {};
  \node[vtx, label=0:$\nu$] (o2) at (4.5,-1) {};
  \graph[use existing nodes]{
     i1 --[photon] v1;
     v1 --[majorana,edge label=$\chi_1$] i2;
     v2 --[majorana] o2;
     i2 --[majorana] o1;
     v2 --[majorana, edge label=$\chi_1$] v1;
     v2--[scalar,edge label'=$\eta$]i2;
  };
  \end{tikzpicture}
\end{minipage}
\begin{minipage}{0.3\textwidth}
\begin{tikzpicture}[thick,scale=0.8, every node/.style={transform shape}]
  \node[vtx, label=180:$Z$] (i1) at (0,1) {};
  \node[vtx] (i2) at (3,2) {};
  \coordinate[] (v1) at (1.5,1) {};
  \node[vtx] (v2) at (3.0,0) {};
  \node[vtx,label=0:$\nu$] (o1) at (4.5,3) {};
  \node[vtx, label=0:$\nu$] (o2) at (4.5,-1) {};
  \graph[use existing nodes]{
     i1 --[photon] v1;
     v1 --[scalar,edge label=$\eta$] i2;
     v2 --[majorana] o2;
     i2 --[majorana] o1;
     v1 --[scalar, edge label'=$\eta$] v2;
     v2--[majorana,edge label'=$\chi_1$]i2;
  };
  \end{tikzpicture}
\end{minipage}
\begin{minipage}{0.3\textwidth}
\begin{tikzpicture}[thick,scale=0.8, every node/.style={transform shape}]
  \node[vtx, label=180:$Z$] (i1) at (0,1) {};
  \coordinate[] (v1) at (1.5,1) {};
  \node[vtx] (v2) at (3,0) {};
  \node[vtx] (a) at (2.5,1.67) {};
\node[vtx] (b) at (3.5,2.3) {};
\node[vtx, label=0:$\eta$] (k) at (3.4, 1.4) {};
  \node[vtx, label=0:$\nu$] (o2) at (4.5,-1) {};
  \node[vtx, label=0:$\nu$] (o1) at (4.5,3) {};
  \graph[use existing nodes]{
     i1 --[photon] v1;
     b--[majorana, edge label'= $\chi_1$]a;
     v1--[majorana, edge label= $\nu$]a;
     o2--[majorana]v1;
     b--[majorana]o1;
  };
    \draw[thick, dashed]  (2.5,1.67) arc (-140:30:0.6);
  \end{tikzpicture}
\end{minipage}
\caption{The leading contributions to $Z\to \ell \ell $ and $Z\to \nu \nu$ in our model. Fermion arrows omitted for brevity, such that each diagram can be associated with multiple flow assignments.}
\label{Zchidiags}
\end{figure}

The leptoquarks $\phi$ and $\varphi$ will modify the $Z$ coupling to leptons through one-loop diagrams involving SM quarks and the vector-like quark $\chi$. For the contributions involving leptoquarks and SM fermions we use the results of ref.~\cite{Arnan:2019olv}, which include corrections due to the external momenta of the $Z$. The additional diagrams with the vector-like quark in the loop are shown in figure~\ref{Zchidiags}. We find the contributions to the leptonic $Z$ couplings from these to be
\begin{align}
    \delta g_{L}^{i j} &= \frac{y^{\chi\varphi}_{j} y^{\chi\varphi *}_{i}}{768 \pi^2 x (x - 1)^4} [x_Z f(x) + x_Z^2 g(x)]
\end{align}
where $x \equiv m_\chi^2 / m_\varphi^2$, $x_Z \equiv m_Z^2 / m_\varphi^2$ and the functions $f(x)$ and $g(x)$ are
\begin{align}
\begin{split}
    f(x) &= 3 x (x - 1) \left[ (4x^3 - 30x + 20) - (x - 1) (19x^2 - 53x + 28) \log x \right]\\ &\quad + 6x\cos^2 \theta_W(x-1) \left[ (x-1)(x^2 - 17x + 10) + 2(x^3 + 6x - 4) \log x \right] \end{split} \\
    \begin{split}
    g(x) &= 5(x-1)(x^3 - 5x^2 + 13x + 3) + 60 x \log x\\ &\quad + \cos^2\theta_W \left[4(x-1)(x^3 - 5x^2 + 13x + 3) - 48 x \log x  \right].
    \end{split}
\end{align}
The couplings $y^{\chi \varphi}_i$ are inversely proportional to $\zeta$, and thus we expect these contributions to be suppressed when $\zeta$ and the $\chi$-$b$ mixing parameter $Y_{b}$ are sizeable.
\subsection{Charm meson decays}
Since couplings to up-type quarks and charged leptons cannot be avoided for the leptoquark that couples to $\chi$, the physics of operators of the form $O_{ijkl} \sim (u_i \Gamma u_j)(\ell_k \Gamma \ell_l)$ is important to study. Here, we consider the leptonic decays of the $D^0$ meson, since a sizeable coupling to the charm quark can assist in the explanation of the large effects seen in the charged current anomalies \cite{Fajfer:2015mia}. The isosinglet LQ $\phi$ generates the entire spectrum of operators which can in principle contribute to the leptonic decays of the $D^0$, since it interacts with both left- and right-chiral SM fermions. Concretely, the dimension-6 Lagrangian
\begin{equation}
  \begin{split}
  \mathcal{L}_{u_i u_j \ell_k \ell_l}
    &= \frac{4 G_F}{\sqrt{2}} \bigg[  C^{ijkl}_{D,V_{R}}  (\bar{u}_i \gamma_\mu P_R u_j)(\bar{\ell}_k \gamma^\mu P_R \ell_l) + C_{D,V_{L}}^{ijkl} (\bar{u}_i \gamma_\mu P_L u_j)(\bar{\ell}_k \gamma^\mu P_L \ell_l)\\ &\quad + C_{D,T}^{ijkl} (\bar{u}_i \sigma_{\mu\nu} P_R u_j)(\bar{\ell}_k \sigma^{\mu\nu} P_R \ell_l) + C_{D,S_L}^{ijkl} (\bar{u}_i P_L u_j)(\bar{\ell}_k P_L \ell_l)\\ &\quad + C_{D,S_R}^{ijkl}(\bar{u}_i P_R u_j)(\bar{\ell}_k P_R \ell_l) + \text{h.c.} 
    \bigg],
  \end{split}
\end{equation}
is generated with tree-level contributions from both leptoquarks:
\begin{align}
  C_{D,V_L}^{ijkl} &= \frac{1}{2\sqrt{2} G_F} \left(\frac{y^{L\phi}_{kj} y^{L\phi*}_{li}}{2m_\phi^2} + \frac{y^{L\varphi}_{kj} y^{L\varphi*}_{li}}{m_\varphi^2} \right),\\
  C_{D,V_R}^{ijkl} &= \frac{1}{4\sqrt{2} G_F} \frac{y^{R\phi*}_{kj} y^{R\phi}_{li}}{m_\phi^2},\\
  C_{D,S_L}^{ijkl} &= \frac{1}{4\sqrt{2} G_F} \frac{y^{L\phi}_{kj}y^{R\phi}_{li}}{m_\phi^2},\\
  C_{D,S_R}^{ijkl} &= \frac{1}{4\sqrt{2} G_F} \frac{y^{R\phi*}_{kj} y^{L\phi*}_{li}}{m_\phi^2},\\
  C_{D,T}^{ijkl} &= -\frac{1}{4} C^{ijkl}_{D,S_L}.
\end{align}

\noindent As highlighted in ref.~\cite{Fajfer:2015mia}, the strongest experimental constraints on these coefficients come from measurements of the process $D^0 \to \ell_i \ell_i$, $i \in \{1, 2\}$. For these decays, we use~\cite{Dorsner:2016wpm}
\begin{equation}
  \begin{aligned}
      \Gamma(D^0 \rightarrow \ell_i \ell_i) &= \frac{f_D^2 m_D^3 G_F^2}{32\pi}\left(\frac{m_D}{m_c}\right)^2\beta_{\ell_i}
      \Bigg[ \left| C_{D,S_L}^{21ii}-C_{D,S_R}^{21ii}\right|^2 \beta_{\ell_i}^2  \\
      & \quad + \left| C_{D,S_L}^{21ii}+ C_{D,S_R}^{21ii} -\frac{2 m_\mu m_c}{m_D^2} (C_{D,V_L}^{21ii}+C_{D,V_R}^{21ii})\right|^2 \Bigg]
    \end{aligned}
\end{equation}
where $\beta_{\ell_i} = (1 - 4m_{\ell_i}^2/m_D^2)^{1/2} \approx 0.99$, $f_D = 212(2)
\text{ MeV}$~\cite{Aoki:2016frl} and $\eta_D = C_{D,S_{L}}^{21ii}(\overline{m_c}) / C_{D,S_{L}}^{21ii}(\Lambda)$. We impose the experimental upper limit $\text{Br}(D^0 \rightarrow
\mu\mu) < 7.6 \cdot 10^{-9}$~\cite{ Aaij:2013cza}. Contributions to the electronic mode from the vector operators are helicity suppressed and we ensure $|y^{R\phi}_{1i}| \ll 1$ in all numerical scans to avoid contributions to the electronic scalar and tensor operators. 

\subsection{Bottom meson decays}
\label{sec:bottommesondecays}
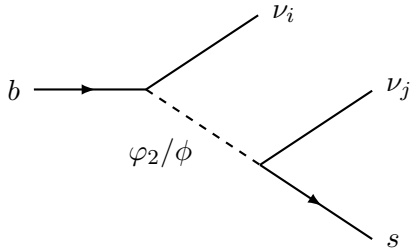
\begin{figure}[t]
\begin{center}
      \hspace{0.5cm}
        \begin{tikzpicture}
  \node[vtx, label=180:$b$] (i1) at (0,1) {};
  \node[vtx, label=0:$\nu_i$] (i2) at (3,2) {};
  \coordinate[] (v1) at (1.5,1) {};
  \coordinate[] (v2) at (3.0,0) {};
  \node[vtx, label=0:$\nu_j$] (o1) at (4.5,1) {};
  \node[vtx, label=0:$s$] (o2) at (4.5,-1) {};
  \graph[use existing nodes]{
     i1 --[fermion] v1;
     i2 --[majorana] v1;
     v2 --[fermion] o2 ;
     v2--[majorana] o1;
     v1 --[scalar, edge label'=$\varphi_2/\phi$] v2;
  };
  \end{tikzpicture}
    \caption{Dominant BSM contribution to $b\to s\nu \nu$ process in this model arises at tree-level.}\label{bsnunu}
      \end{center}
\end{figure}

The $b\to s \nu \nu$ transition provides one of the theoretically cleanest FCNC processes. Predictions for this transition are devoid of hadronic uncertainty, beyond the form-factors, unlike the $b\to s \ell\ell$ transition. This makes measurements of $B\to K \nu \nu$  highly useful for constraining BSM in the flavour sector. 

In this model, there are two types of contribution to this process, depicted in figure~\ref{bsnunu}: one with two neutrinos in the final state, and one with a neutrino and its charge-conjugate.  Assuming the $\chi$-quark mixing angle is small, then the dominant contribution is to:
\begin{equation}
\label{eq:onunu}
\mathcal{O}^{ij}_{\nu\nu} =\frac{8 G_F}{\sqrt{2}} \frac{\alpha}{4\pi} V_{tb} V_{ts}^*[\overline{\nu}_i  \gamma_\mu P_L \nu_j][\overline{s} \gamma^\mu P_L b],
\end{equation}
parameterised by the BSM coefficients:
$$w_{\nu\nu}^{ij}
\sim \frac{c_{\theta_L}}{2\sqrt{2} G_F V_{tb} V_{ts}^*} \frac{4\pi}{\alpha}\left(\frac{x^{L\phi}_{i3} x^{L\phi*}_{j2} }{m_{\phi}^2} +\frac{x_{i3}^{L\varphi}   x^{L\varphi*}_{j2} }{2m_{\varphi}^2}\right).$$

We do not presume that the final-state leptons are of the same flavour --  all combinations of $i$ and $j$ can be incorporated into a fit using the \texttt{Flavio} software. However, as was noted earlier, of particular importance are the contributions from the lepton-number conserving processes which can interfere with the SM contributions. Additionally, the recovery of lepton-number as a global symmetry in the large-mass limit for exotics motivates this parameter choice.

Contrasting the contributions to $w_{\nu\nu}$ with the structure of the BSM contributions to other anomalies, we  observe significant overlap in the relevant observables for fitting measurements of $B\to K \nu \nu$ and in explaining the $R_D$ and $R_{D^*}$ anomalies. Consequentially, we expect $b\to s \nu\nu$ measurements to provide some of the most stringent constraints on free parameters relevant for BSM in the $b\to c\tau \nu$ transition.

Experimental upper bounds exists for the branching fractions of $B\to K^{(*)} \nu \overline{\nu}$ from the Belle collaboration~\cite{Grygier:2017tzo}:
\begin{align}
\text{Br}(B\to K \nu \overline{\nu})< 1.6\times 10^{-5} \;\; (90\% ~\text{confidence limit}),\\
\text{Br}(B\to K^* \nu \overline{\nu})< 2.7\times 10^{-5} \;\; (90\% ~\text{confidence limit}),
\end{align}
which correspond to the following ratios, again within a $90\%$ confidence interval
\begin{align}
r_K^{\nu\nu} \equiv \frac{[\text{Br}(B\to K \nu \overline{\nu})]_{\text{BSM}+\text{SM}}}{[\text{Br}(B\to K \nu \overline{\nu})]_{\text{SM}} } < 3.9,\\
r_{K^*}^{\nu\nu} \equiv \frac{[\text{Br}(B\to K^* \nu \overline{\nu})]_{\text{BSM}+\text{SM}}}{[\text{Br}(B\to K^* \nu \overline{\nu})]_{\text{SM}} } < 2.7.
\end{align}

\subsection{Meson mixing}
\label{sec:mesonmixing}

The process of $B_s$--$\bar{B}_s$ mixing provides a complementary constraint on the same couplings involved in the $b \to s \nu \nu$ processes discussed in section~\ref{sec:bottommesondecays}. It was found in refs.~\cite{Bauer:2015knc, Cai:2017wry} that the process leads to a weaker constraint than $B \to K \nu \nu$ and $B \to K^* \nu \nu$ for low $\phi$ masses, but becomes relevant for masses larger than a few TeV. In our model, we have contributions from both the isosinglet and isotriplet LQs through box diagrams with neutrinos and charged leptons in the loop. These contribute to the operator $C^{bs}_1$,
\begin{equation}
\mathcal{L}_{\Delta B = 2} \supset  C^{bs}_1 (\bar{b} \gamma_\mu P_L s)(\bar{b} \gamma^\mu P_L s),
\end{equation}
where colour indices are contracted within parentheses. The combination $C_{B_s} \exp{2 i \phi_{B_s}} = \Delta m_s^{\text{exp}} / \Delta m_s^{\text{SM}}$ is calculated using \texttt{SPheno} \cite{Vicente:2015zba, Porod:2003um, Porod:2011nf}, and we impose the UT\textit{fit} collaboration's result~\cite{Bona:2007vi}
\begin{equation}
C_{B_s} = 1.110 \pm 0.090
\end{equation}
in our numerical scans. We will work with small imaginary parts for the couplings fixed by the neutrino mass and we maintain $\phi_{B_s} \approx 0$ for the phase, consistent with UT\textit{fit}'s result.

\subsection{Summary of constraints}
\label{sec:summaryofconstraints}

In tables \ref{table:muonelectron} and \ref{tab:summary} we present summaries of the constraints of this section. The tables contain the observables we consider in our later phenomenological analysis as well as the limits we require.
\begin{table}[t]
    \centering
    \begin{tabular}{c|c|c}
    Process & Quantity & Requirement \\
    \hline
       $ss \to \mu \mu$                 &  $|x_{22}^{L\varphi}|$ & $< 0.41 m_{\varphi} / \text{ TeV}$~\cite{Angelescu:2018tyl} \\
       $bb \to \mu \mu$                 &  $|x_{23}^{L\varphi}|$ & $< 0.58 m_{\varphi} / \text{ TeV}$~\cite{Angelescu:2018tyl} \\
       $ss \to \tau \tau$                 &  $|x_{32}^{L\varphi}|$ & $< 0.54 m_{\varphi} / \text{ TeV}$~\cite{Angelescu:2018tyl} \\
       $bb \to \tau \tau$                 &  $|x_{33}^{L\varphi}|$ & $< 0.80 m_{\varphi} / \text{ TeV}$~\cite{Angelescu:2018tyl} \\
       $bb \to e e$                 &  $|x_{13}^{L\varphi}|$ & $< 0.44 m_{\varphi} / \text{ TeV}$~\cite{Greljo:2017vvb} \\
       $Z \to b b$                      &  $C_{Hd}^{33}$ & $\in [-0.38, 0.03]$~\cite{Ciuchini:2013pca} \\
       $\tau \to \eta e$                &  Br  & $< 9.2 \cdot 10^{-8}$ \\
       $\tau \to \pi e$                 &  Br  & $< 8.0 \cdot 10^{-8}$ \\
       $\tau \to \phi \mu$              &  Br  & $< 8.4 \cdot 10^{-8}$ \\
       $Z \to e^{\pm} \mu^{\mp}$                    &  Br  & $< 7.5 \cdot 10^{-7}$ \\
       $Z \to e^{\pm} \tau^{\mp}$                   &  Br  & $< 9.8 \cdot 10^{-6}$ \\
       $Z \to \mu^{\pm} \tau^{\mp}$                 &  Br  & $< 1.2 \cdot 10^{-5}$ \\ 
       $Z \to \ell_i \ell_i$                      &  $g_{L}$  & $\in [-8.5, 12] \cdot 10^{-4}$ \\
       $Z \to \ell_i \ell_i$                      &  $g_{R}$  & $\in [-5.4, 6.7] \cdot 10^{-4}$ \\
       $Z \to \nu_i \nu_i$              &  $N_\nu$  & within $2.9840 \pm 0.0164$ \\
       $D^0 \to \mu \mu$                &  Br  & $< 7.6 \cdot 10^{-9}$~\cite{Aaij:2013cza} \\
       $B^+ \to K^+ e^{\pm} \mu^{\mp}$                  &  Br  & $< 9.1 \cdot 10^{-8}$ \\
       $B^0 \to K^{0*} e^{\pm} \mu^{\mp}$                &  Br  & $< 1.8 \cdot 10^{-7}$ \\
       $B_s \to \mu^{\pm} e^{\mp}$                  &  Br  & $< 5.4 \cdot 10^{-9}$ \\
       $B \to D \ell \nu$               &  $R_{D}^{\mu / e} = \frac{\text{Br}(B \to D \mu \nu)}{\text{Br}(B \to D e \nu)}$  & within $0.995 \pm 0.090$~\cite{Glattauer:2015teq} \\
       $B \to D^* \ell \nu$             &  $R_{D^*}^{e / \mu} = \frac{\text{Br}(B \to D^* e \nu)}{\text{Br}(B \to D^* \mu \nu)}$  & within $1.04 \pm 0.10$~\cite{Abdesselam:2017kjf} \\
       $B_s$--$\bar{B}_s$ mixing        &  $C_{B_s}$  & $\in [0.942, 1.288]$~\cite{Bona:2007vi} \\
       $B \to K \nu \nu$                 &  $r_K^{\nu\nu} = \frac{\text{Br}}{\text{Br}_{\text{SM}}}$ & $< 3.9$~\cite{Grygier:2017tzo} \\
       $B \to K^* \nu \nu$                 &  $r_{K^*}^{\nu\nu} = \frac{\text{Br}}{\text{Br}_{\text{SM}}}$ & $< 2.7$~\cite{Grygier:2017tzo}\\
       $b \to s \gamma$                 &  Br & $\in [-0.17, 0.24]$~\cite{Amhis:2014hma} \\
       $B_c \to \tau \nu$                 &  Br & $< 30 \%$~\cite{Alonso:2016oyd} \\
       $K \to \ell \nu$                 &  $r_K^{\mu/e} = \frac{\text{Br}(K \to e \nu)}{\text{Br}(K \to \mu \nu)}$  & within $(2.488 \pm 0.018) \cdot 10^{-5}$
    \end{tabular}
    \caption{The table is a summary of the constraints considered in this section, not also mentioned in table~\ref{table:muonelectron}. In cases where opposite-sign lepton pairs can have differing flavour, we choose the observable with both combinations of signs averaged. For the rare tau decays not elsewhere referenced, we have included only those which we found gave most competitive constraints.} Where citations are omitted the requirements are taken from ref.~\cite{PhysRevD.98.030001}.
    \label{tab:summary}
\end{table}

\section{Results and Discussion}
\label{sec:results}

Below we explore the extent to which this model can accommodate the charged- and neutral-current anomalies, the anomalous magnetic moment of the
muon and neutrino mass in light of the constraints presented in the previous section. 

First, we review the minimal setup introduced in section~\ref{sec:constraints} and present the results of our Monte Carlo analysis. We comment briefly on non-minimal scenarios in section~\ref{sec:vectorOp}.

\subsection{Monte Carlo analysis}

In the minimal scenario the deviations in $R_{D^{(*)}}$ are explained by the
isosinglet leptoquark $\phi$ with contributions in the direction $C_{S_L}(m_\phi) = -4
C_T(m_\phi)$, implying $\mathcal{O}(1)$ values for the couplings
$x^{L\phi}_{33}$ and $y^{R\phi}_{32}$~\cite{Cai:2017wry, Angelescu:2018tyl, Feruglio:2018fxo}. Contributions to the vector operator are
more heavily constrained since they necessarily imply large effects in $B \to
K^{(*)} \nu\nu$ and $B_s$--$\bar{B}_s$ mixing, in the absence of any kind of
cancellation (see section~\ref{sec:vectorOp}). The $\phi$ particle also explains
the anomalous magnetic moment of the muon with the values of $y^{R\phi}_{23}$
and $y^{L\phi}_{23} = x_{23}^{L\phi}$ fixed according to eq.~\eqref{eq:amu}. The
limits derived in section~\ref{sec:reg2} suggest the extent to which $\phi$ can
contribute to the generation of neutrino masses is small. Since we consider
suppressed LQ mixing, this means there is no connection between the neutrino
mass mechanism and the anomalies in $R_{D^{(*)}}$ and $(g-2)_{\mu}$ in this
model. For this reason, we fix $m_\phi$ and the couplings involved in
eq.~\eqref{eq:amu} and eq.~\eqref{eq:csl} to meet the respective central values
to explain these deviations. Explicitly, the conditions
\begin{equation}
\text{Re}(x_{23}^{L\phi} y_{23}^{R\phi}) \approx \frac{0.004 \hat{m}_\phi^2}{1 + \log \hat{m}_\phi} \quad \text{ and } \quad x_{33}^{L\phi} y_{32}^{R\phi} \approx 2.7 C_{S_L} \hat{m}_\phi^2,
\end{equation}
with $\hat{m}_\phi = m_\phi / \text{TeV}$, are met with $m_\phi = 2 \text{ TeV}$, $C_{S_L} = 0.14$, $x_{33}^{L\phi} = 0.7$, $y^{R\phi}_{32} = 2.15$, $y^{R\phi}_{23} = 0.5$ and $x_{23}^{L\phi} = 0.02$ in all results presented in this section. Many of the
implications of explaining $R_{D^{(*)}}$ with $C_{S_L} (\Lambda) = -4 C_T(\Lambda)$ have been discussed
in the literature~\cite{Feruglio:2018fxo, Asadi:2018sym, Alok:2019uqc}. Here we expand briefly on some of these. 

The fit we present in section~\ref{sec:bctaunu} does not include the less-precisely measured observables $R_{J/\psi}$, $f_L^{D^*}$ and $\mathcal{P}_\tau^*$, introduced in section.~\ref{sec:chargedcurrentprocesses}. We instead use the preferred values from our fit to make predictions for these observables, concentrating on the scalar--tensor solution, since this is the easiest to accommodate with the $\phi$ LQ. We note that this solution gives negligible efficiency variation from the SM for the measurement in the $D^*$ mode~\cite{Sato:2016svk}, predicts $\text{Br}(B_c \to \tau \nu) \lesssim 30\%$~\cite{Cai:2017wry, Angelescu:2018tyl, Bardhan:2019ljo} and displays a $q^2$ spectrum that agrees well with experiment~\cite{Freytsis:2015qca}. 

\begin{figure}[t] 
  \centering
  \includegraphics[width=\textwidth]{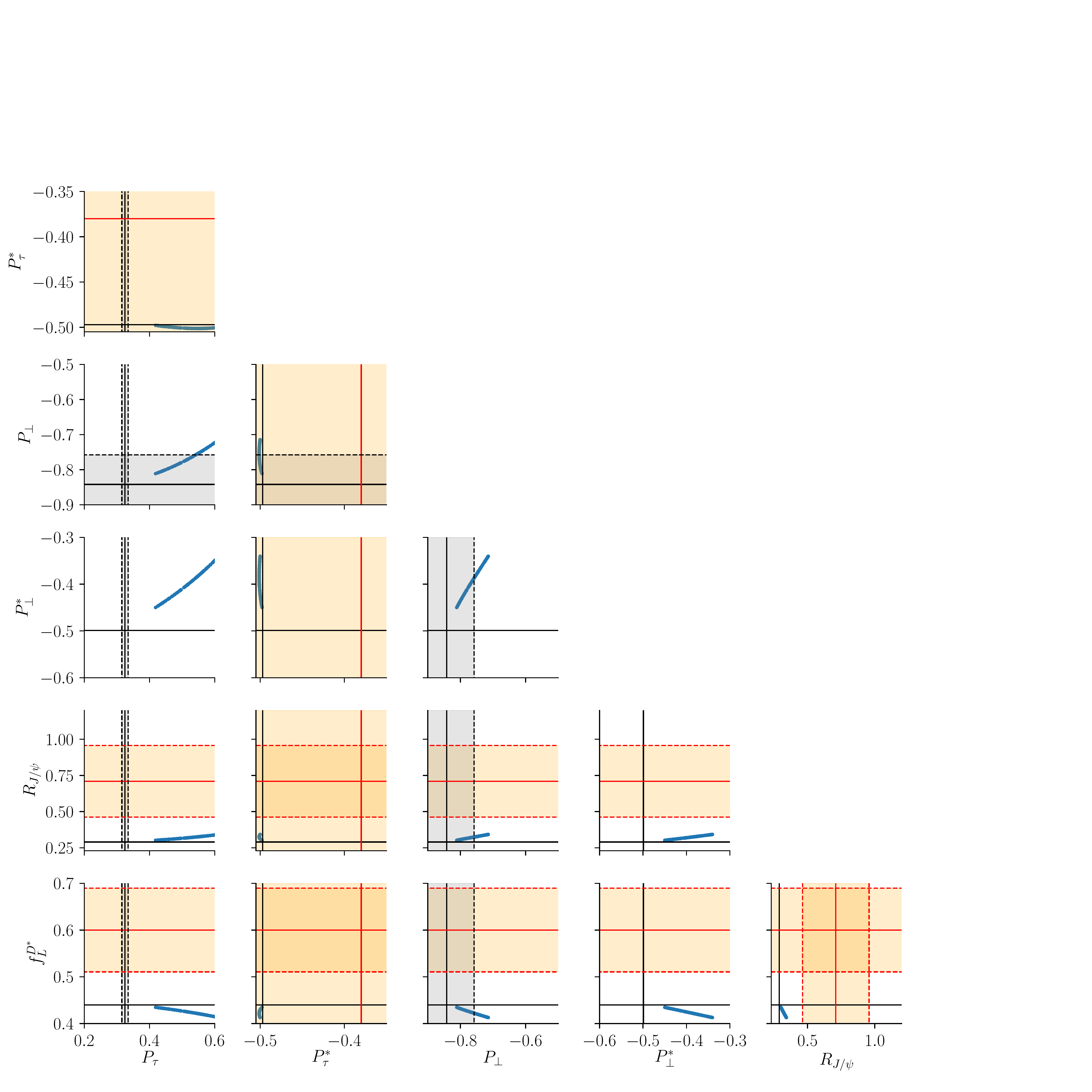}
  \caption{A grid plot of the various $b \to c$ related observables in addition to $R_{D}$ and $R_{D^*}$ considered in our analysis. Solid black lines represent the SM predictions around which the grey shaded regions are the Belle II $50 \text{ ab}^{-1}$ sensitivities~\cite{Alonso:2017ktd}, bordered by the black dashed lines. Red lines are current measurements and orange regions are their $1\sigma$ errors. Where the Belle II sensitivity is unavailable we present only the SM prediction without a shaded region. The blue points explain $R_{D^{(*)}}$ to $2 \sigma$.}
  \label{fig:btocpredictions}
\end{figure}

In figure~\ref{fig:btocpredictions} we project the $2\sigma$ preferred region for $C_{S_L}$ (see Table~\ref{tab:fitresults}) onto combinations of $b \to c$ related observables to illustrate the ability of combined measurements to close in on this scenario. Were possible, we have also shown Belle II $50 \text{ ab}^{-1}$ sensitivity~\cite{Alonso:2017ktd} in grey centred around the SM prediction in black. Current measurements are shown in red with their $1 \sigma$ errors in orange. With contributions in the scalar--tensor direction, the $\phi$ leptoquark's contributions to $f_L^{D^*}$ are in the opposite direction to current measurements, although still within the $2\sigma$ region. If the central value of $f_L^{D^*}$ stays close to where it is, or moves down slightly, the model would then predict $\mathcal{P}_\tau \approx 0.4$, which compromises the potential mild improvement in $R_{J/\psi}$ the model can offer.  This scenario leads to a SM-like $\mathcal{P}_{\tau}^*$, but potentially large deviations in the $\mathcal{P}_{\perp}^{(*)}$ observables.

The isotriplet scalar $\varphi$ explains the neutral-current anomalies and
participates in the neutrino-mass generation. Thus, the couplings entering the
expression for $C^{\mu\mu}_{9} = - C^{\mu\mu}_{10}$ [eq.~\eqref{eq:c9c10}] are fixed
by the Casas-Ibarra parametrisation, itself following from the structure of the
neutrino-mass matrix. A consequence of this is that the $x_{i3}^{L\varphi}$ take
complex values and in general $\text{Im}(C^{\mu\mu}_{9}) \neq 0$. Indeed, for
$\zeta \in \mathbb{R}$ the imaginary part of $C^{\mu\mu}_{9}$ is much larger than
the real part, since $\text{Re}(x_{23}^{L\varphi}) = \sqrt{m_{2}/m_{3}}
\text{Im}(x_{23}^{L\varphi})$ from eq.~\eqref{eq:yuknux}. Although $\text{Im}
C^{\mu\mu}_{9} > \text{Re} C^{\mu\mu}_{9}$ may lead to an acceptable explanation of
the $b \to s$ anomalies (see, for example, Appendix C of ref.~\cite{Altmannshofer:2014rta}), most fits in the literature
assume $\text{Im} C^{\mu\mu}_{9} = 0$ and we aim to reproduce this in our model as
well. The simplest way to do this is to assume $\arg \zeta \approx \pi /
2$, so that $\zeta$ is mostly imaginary. This now implies
$\text{Re}(x_{23}^{L\varphi}) = \sqrt{m_{3}/m_{2}}
\text{Im}(x_{23}^{L\varphi})$, and so the muonic couplings of $\varphi$ are
mostly real.

\begin{figure}[t] 
  \centering
  \includegraphics[width=0.49\textwidth]{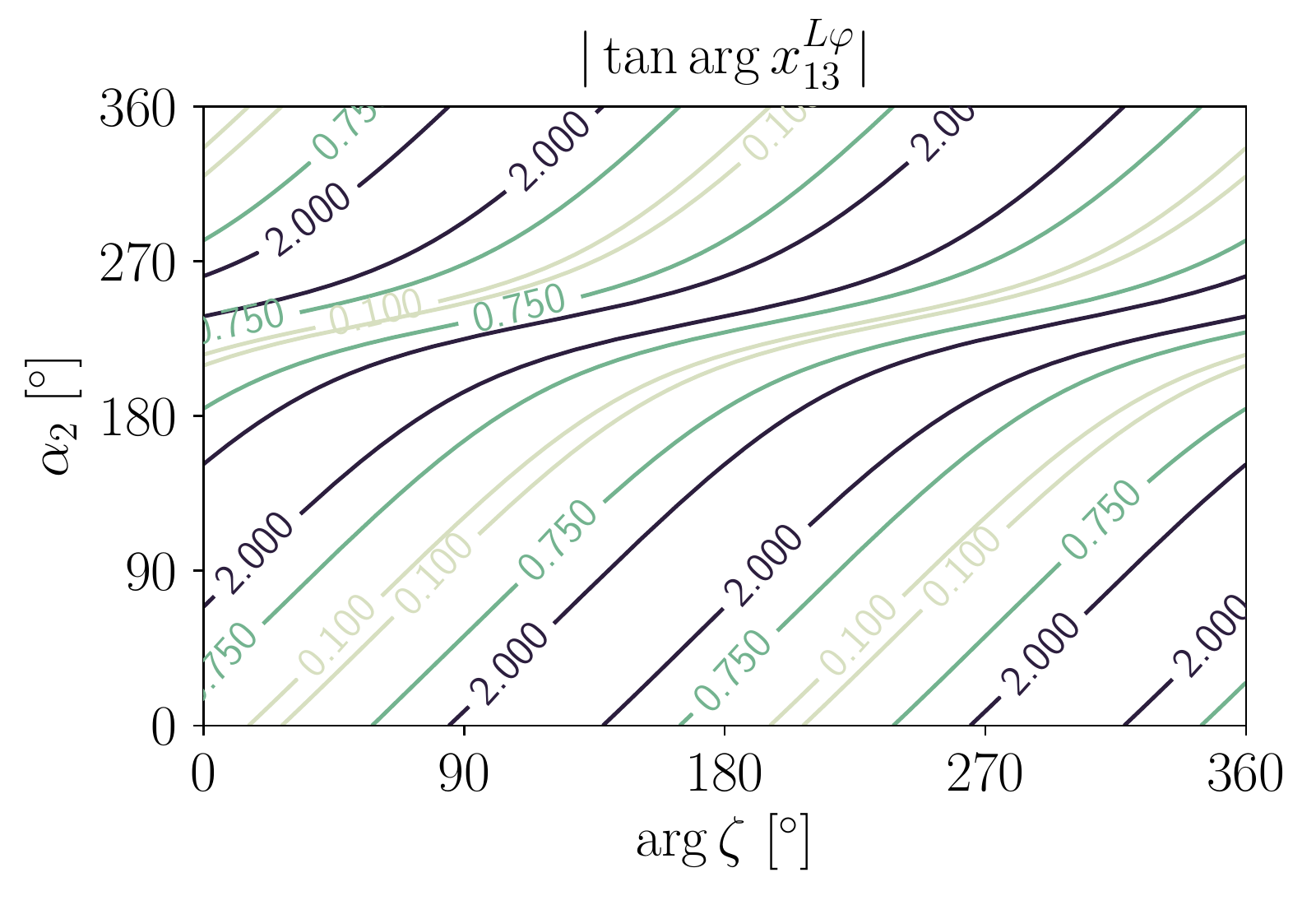}
  \includegraphics[width=0.49\textwidth]{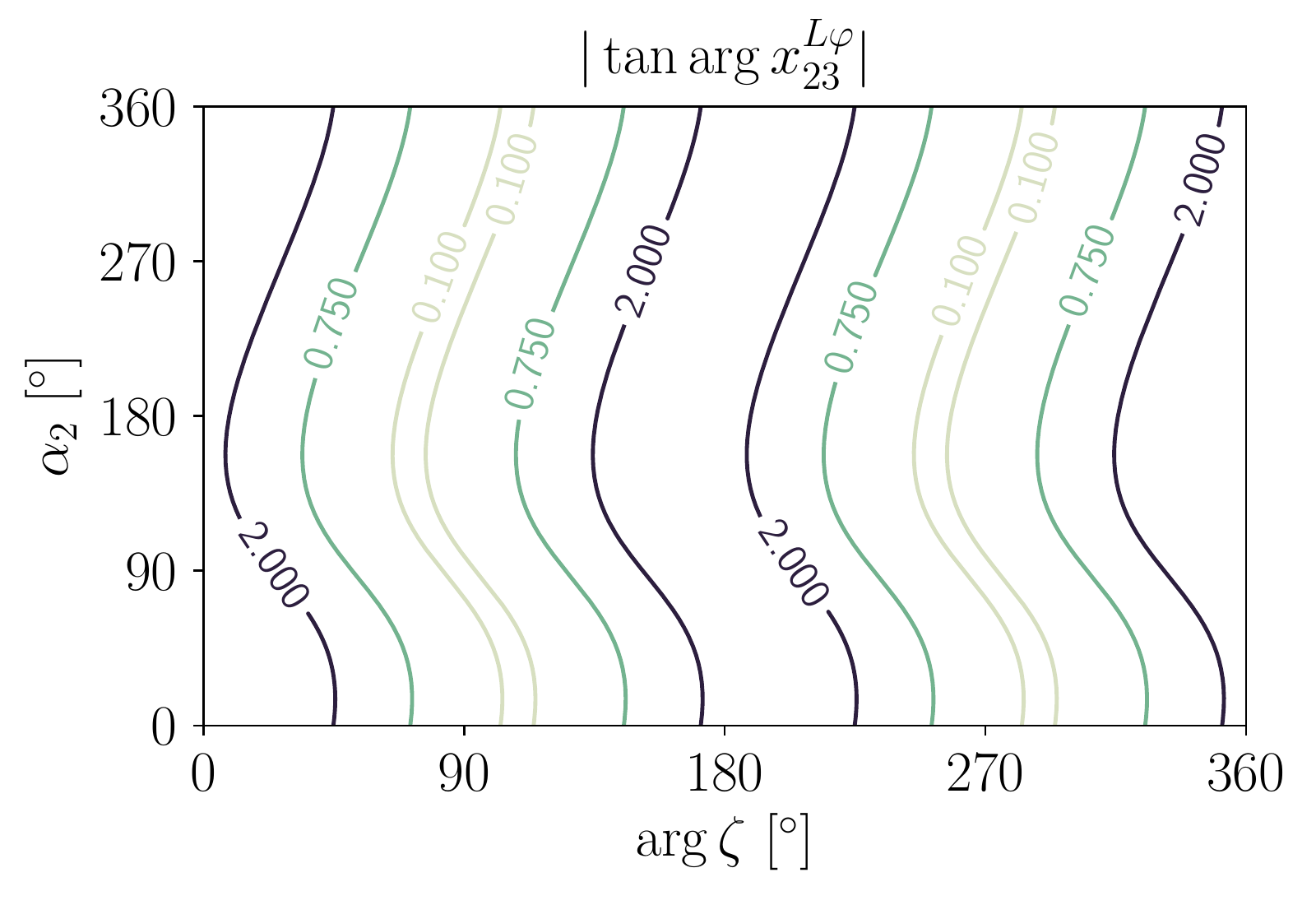}
  \includegraphics[width=0.49\textwidth]{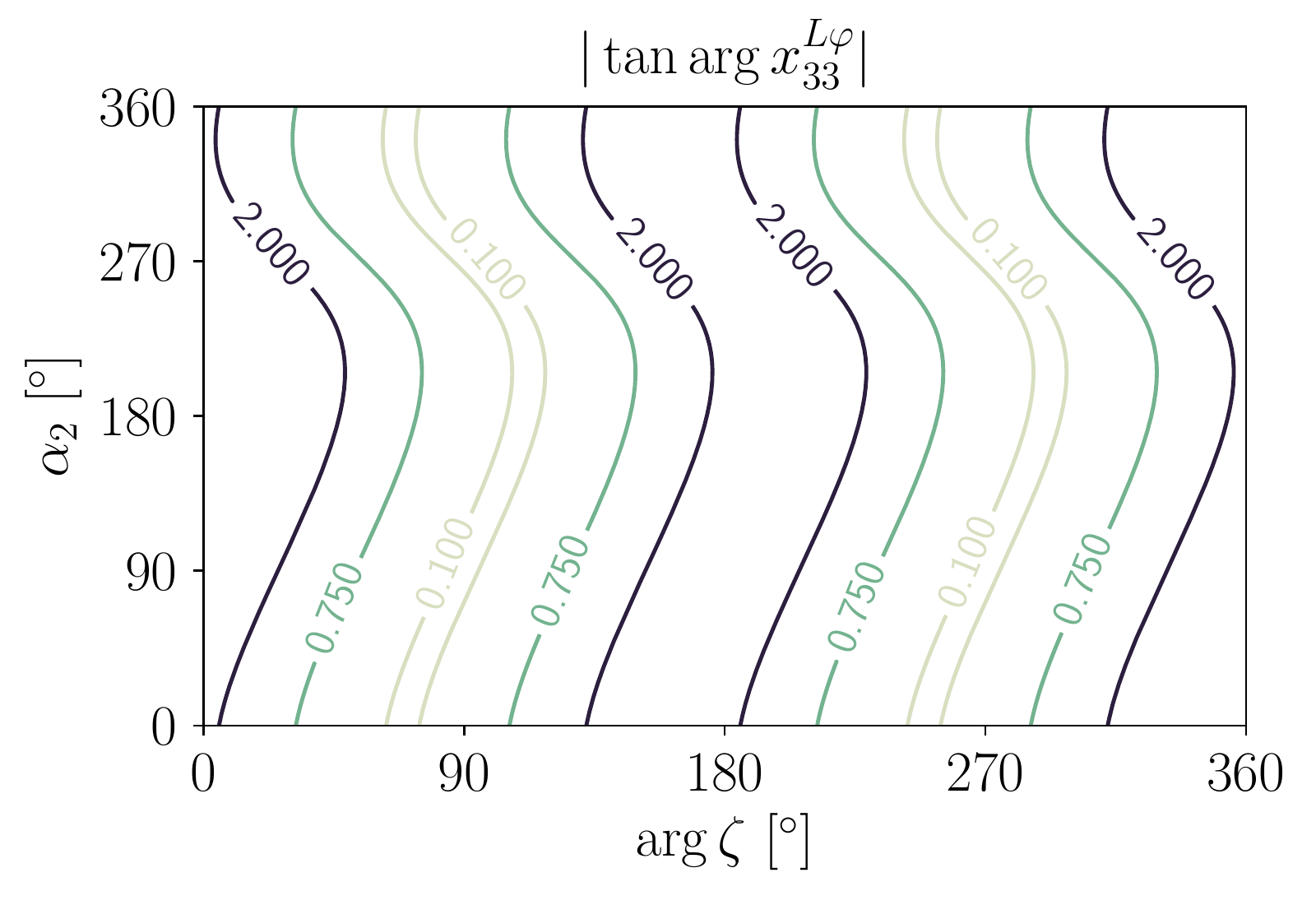}
  \caption{Contours of $|\tan \arg x_{i3}^{L\varphi}| = |\text{Im}(
    x^{L\varphi}_{i3}) / \text{Re}( x^{L\varphi}_{i3})|$ with varying $\arg
    \zeta$ and Majorana phase $\alpha_{2}$. The index $i$ enumerates over
    charged-lepton flavours. It is clear that the choice $\arg \zeta \approx
    \pi/2$ ensures $\text{Im}x^{L\varphi}_{i3} \ll \text{Re}x^{L\varphi}_{i3}$
    for the muon and tau couplings ($i = 2, 3$). The ratio of the imaginary and
    real parts of the electron coupling ($i = 1$) varies significantly with
    $\alpha_{2}$. The Dirac phase and all other neutrino parameters have been
    set to their central values.}
  \label{fig:c9rat}
\end{figure}

One may worry that the central value of $\delta_{CP}$ (used in our numerical
analysis) or a non-zero value for the Majorana phase $\alpha_{2}$ will spoil the
desired $\text{Im} C^{\mu\mu}_{9} \ll \text{Re} C^{\mu\mu}_{9}$. Using
$\text{Im}C^{\mu\mu}_{9} / \text{Re}C^{\mu\mu}_{9} = \tan \arg C_{9}^{\mu\mu}$ as a
measure of the relative size of imaginary part of $C_{9}^{\mu\mu}$, we find
\begin{equation}
\small
  \label{eq:c9rat}
    \tan \arg C_{9}^{\mu\mu} \approx - \cot \arg \zeta + \sqrt{\frac{m_{2}}{m_{3}}}[0.085 \cos (\alpha_{2} + \delta_{CP}) - 0.72 \cos\alpha_{2}] \csc^{2}\arg \zeta+ \mathcal{O}\left(\frac{m_{2}}{m_{3}}\right),
\end{equation}
\normalsize
for our model, derived from eq.~\eqref{eq:yuknux} and eq.~\eqref{eq:c9c10}. This clarifies that the effects of the phases are subleading in
$\sqrt{m_{2} / m_{3}}$. In figure~\ref{fig:c9rat} we plot $\text{Im}(
x^{L\varphi}_{i3}) / \text{Re}( x^{L\varphi}_{i3})$ for $i = 1, 2, 3$ as
contours with varying $\arg \zeta$ and $\alpha_{2}$. This illustrates the
behaviour discussed above but also investigates the effect on the other leptonic
couplings. It is evident that the choice $\arg \zeta \approx \pi/2$ also leaves
the tau coupling mostly real, although this cannot be said for the electron
coupling where the dependence on $\alpha_{2}$ is significant. We nevertheless
proceed with the choice $\arg \zeta = \pi / 2$ in our numerical analysis and
account for the possibility of a large imaginary part in the coupling
$x_{13}^{L\varphi}$.

To explore the extent to which this scenario can explain the flavour anomalies
and neutrino mass, we perform a random scan over the 5 free parameters of the
setup: $|\zeta|$, $x^{L\varphi}_{22}$, $m_{\varphi}$, $m_{\chi}$, $\alpha_{2}$.
Random values are drawn uniformly over the intervals defined for these
parameters in table~\ref{tab:scenarioi}. Notably, the Yukawa coupling $Y_{b}$
is fixed to $0.25 m_{\chi} / \text{TeV}$, the lower edge of the $2\sigma$ region
from eq.~\eqref{eq:bmixinglimit} needed to explain the small discrepancy in $Z
\to b\bar{b}$. We generate $2 \cdot 10^6$ points which are filtered through all of the constraints presented in section.~\ref{sec:constraints}.
\begin{table}[t]
  \centering
  \begin{tabular}{l||c|c|c|c|c}
    Parameters & $|\zeta|$ & $x^{L\varphi}_{ 
22}$ & $m_{\varphi}$ & $m_{\chi}$ & $\alpha_{2}$ \\
    \hline        
    Interval & $[1, 600]$ & $[0, \sqrt{4\pi}]$ & $[1, 30] \text{ TeV}$ & $[1, 10] \text{ TeV}$ & $[0, 2\pi]$
  \end{tabular}
  \caption{The table shows the intervals from which the corresponding free
    parameters are randomly drawn for our Monte Carlo analysis. All other parameters are fixed, see the text for details.}
  \label{tab:scenarioi}
\end{table}
 
 \begin{figure}[t]
  \centering
  \includegraphics[width=0.495\linewidth]{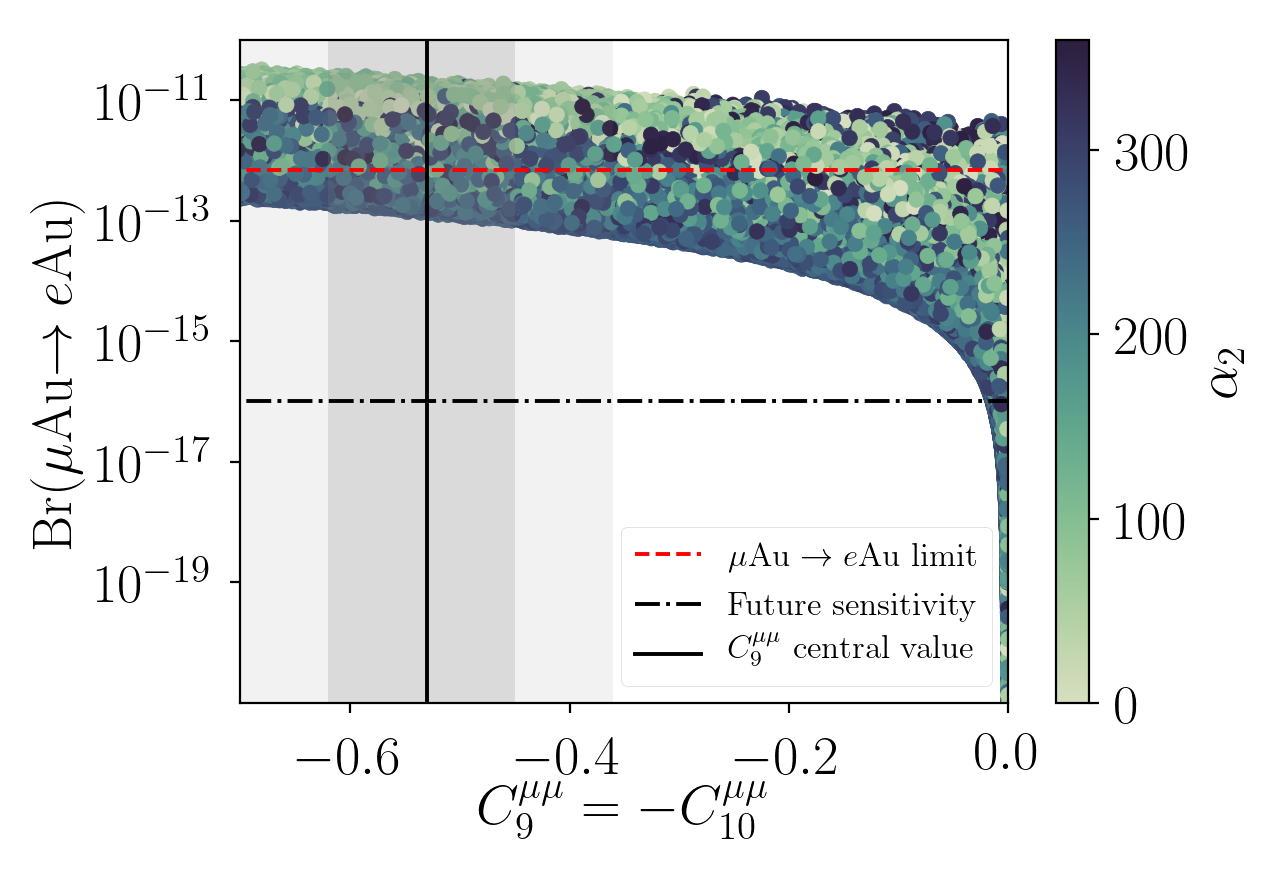}
  \includegraphics[width=0.495\linewidth]{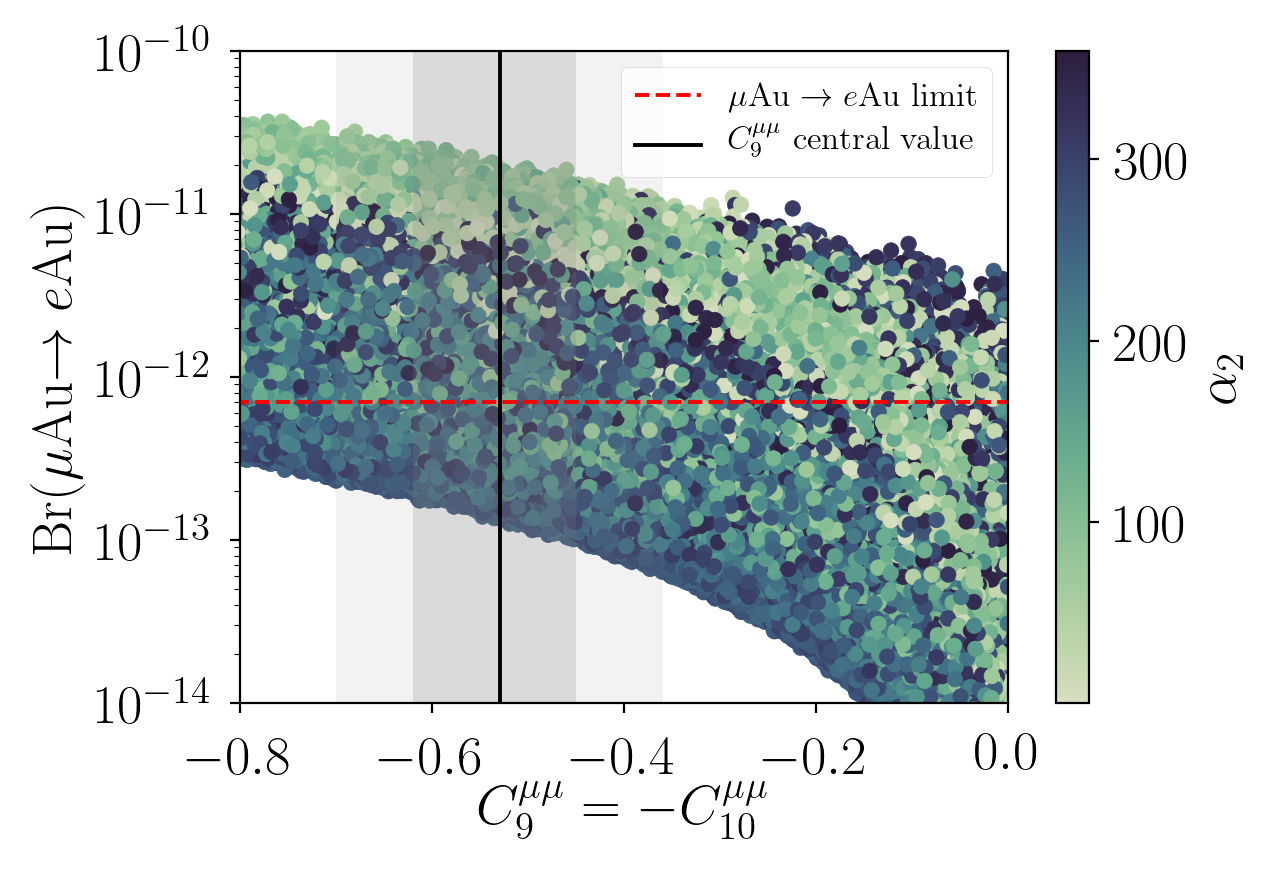}
  \caption{The results of the random scan projected onto $\text{Br}(\mu \text{Au} \to e \text{Au})$ and
    $C_{9}^{\mu\mu}$. All constraints except muon--electron conversion in Gold
    have been applied. The solid black line represents the central value of
    the fit of ref.~\cite{Aebischer:2019mlg} to the anomalous $b \to s$ data, and the heavier
    and lighter shaded regions are the $1$ and $2\sigma$ regions. The dashed
    red line corresponds to the current most stringent limit on $\text{Br}(\mu
    \text{Au} \to e \text{Au})$ from SINDRUM II~\cite{Bertl:2006up} and the black dot-dashed line is a representation of the
    projected sensitivity of future experimental reach \cite{KURUP201138, Cui:2009zz, Wu:2017zwh, Adamov:2018vin, Bartoszek:2014mya, Pezzullo:2018fzp, Bonventre:2019grv}. The plot on the right is an
    enlarged look at the interesting region of the plot on the left. The
    colour axis represents the value of the Majorana phase $\alpha_{2}$.}
  \label{fig:m2ecVSc9}
\end{figure}
\begin{figure}[b!]
  \centering
  \begin{subfigure}{0.495\linewidth}
    \centering
    \includegraphics[width=\textwidth]{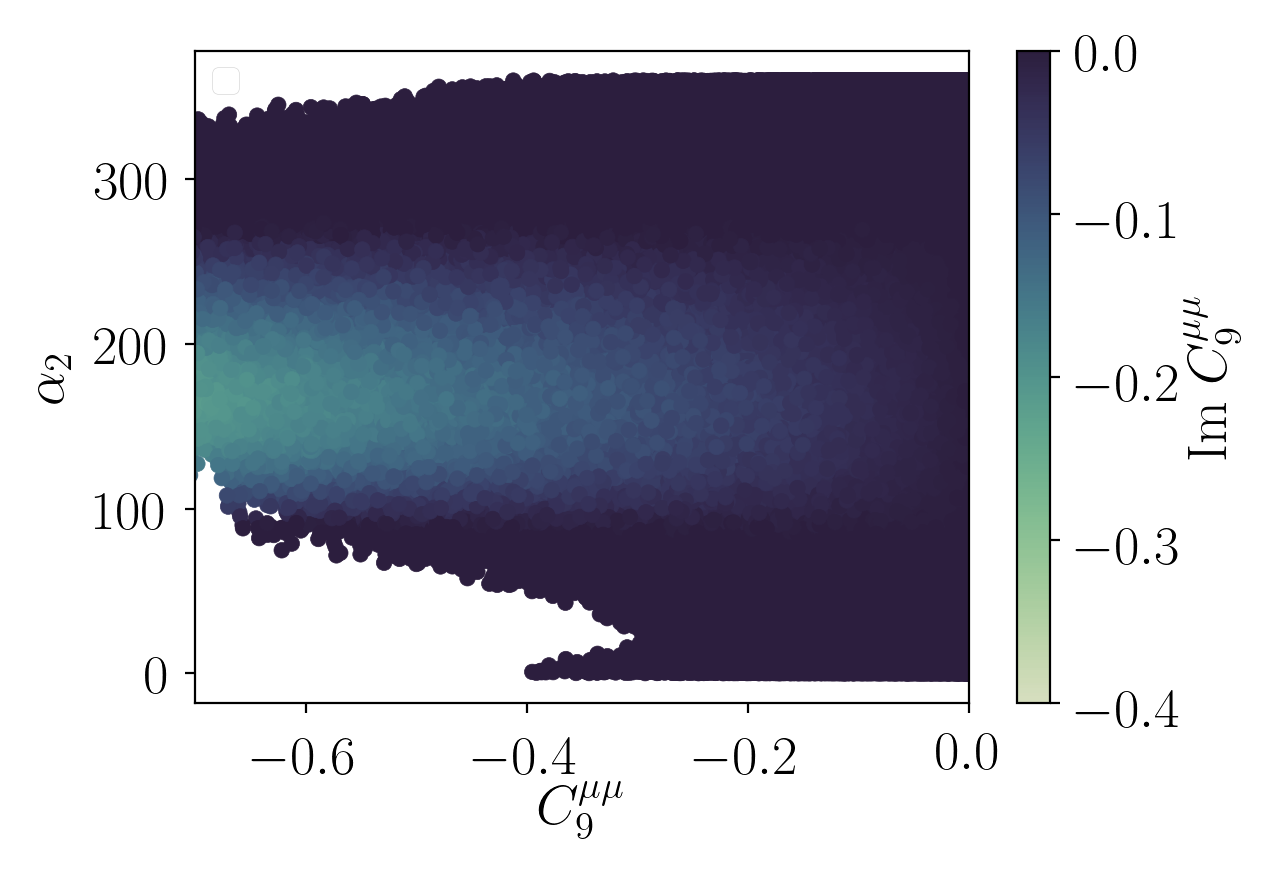}
    \caption{}
    \label{fig:subfig1}
  \end{subfigure}
  \begin{subfigure}{0.495\linewidth}
    \centering
    \includegraphics[width=\textwidth]{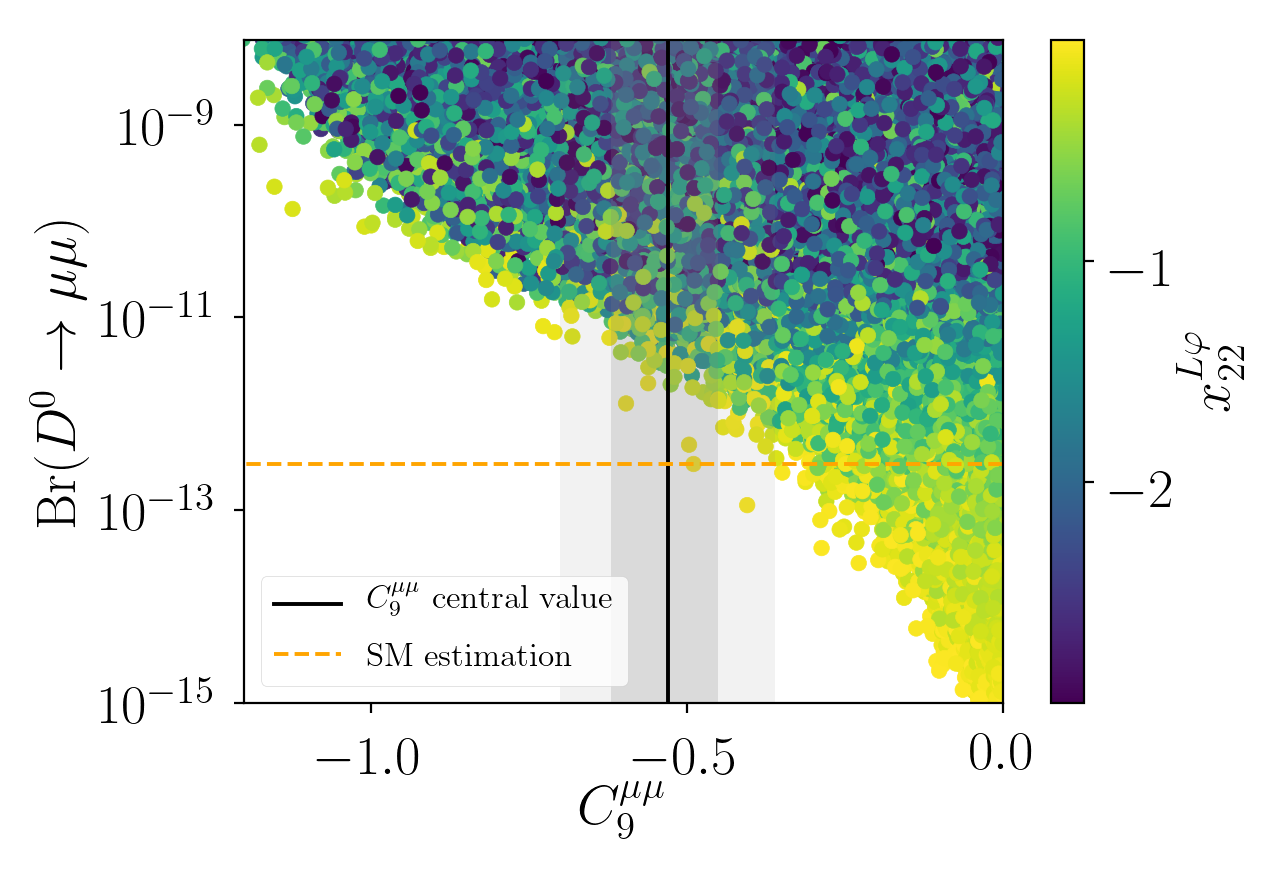}
    \caption{}
    \label{fig:subfig2}
  \end{subfigure}
  \begin{subfigure}{0.497\linewidth}
    \centering
    \includegraphics[width=\textwidth]{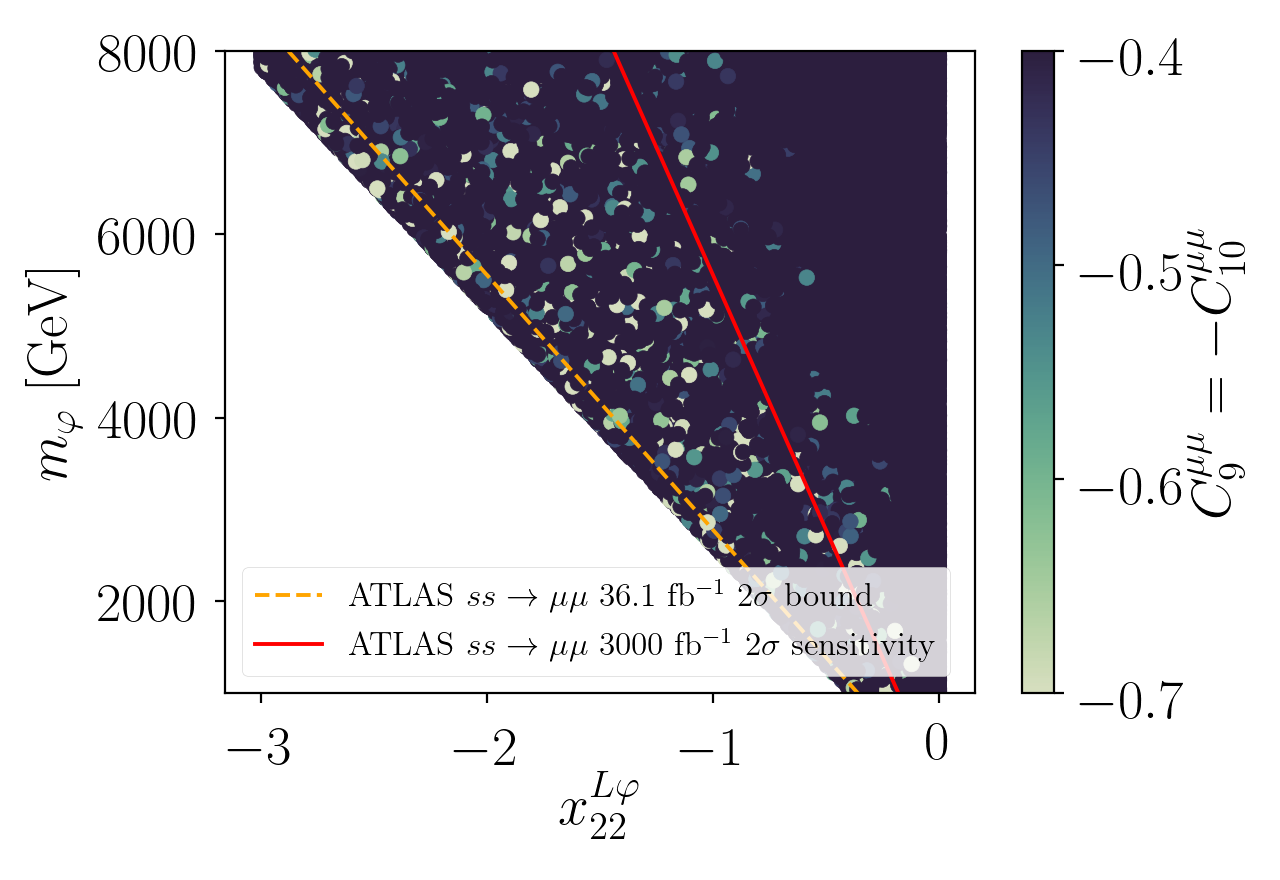}
    \caption{}
    \label{fig:subfig3}
  \end{subfigure}
  \caption{The other interesting results of our Monte Carlo analysis. (a) The relation
    between $C_{9}^{\mu\mu}$ and the Majorana phase $\alpha_{2}$. The points shown
    pass all of the constraints considered in our analysis. The colour axis
    represents the imaginary part of $C_{9}^{\mu\mu}$. The plot shows the
    preference away from a vanishing $\alpha_{2}$, driven by the constraint
    $\text{Br}(\mu \text{Au} \to e \text{Au})$, and the consistency of the
    available parameter space with a small imaginary part of $C_{9}^{\mu\mu}$. (b)
    The results of the random scan projected onto $\text{Br}(D^{0} \to \mu
    \mu)$ and $C_{9}^{\mu\mu}$. Points shown pass all constraints. Our model
    predicts $\text{Br}(D^{0} \to \mu \mu) \gtrsim 10^{-12}$, about an order of
    magnitude larger than the SM estimate from ref.~\cite{Burdman:2001tf}. We note that our
    calculation is not valid below the dashed orange line since it only
    represents the new-physics contribution. (c) The plot shows the influence of
    the ATLAS $ss \to \mu\mu$ limits on the parameter space of our model.
    Coloured points lie in the $2\sigma$ region of the $b \to s$ fit we use.
    Dark blue points cannot explain the $b \to s$ data. The dashed orange line
    corresponds to the current ATLAS limit, while the solid red line is the
    $3000 \text{ fb}^{-1}$ projection. The abrupt absence of points in the
    bottom left of the plot is due to the constraint $D^{0} \to \mu \mu$.}
  \label{fig:otherscan1results}
\end{figure}                 
As discussed at length in section~\ref{sec:leptonic}, the leptoquark $\varphi$
mediates highly-constraining processes of muon--electron conversion in nuclei at
tree-level, and the couplings involved are directly related to those that
explain the neutrino masses and the $b \to s$ anomalies. 

We find that only about
$12\%$ of the points in our numerical scan are rejected on the basis of a
constraint, but from among these almost all are disallowed because they violate
the muon--electron conversion bound given in table~\ref{table:muonelectron}. In
figure~\ref{fig:m2ecVSc9} we present the results of our random scan with
slices through the parameter space and various projections. We find that our
model requires muon--electron conversion in Gold nuclei at a rate no less than
$2 \cdot 10^{-13}$ to accommodate the preferred value of $C_{9}^{\mu\mu}$.
The COMET~\cite{KURUP201138, Cui:2009zz, Wu:2017zwh, Adamov:2018vin} and Mu2e~\cite{Bartoszek:2014mya, Pezzullo:2018fzp, Bonventre:2019grv} experiments have projected sensitivities of $\text{Br}(\mu \text{Al} \to e\text{Al}) \lesssim 10^{-16}$ at $90\%$ confidence\footnote{Although COMET and Mu2e will measure muon--electron conversion in Aluminium, we nevertheless display the result on the same plot since we find that the calculations in Gold and Aluminium differ by less than an order of magnitude.}. These will provide an improvement on the current limit by four orders of magnitude, and will test and potentially falsify this scenario.  Interestingly, our model cannot simultaneously
avoid the muon--electron conversion bound and explain the $b \to s$ anomalies
with a vanishing Majorana phase $\alpha_{2}$, a result made clear in
figure~\ref{fig:subfig1}. There, it is also apparent that the constraint can be
avoided for $100^{\circ} \lesssim \alpha_{2} \lesssim 300^{\circ}$, a region
that overlaps with that shown in figure~\ref{fig:c9rat}, needed for a small
imaginary part for the electron couplings $x^{L\varphi}_{13}$. We find that an
additional two constraints cut into the parameter space significantly: bounds
from $D^{0} \to \mu \mu$ and the ATLAS measurement of $ss \to \mu\mu$. Our model
predicts the $D^{0} \to \mu \mu$ rate to be an order of magnitude larger than
estimates of the SM contribution $\text{Br}(D^0 \to \mu \mu)_{\text{SM}} \sim 3 \cdot 10^{-13}$~\cite{Burdman:2001tf} (see figure~\ref{fig:subfig2}), while the ATLAS $3000 \text{ fb}^{-1}$
projected limit from $ss \to \mu\mu$ indicates that a non-observation would
almost entirely rule out the model for low LQ masses (see figure~\ref{fig:subfig3}).

\subsection{Comments on explaining \texorpdfstring{$R_{D^{(*)}}$}{RD(*)} with the vector operator}
\label{sec:vectorOp}

In our analysis above we consider only contributions in the scalar--tensor direction to explain the charged-current anomalies in $R_D$ and $R_{D^*}$, necessitating the inclusion of $\phi$ in this model to generate these contributions. This choice is made to avoid the dangerous contributions to $B \to K^{(*)} \nu \nu$, which necessarily exist in the presence of a large $C_{V_L}$. We explored two ways these constraints could be avoided in the context of our model:
\begin{enumerate}
    \item As discussed previously in the literature~\cite{Cai:2017wry,Deshpand:2016cpw, Altmannshofer:2017poe}, one way to avoid the constraints from $B \to K^{(*)}\nu\nu$ and $B_s$--$\bar{B}_s$ mixing is to explain the $R_{D^{(*)}}$ anomalies with a large $x_{33}^{L\phi}$ while ensuring $x_{32}^{L\phi} \approx 0$. The coupling $y_{32}^{L\phi}$ required to explain $R_{D^{(*)}}$ is generated through eq.~\ref{eq:ckmmixing}, while keeping the strange-quark coupling to the neutrinos zero. Combining eq.~\ref{eq:ckmmixing} and eq.~\ref{eq:cvl} with $x_{33}^{L\phi} \gg x_{33}^{L\varphi}$ gives
    \begin{equation}
    C_{V_L} = \frac{\cos \theta_L}{4\sqrt{2} G_F V_{cb}} \frac{|x_{33}^{L\phi}|^2 V_{ts}}{m_\phi^2},
    \end{equation}
    which implies $1.7 \lesssim |x_{33}^{L\phi}| / (m_\phi \text{ TeV}) \lesssim 7.2$ for $\cos \theta_L \approx 1$ to explain $R_{D^{(*)}}$ according to our fit to $C_{V_L}$ (see table~\ref{tab:fitresults}). We note here that even saturating the lower $2\sigma$ bound on $C_{V_L}$ leads to contributions to $Z \to \tau \tau$ that disagree with experiment, despite the reduction of the global average driven by the latest Belle result.
    
    \item  Ref.~\cite{Crivellin:2017zlb} proposed that the $R_{D^{(*)}}$ anomalies could be explained through the vector operator by considering a cancellation between the $\phi$ and $\varphi$ particles in this model to the processes $B \to K^{(*)}\nu\nu$. This was further studied in ref.~\cite{Buttazzo:2017ixm} and ref.~\cite{Marzocca:2018wcf}.
    We have investigated this suggestion in considerable detail for this model, and could not find any parameter space that could simultaneously resolve the $R_{D^{(*)}}$ anomalies and be consistent with constraints from $B_s$--$\bar{B}_s$ mixing.  Our findings are in agreement with ref.~\cite{Marzocca:2018wcf}. We note that ref.~\cite{Marzocca:2018wcf} proposed some lines of investigation, such as the use of complex-valued couplings constants, that could potentially alter this conclusion, but an investigation of such a scenario is beyond the scope of this paper.
\end{enumerate}

\section{Conclusions}
\label{sec:con}
In this work, we have established a near-minimal scalar leptoquark model capable of producing significant flavour-specific BSM effects, and radiatively generating Majorana neutrino masses. Combining two existing completions of D7 $\Delta L=2$ effective operators, this model consists of two scalar LQs, $\varphi \sim (\mathbf{3},\mathbf{3},-1/3) $ and $\phi \sim (\mathbf{3},\mathbf{1},-1/3) $, and the vector-like quark doublet $\chi \sim (\mathbf{3},\mathbf{2},-5/6) $. We developed the structure of this model, including mixing between the vector-like exotic $\chi$ and the SM $b$-quark, and explored the significance of this effect for one-loop generation of radiative neutrino masses in two distinct phenomenological regimes: Regime 1 ($\varphi$ contribution) and Regime 2 ($\phi$ contribution). The tree-level contributions of each of these scalar LQs to the anomalous processes $b \to c \tau \nu$ and $b \to s \mu \mu$ were established, within the context of an EFT framework.

We then discussed the experimental constraints on this model for each mass-generation regime. Regime 2 was found to be significantly constrained by a combination of $\mu-e$ conversion in nuclei and the required contribution to ameliorate anomalies in $b \to c \tau \nu$ processes. Meanwhile, it was found that Regime 1 is capable of explaining neutrino masses, charged- and neutral-current anomalies, $(g-2)_\mu$ and the minor deviation in $Z \to bb $, whilst also avoiding other notable constraints. Therefore, we chose to concentrate our analysis on Regime 1, where $\varphi$ was primarily involved in neutrino mass generation and the couplings of $\phi$ were free to accommodate anomalies in $b \to c \tau \nu$ processes.

In order to avoid strong constraints from lepton-flavor violating processes, whilst also accommodating the neutral-current anomalies via tree-level contributions from the triplet $\varphi$, we needed to float the  Majorana phase $\alpha_2$. To avoid the constraint from $\mu-e$ conversion in nuclei, this model showed an interesting preference for the region in which the generated Yukawa couplings were mostly real-valued, with $100^{\circ} \lesssim \alpha_{2} \lesssim 300^{\circ}$.  

The established scenario was found to be highly predictive and extremely testable. For $\mu-e$ conversion in nuclei, we predict a rate of no less than $2 \times 10^{-13}$, a regime testable by the projected sensitivity of the COMET and Mu2e experiments. We also predict rates of $D^0 \to \mu \mu$ an order-of-magnitude larger than current estimates of the SM contribution. This model is also testable at the LHC, where dimuon searches at high $p_T$ provide strong limits. Additionally, future measurements of the asymmetry observable $P_\tau$ by Belle II can test this model, which prefers $P_\tau \approx 0.4$, assuming the central values of $f_L^{D^*}$ and $R_{D^{(*)}}$ remain constant.

\acknowledgments
 This work was supported in part by the Australian Research Council and the Australian Government Research Training Program Scholarship initiative. We also acknowledge useful discussions with Joshua P Ellis, Iulia Popa-Mateiu, Michael A Schmidt, and Phillip Urquijo. IB would also like to thank F. Staub and A. Vicente for their support on the \texttt{SARAH} forum, and Denner et al. for ref.~\cite{Denner:1992vza} which was useful for analytic calculations throughout.

\appendix

\section{Structure of the Scalar potential: \texorpdfstring{$\phi$-$\varphi$}{phivarphi} mixing}
\label{appendix1}
The scalar potential contains the following terms:
\begin{align}\mathcal{V} \supset \mu_\varphi^2 |\varphi|^2+  \mu_\phi^2 |\phi|^2 +\lambda_{\rm mix}\, \phi \varphi^\dagger (H \tilde H)_\mathbf{3}.\label{potential} \end{align}
Of particular importance is the final term which induces $\phi$-$\varphi$ mixing. We analyse this in the following section. Furthermore, whilst the tree-level masses for each component of $\varphi$ are equal, higher-order corrections will break the degeneracy with a pattern driven by the relevant couplings strengths. For example, the $\varphi_2$ component couples less strongly to the photon than the $\varphi_3$ component, so the associated radiative correction to these masses will push $m_{\varphi_3}>m_{\varphi_2}$. However, since these higher-order effects are loop-suppressed, we assume a degenerate mass spectrum for $\varphi$: $m_{\varphi_1}=m_{\varphi_2}=m_{\varphi_3}=m_{\varphi}.$

\subsection*{Parameterising Leptoquark Mixing}

To analyse the structure of the $\phi$-$\varphi$ LQ mixing driven by \eqref{potential}, we construct the isotriplet combination $(H \tilde H)_{\mathbf{3}}$ from
 \begin{align}
 H \sim \binom{H^+}{ H^0}, \hspace{2cm}
 \tilde H \sim \binom{H^{0*}}{ -H^-}.
 \end{align}
and insert it into the third term in \eqref{potential} to obtain
\begin{align}
 \mathcal{L} \supset \lambda_{\rm mix}\, \phi \big(\varphi_1^*H^+ H^{0*}+\frac{ \varphi_2^*}{\sqrt{2}}(H^0 H^{0*}- H^+ H^-)-\varphi_3^* H^0 H^-\big).\label{higgsmix}
 \end{align}
After EWSB, the term proportional to $|H_0|^2$ in eq.~\eqref{higgsmix} generates mass-mixing between $\phi$ and $\varphi_2$. We may parameterise this through:
 \begin{equation}
 \begin{pmatrix} \phi \\ \varphi_2 \end{pmatrix}= \mathbf{R}  \begin{pmatrix} \eta_1\\ \eta_2 \end{pmatrix} \equiv
\begin{pmatrix}  \cos\theta_m & -\sin\theta_m \\  \sin\theta_m & \cos\theta_m \end{pmatrix}
  \begin{pmatrix} \eta_1\\ \eta_2 \end{pmatrix}, 
 \end{equation}
where $\theta_m$ is the scalar mixing angle, $\mathbf{R}$ is the rotation matrix, and $\eta_{1}$, $ \eta_2$ are the mass eigenstates. In order to derive their masses, we begin by re-expressing the relevant Lagrangian terms in matrix form:
\begin{align}
\mathcal{L} \supset \;\;
\begin{pmatrix}
\phi^\dagger  \varphi_2^{\dagger}
\end{pmatrix}
\mathbf{M_s^2}
\begin{pmatrix}
\phi\\
\varphi_2
\end{pmatrix}
&\equiv
\begin{pmatrix}
\phi^\dagger & \varphi_2^{\dagger}
\end{pmatrix}
 \begin{pmatrix}
\mu_\phi^2 & \kappa\\
\kappa^* & \mu_\varphi^2\\
\end{pmatrix}
\begin{pmatrix}
\phi\\
\varphi_2
\end{pmatrix}=
 \begin{pmatrix}
\eta^\dagger_1 &  \eta^\dagger_2
\end{pmatrix} R M_s^2 R^\dagger 
\begin{pmatrix}
\eta_1\\
\eta_2
\end{pmatrix},
\end{align}
\normalsize
\begin{align}
\kappa\equiv \frac{\lambda_{mix} \langle H_0 \rangle^2}{\sqrt{2}}.
\end{align}
Requiring $R M_s^2 R^\dagger$ to be diagonal gives the following mixing parameters (here we have assumed, for simplicity, that the parameter $\kappa$ is real-valued):
\begin{align}
&\tan{2\theta_m}= \frac{2\kappa}{\mu_\varphi^2 - \mu_\phi^2},\label{MixingLQ}
& m_{\eta_{1/2}}^2= \frac{1}{2} \left( \mu_\phi^2+ \mu_\varphi^2\mp |\mu_\phi^2-\mu_\varphi^2 | \sqrt{1+\left(\frac{2\kappa}{\mu_\phi^2 -\mu_\varphi^2}\right)^2} \right).
\end{align}
\normalsize
Nevertheless, although this mixing can produce additional contribution to observable effects, the mass-insertion is likely to render this sub-dominant to contributions that are independent of $\phi$-$\varphi$ mixing.

\section{Fierz transforms and the \texorpdfstring{$C$}{C}-operator}
\label{appendix2}
The following field rearrangements will be useful in calculations throughout this work. The general Fierz transform for anti-commuting fields~\cite{Nishi:2004st} is:
\begin{equation}
(\Gamma^A)[\Gamma^B]= -\frac{1}{4} Tr[\Gamma^A\Gamma_C \Gamma^B \Gamma_D](\Gamma^D][\Gamma^C),
\end{equation}
where here each bracket represents the respective fields that sandwich the operators:
\begin{equation}
\{\Gamma_A\}= \{P_R, P_L, P_L \gamma_\mu, P_R \gamma_\mu, \frac{1}{2}\sigma_{\mu \nu}\}, \hspace{1cm} \{\Gamma^A\}= \{P_R, P_L, P_R \gamma^\mu, P_L \gamma^\mu, \sigma^{\mu \nu}\}.
\end{equation}
Useful chiral Fierz transforms for this work include:
\small
\begin{align}
(P_L)[P_R]&= -\frac{1}{2}(P_L \gamma_\mu][P_R \gamma^\mu) \\
(P_R)[P_L]&=-\frac{1}{2}(P_R \gamma_\mu][P_L \gamma^\mu) \\
(P_R)[P_R]&=-\frac{1}{2}(P_R][P_R)-\frac{1}{4}(\sigma_{\mu\nu}][\sigma^{\mu\nu})-\frac{i}{8} \epsilon_{\mu \nu \alpha \beta}(\sigma^{\alpha \beta}][\sigma^{\mu\nu}) =-\frac{1}{2}(P_R][P_R)+\frac{1}{8}(P_R\sigma_{\mu\nu}][P_R\sigma^{\mu\nu})\\
(P_L)[P_L]&=-\frac{1}{2}(P_L][P_L)- \frac{1}{4}(\sigma_{\mu\nu}][\sigma^{\mu\nu})+\frac{ i }{8}\epsilon_{\mu \nu \alpha \beta}(\sigma^{\alpha\beta}][\sigma^{\mu\nu})=-\frac{1}{2}(P_L][P_L)- \frac{1}{8}(P_L\sigma_{\mu\nu}][P_L\sigma^{\mu\nu})
\end{align}
\normalsize
We will also employ the actions of charge conjugation on certain operators, as outlined below:

\begin{align}
[\overline{\nu^C_3}P_L \gamma_\mu \ell_3^C]= -[\overline{\ell_3}P_R \gamma_\mu \nu_3], \hspace{0.5cm}
[\overline{\nu^C_3}P_L\ell_3^C]=[\overline{\ell_3}P_L\nu_3],\hspace{0.5cm}
[\overline{\nu^C_3}P_L\sigma_{\mu\nu}\ell_3^C]=-[\overline{\ell_3}P_L\sigma_{\mu\nu}\nu_3].
\end{align}

\bibliography{manuscript}

\end{document}

%% file: tikzdefs.tex
\usepackage{tikz}

\usetikzlibrary{arrows,calc,graphs,patterns,positioning}
\usetikzlibrary{decorations,decorations.markings,decorations.pathmorphing,decorations.pathreplacing}
\usetikzlibrary{shapes, shapes.geometric}

\tikzset{/pgf/decoration/.cd,
    number of sines/.initial=5,
    angle step/.initial=20,
}
\newdimen\tmpdimen

\pgfdeclaredecoration{complete sines}{initial}
{
    \state{initial}[
        width=+0pt,
        next state=move,
        persistent precomputation={
            \pgfmathparse{\pgfkeysvalueof{/pgf/decoration/angle step}}%
            \let\anglestep=\pgfmathresult%
            \let\currentangle=\pgfmathresult%
            \pgfmathsetlengthmacro{\pointsperanglestep}%
                {(\pgfdecoratedremainingdistance/\pgfkeysvalueof{/pgf/decoration/number of sines})/360*\anglestep}%
        }] {}
    \state{move}[width=+\pointsperanglestep, next state=draw]{
        \pgfpathmoveto{\pgfpointorigin}
    }
    \state{draw}[width=+\pointsperanglestep, switch if less than=1.25*\pointsperanglestep to final, 
        persistent postcomputation={
        \pgfmathparse{mod(\currentangle+\anglestep, 360)}%
        \let\currentangle=\pgfmathresult%
    }]{%
        \pgfmathsin{+\currentangle}%
        \tmpdimen=\pgfdecorationsegmentamplitude%
        \tmpdimen=\pgfmathresult\tmpdimen%
        \divide\tmpdimen by2\relax%
        \pgfpathlineto{\pgfqpoint{0pt}{\tmpdimen}}%
    }
    \state{final}{
        \ifdim\pgfdecoratedremainingdistance>0pt\relax
            \pgfpathlineto{\pgfpointdecoratedpathlast}
        \fi
   }
}

\tikzset{
    scalar/.style={draw=black, thick, dashed},
    cscalar/.style={draw=black, postaction={decorate}, decoration={markings, mark=at position .55 with {\arrow[]{latex}}}, thick, dashed},
    fermion/.style={draw=black, postaction={decorate}, decoration={markings, mark=at position .55 with {\arrow[]{latex}}}, thick},
    antifermion/.style={draw=black, postaction={decorate}, decoration={markings, mark=at position .55 with {\arrow[>=latex]{<}}}, thick},
    majorana/.style={draw=black, thick},
    semiloop/.style={draw, thick, dashed, shape=semicircle, minimum size=1cm, outer sep=9pt, postaction={decorate}, decoration={markings, mark=between positions 0.32 and 1 step 1/2. with {\arrow[]{latex}}}},
    photon/.style={draw=none, outer ysep=3pt, postaction={draw=black, thick, decorate}, decoration={complete sines, amplitude=7pt}},
    photonloop/.style={draw=none, outer ysep=3pt, postaction={draw=black, thick, decorate}, decoration={complete sines, number of sines=15, amplitude=5pt}},
    gluon/.style={draw=none, outer ysep=3pt, postaction={draw=black, thick, decorate}, decoration={coil, amplitude=3pt, segment length=5pt}}, 
    composite/.style={draw=gray!70!, line width=5pt},
    vtx/.style={inner sep=0pt},  
    mass/.style={cross out, draw, minimum size=5pt, inner sep=0pt},
    ins/.style={draw, cross, shape=circle, minimum size=5pt, inner sep=0pt}, 
    cross/.style={path picture={\draw[black] (path picture bounding box.south east) -- (path picture bounding box.north west) (path picture bounding box.south west) -- (path picture bounding box.north east);}},
    vertex/.style={draw, shape=circle, fill=black, minimum size=3pt, inner sep=0pt},
    blob/.style={draw, shape=circle, preaction={fill,white}, pattern = north west lines, minimum size=15pt, inner sep=0pt},        
    loop/.style={shape=circle, minimum size=1.7cm, outer sep=9pt},
    site/.style={draw, shape=circle, minimum size=1.5cm, inner sep=0pt},
    brane/.style={trapezium, draw, trapezium stretches=true, minimum height=4cm, minimum width=3.5cm, inner sep=0, trapezium left angle=35, trapezium right angle=145, shape border uses incircle, shape border rotate=17.5},
    momentum/.style={->, dist/.store in=\segDistance, pos/.store in=\segPos, len/.store in=\segLength,
      to path={
      ($(\tikztostart)!\segPos!(\tikztotarget)!\segLength/2!(\tikztostart)!\segDistance!90:(\tikztotarget)$) -- 
      ($(\tikztostart)!\segPos!(\tikztotarget)!\segLength/2!(\tikztotarget)!\segDistance!-90:(\tikztostart)$)  \tikztonodes
      }, 
      pos=.5,
      len=7mm,
      dist=2mm
    },    
}
